\documentclass[11pt]{article}

\usepackage{amsmath,jheparxiv}

\usepackage{enumitem}

\usepackage[T1]{fontenc} 
\usepackage{hyperref}
\hypersetup{
    colorlinks,
    citecolor=black,
    filecolor=black,
    linkcolor=black,
    urlcolor=black
}
\usepackage[english]{babel}
\usepackage{tikz}
\usepackage{tikz-cd}
\usepackage{tensor}
\usepackage{amsfonts}
\usepackage{amsmath}
\usepackage{amssymb}
\usepackage{amsthm}
\usepackage{mathtools}
\usepackage{tensor}
\usepackage{slashed}
\usepackage{mathrsfs}
\usepackage{bm}
\usepackage{setspace}
\usepackage{amssymb}

\usepackage{bbm}
\usepackage{graphicx}
\usepackage{stmaryrd}
\usepackage{subcaption}
\usepackage{xcolor}
\usepackage{cancel}
\usepackage{bigints}

\usepackage{setspace}
\def\bea{\begin{eqnarray}}
\def\eea{\end{eqnarray}}
\newcommand{\beq}{\begin{eqnarray}}
\newcommand{\eqq}{\end{eqnarray}}
 \newcommand{\badat}{\begin{alignedat}}
 \newcommand{\eadat}{\end{alignedat}}

\newcommand{\eal}[1]{\be \begin{aligned} #1 \end{aligned}\end{equation}} 
\newcommand{\eqn}[1]{\be #1 \end{equation}} 
\newcommand{\eqa}[1]{\bea  #1\end{eqnarray}}
\newcommand{\PT}{\mathbb{PT}}

\newcommand{\cO}{\mathcal{O}}

\newcommand{\p}{\partial}

\newcommand{\scri}{\mathscr{I}}

\newcommand{\e}{\epsilon}

\newcommand{\bo}{\mathbf{o}}
\newcommand{\bi}{\boldsymbol{\iota}}

\long\def\new#1\endnew{{\bf #1}}		
\long\def\del#1\enddel{}

\def\del{\partial}

\usepackage{xcolor}
\definecolor{oldmauve}{rgb}{0.4, 0.19, 0.28}
\definecolor{pansypurple}{rgb}{0.47, 0.09, 0.29}
\definecolor{burgundy}{rgb}{0.5, 0.0, 0.13}
\definecolor{carminepink}{rgb}{0.92, 0.3, 0.26}
\definecolor{blue(pigment)}{rgb}{0.2, 0.2, 0.6}
\definecolor{darkseagreen}{rgb}{0.56, 0.74, 0.56}
\definecolor{darkspringgreen}{rgb}{0.09, 0.45, 0.27}
\definecolor{ceruleanblue}{rgb}{0.16, 0.32, 0.75}

\newcommand{\be}{\begin{eqnarray}}
\newcommand{\en}{\end{eqnarray}}

\newtheorem{propn}{Proposition}

\author{}
\numberwithin{equation}{section} 

\begin{document}

\begin{titlepage}
  \thispagestyle{empty}

 \begin{flushright}
 \end{flushright}


  \begin{center}  
{\LARGE\textbf{Quasi-Local Celestial Charges and Multipoles}}


\vskip1cm
Adam Kmec$^*$\footnote{\fontsize{8pt}{10pt}\selectfont\ \href{mailto:adam.kmec@maths.ox.ac.uk}{adam.kmec@maths.ox.ac.uk}},
Lionel Mason$^*$\footnote{\fontsize{8pt}{10pt}\selectfont\ \href{mailto:Lionel.Mason@maths.ox.ac.uk}{lionel.mason@maths.ox.ac.uk}},
Romain Ruzziconi$^\dagger$\footnote{\fontsize{8pt}{10pt}\selectfont\ \href{mailto:romainruzziconi@fas.harvard.edu}{romainruzziconi@fas.harvard.edu}}
\vskip0.5cm

\normalsize
\medskip

$^*$\textit{Mathematical Institute, University of Oxford, \\ Andrew Wiles Building, Radcliffe Observatory Quarter, \\
Woodstock Road, Oxford, OX2 6GG, UK}

$^{\dagger}$\textit{Center for the Fundamental Laws of Nature, Harvard University \\
17 Oxford Street, Cambridge, MA 02138, USA} \\

\vspace{2mm}

$^{\dagger}$\textit{Black Hole Initiative, Harvard University \\
20 Garden Street, Cambridge, MA 02138, USA}

\end{center}

\vskip0.5cm

\begin{abstract}

\begin{center}
    \textbf{Abstract} 
\end{center} \vspace{-0.4cm}
\noindent We extend Penrose’s quasi-local mass definition to include higher-spin charges associated with the celestial $Lw_{1+\infty}$ symmetries and relate them to traditional definitions of multipoles. The resulting formulae provide explicit expressions that can be computed on finite 2-surfaces, given a choice of null hypersurface. They yield a geometric definition of celestial symmetries and multipoles in generic spacetimes in terms of higher-valence solutions to the twistor equations. This, in turn, gives rise to natural flux-balance laws along the null hypersurface. We also present a first-principles phase-space derivation of these charges, starting from a twistor space action for self-dual gravity that can be identified with the standard gravitational asymptotic phase space at null infinity, performing a large gauge transformation analysis and using the Penrose transform to connect with the corresponding spacetime expressions. Finally, we formulate the spacetime analysis in the Plebanski gauge and relate the celestial symmetries to the integrability of self-dual gravity in the case of a self-dual background.
\end{abstract}

\end{titlepage}

\setcounter{tocdepth}{2}

\tableofcontents

\section{Introduction}

Definitions of local stress-energy  and multipole expansions more generally form the basis of our understanding of physical theories.   In general relativity  one should  incorporate the local gravitational contributions also because energy is carried by gravitational  waves.  However such definitions, as in \cite{Einstein:1916cd,Landau:1975pou}, are notoriously coordinate-dependent and violate general covariance. Nevertheless, because  in the Hamiltonian formalism  diffeomorphisms are a gauge symmetry generated by the (vanishing) constraint equations of general relativity, the total Hamiltonian for gravity plus matter is an exact 3-form so that a quasi-local notion of energy-momentum defined inside a 2-surface should depend only on choices made at the bounding 2-surfaces.
There are two challenges to a Noether or Hamiltonian argument. One is to define the putative symmetries for the Hamiltonian to generate (i.e., to feed into Noether’s theorem), and the second is to characterize the appropriate boundary term for the Hamiltonian; see, for example \cite{Dirac:1958jc,Dirac:1959qwq,Arnowitt:1959ah,Regge:1974zd} for discussion. These issues were sidestepped by Penrose \cite{Penrose:1982wp,penrose1984new,penrose1985suggested,penrose1984spinors}, who used 2-surface twistors to reduce the spin of the gravitational field to one, so that a normal Maxwell charge integral could be used. However, the Penrose definition can be given Hamiltonian and Noether interpretations as in \cite{Mason_1989,Mason:1990Fra}. There it was shown that the appropriate boundary term for the gravitational Hamiltonian arises from the so-called Sen–Witten identity underlying Witten’s positive energy proof \cite{Witten:1981mf}, and that the 2-surface concept of symmetries should be given by solutions to the 2-surface twistor equation;  this can in turn be related to conformal embeddings of the intrinsic and extrinsic data of the 2-surface into Minkowski space. Different definitions, both from the action and from other considerations, were subsequently given in \cite{Bartnik:1989zz,Dougan:1991zz,Brown:1992br,Kijowski1997,Liu:2003bx,Wang:2008jy}; see \cite{Szabados:2009eka} for a review.
The covariant phase space formulation of \cite{Iyer:1994ys,Wald:1999wa,Barnich:2001jy,Barnich:2003xg} was developed to construct surface charges in diffeomorphism-invariant theories more generally but faces similar issues.

Recently, new symmetries have been uncovered via studies of the collinear sector of gravity amplitudes within the framework of celestial holography \cite{Guevara:2021abz,Strominger:2021mtt}. They form the so-called celestial $Lw_{1+\infty}$ symmetry algebra,\footnote{\label{celestial algebra name} More precisely, the celestial algebra for gravity is given by $LHam(\mathbb{C}^2)$, the loop algebra of the canonical transformations on $\mathbb{C}^2$ equipped with its standard holomorphic Poisson bracket. This celestial algebra is often commonly referred to as the $Lw_{1+\infty}$ algebra, and we will use both terminologies in this paper.  
} whose commutators are given by
    \begin{equation} \label{algebraIntro}
     \{w^p_{m,a} , w^q_{n,b}\} = 2(m(q-1) - n(p-1))w^{p+q-2}_{m+n,a+b} ,
\end{equation} where the range of indices for the generators $w^p_{m,a}$ is $p= 1, \frac{3}{2}, 2, \frac{5}{2}, \ldots$, $m\pm p \in \mathbb{Z}$, $a \in \mathbb{Z}$ and $|m|< p$. These symmetries admit a natural interpretation on twistor space in the self-dual (SD) sector \cite{ Adamo:2021lrv}, where they arise as the local holomorphic diffeomorphism symmetries of Penrose's non-linear graviton construction \cite{Penrose:1976jq,Penrose:1976js}. The corresponding higher charge integrals and flux-balance laws have been constructed at null infinity \cite{Freidel:2021ytz,Geiller:2024bgf} and further developed in \cite{Cresto:2024fhd,Cresto:2024mne} (see \cite{Cresto:2025fbc} for a comprehensive review), using the Ashtekar–Streubel radiative phase space \cite{Ashtekar:1981bq} and hierarchies of recursion relations. Their derivation from first principles was presented in \cite{Kmec:2024nmu}, employing phase space methods on the twistor space action for self-dual gravity \cite{Mason:2007ct} and translating the result into spacetime data.

While the $Lw_{1+\infty}$ charges have been mostly discussed at null infinity, the associated symmetries are believed to extend to the full self-dual sector of gravity and perhaps beyond to the full theory. One would therefore expect to be able to compute the corresponding charges on any bulk 2-surface. A first construction was proposed in \cite{Ruzziconi:2025fct, Ruzziconi:2025fuy} to evaluate these charges on a finite-distance null hypersurface, such as a black hole horizon. This heuristic construction  imports the arguments of \cite{Freidel:2021ytz,Geiller:2024bgf} to finite distance by connecting to null infinity using a null hypersurface in the bulk.  However, the spacetime interpretation of the underlying symmetries in terms of spacetime diffeomorphisms remains unclear.

Interestingly, some of the structures associated with the celestial charges have been argued to survive beyond the self-dual sector and should play a very important role in gravitational-wave physics. In \cite{Compere:2022zdz}, the Geroch–Hansen multipole moments \cite{Geroch:1970cc,Geroch:1970cd,Hansen:1974zz}, which appear in the metric expansion at null infinity \cite{Blanchet:1985sp,Blanchet:1986dk,Blanchet:1987wq,Blanchet:1992br,Blanchet:2004re,Blanchet:2020ngx}, were related to the $Lw_{1+\infty}$ charge aspects. Furthermore, the multipole moments were given a Hamiltonian interpretation in \cite{Compere:2017wrj} in terms of overleading diffeomorphisms in harmonic gauge, reproducing Thorne’s mass and current multipoles \cite{Thorne:1980ru}, which have been related to the Geroch–Hansen definition in the stationary case \cite{Gursel:1983nkl}.
Our analysis below is closely related to an early and intriguing connection between the higher-valence twistor equation and multipole moments, introduced by Curtis in \cite{Curtis:1978}, and naturally extends to the time-dependent case.

The goal of the present paper is threefold:

 $(i)$ First, we clarify the geometric interpretation of the celestial symmetries on spacetime. We show that they are naturally expressed in terms of conformal Killing spinors satisfying higher-spin twistor equations 
\begin{equation} \label{IntroTwistorEquation}
\nabla^{(\alpha_0}_{\dot\alpha}\xi^{\alpha_1\cdots\alpha_{s+1})} = 0  ,  \quad \xi^{\alpha_1\cdots\alpha_{s+1}} \sim O\left(x^{s+1}\right) , \quad s = -1, 0, 1, 2, \ldots 
\end{equation} on a bulk null hypersurface. These spinors are overleading with respect to the asymptotic expansion. They can be related to the twistor representatives satisfying the $Lw_{1+\infty}$ algebra \eqref{algebraIntro} via the Penrose transform, with $s=2p - 3$. The higher-valence spinors $\xi$ are not generally
identified with components of spacetime vectors
satisfying the ordinary asymptotic Killing equation.
The $s=0,1$ examples reduce to
residual spacetime diffeomorphisms, while higher-spin
transformations generally involve non-local flows
of the self-dual integrable hierarchy.

$(ii)$ Second, we propose a quasi-local spacetime charge expression for $Lw_{1+\infty}$ symmetries on any 2-surface $\mathcal{S}$ embedded in a null hypersurface:
\begin{equation}
     H_s = \oint_{\mathcal{S}} \phi_{\alpha_1 \cdots \alpha_{s+1}\beta\gamma}\xi^{\alpha_1\cdots \alpha_{s+1}}\Sigma^{\beta\gamma}\, ,\quad  \nabla^{\alpha_1}_{\,\,\,\,\dot \alpha}\phi_{\alpha_1\cdots\alpha_{s+1}\beta\gamma} = 0\, ,
\end{equation} where $\phi$ is a zero-rest-mass field built out of the gravitational data. These expressions naturally extend the original construction of the Penrose quasi-local mass \cite{Penrose:1982wp} and follow directly from the application of covariant phase space methods on the self-dual twistor space action and their translation to spacetime. In particular, in the spirit of \cite{Mason:2007ct}, this provides a Hamiltonian derivation of all the $Lw_{1+\infty}$ quasi-local charges. As mentioned above, the generic higher-spin
transformations are not identified with ordinary
spacetime diffeomorphisms, so the standard
construction of diffeomorphism charges does not
directly apply. An independent spacetime derivation would require
a Hamiltonian treatment of the non-local symmetries
of the self-dual integrable hierarchy, which is not provided here. In general, the definition of the higher-spin ingredients is most naturally understood in terms of solving equations on a null hypersurface connecting $\mathcal{S}$ to null infinity, and matching to definitions at $\scri$. It is therefore semi-local rather than quasi-local, in the sense that it requires knowledge of the entire null hypersurface and $\scri$, rather than data specified solely on $\mathcal{S}$. Nevertheless, tentative quasi-local definitions can still be formulated. 

$(iii)$  Third, in the linearized theory, using \cite{Curtis:1978}, we relate our quasi-local charge expressions to the Geroch–Hansen definition of multipoles \cite{Geroch:1970cc,Geroch:1970cd,Hansen:1974zz}, thereby providing a concrete physical interpretation of these celestial charges. Our definition also extends to the full theory of general relativity, when restricted to a null hypersurface. However, their Hamiltonian interpretation in terms of celestial symmetries only holds at $\scri$ or more generally in the self-dual theory.

The rest of the paper is organized as follows. In Section \ref{sec:Penrose's quasi-local mass}, we review the Penrose quasi-local mass construction and its Hamiltonian interpretation in terms of the Witten–Nester $2$-form. The definition extends to higher spin, and in Section \ref{sec:Space-time definitions of celestial charges and multipole moments} we first connect this spacetime expression to the definition of multipole moments for linearized fields, as introduced by Geroch and Hansen for stationary fields and further developed by Curtis. These definitions are also shown to give the $Lw_{1+\infty}$ celestial charges at $\scri$. We then show that our definition naturally extends to the nonlinear theory in spacetime on a chosen null hypersurface. In Section \ref{sec:Celestial charges from twistor space}, we first give the multipole fields in linear theory in twistor space. We go on to review the construction of the twistor action for the self-dual sector of Einstein gravity and derive the general charge algebra associated with large gauge symmetries on twistor space. We explain how the celestial $Lw_{1+\infty}$ algebra emerges as large gauge transformations in this setting. In Section \ref{sec:Quasi-local higher-spin charges}, we translate our results to spacetime and find perfect agreement with the heuristic higher-spin extension of the Penrose quasi-local mass formula obtained on a general space-time earlier. We show that our results connect directly to earlier formulations both at null infinity and on a null hypersurface at finite distance. Finally, in Section \ref{sec:Celestial symmetries of Plebanski's equations}, we present the results in the Plebański gauge on self-dual spacetimes and provide further interpretation of the celestial symmetries. In Appendix \ref{sec:background}, we display our spinor and Newman-Penrose conventions. In Appendix \ref{sec:Proof of Algebra}, we present a proof of the celestial $Lw_{1+\infty}$ algebra on spacetime in Plebański gauge. In Appendix \ref{sec: Linear SD Perturbation} we further discuss the asymptotic growth of the SD perturbation that gives rise to the twistor \ref{IntroTwistorEquation} in the linearised case.

\section{Penrose's quasi-local mass and higher-spin extension}
\label{sec:Penrose's quasi-local mass}

In this section, we introduce  Penrose's quasi-local mass and its Hamiltonian origins  which will serve as the basis of our discussion. We then propose a higher-spin extension of that formula using heuristic arguments. In Appendix \ref{sec:background} we define and review the spinor notations and conventions used here and elsewhere.

\subsection{Penrose's quasi-local mass formula}
\label{sec:Penrose's Quasi-Local Mass Formula}

Penrose defines the quasi-local mass associated with a spacetime region bounded by the compact space-like $2$-surface $\mathcal{S}$ to be \cite{Penrose:1982wp,penrose1984new,penrose1985suggested,penrose1984spinors}:
\begin{equation}\label{PenroseCharge}
    H[f] = \oint_\mathcal{S} *R_{a b} f^{ab}.
\end{equation}
Here  $*$ is the Hodge star, and $R_{ab}$ is the Riemann curvature 2-form, which can be expanded in a co-tetrad basis $\theta^a$ and further decomposed into anti-self-dual (ASD) and self-dual (SD) parts 
\begin{equation}
\begin{split}
     R_{ab} &:= R_{abcd}\,\theta^c \wedge \theta^d
     = R_{\alpha\beta}\epsilon_{\dot\alpha\dot\beta}+\tilde R_{\dot\alpha\dot\beta}\epsilon_{\alpha\beta} .
\end{split}
\end{equation}
The charge $H[f]$ is conserved if the integrated 2-form is closed
\begin{equation}
    d\left(*R_{ab} f^{ab}\right) = 0\, .
\end{equation}
This occurs when the spacetime satisfies the vacuum Einstein equations and $f^{ab}$ satisfies the conformal Killing-Yano equation
\begin{equation}\label{CKY}
    \nabla_a f_{bc} = \nabla_{[a}f_{bc]} -\frac{2}{3}g_{a[b} \nabla^d f_{c] d} .
\end{equation}
In terms of  spinors, we can decompose $f^{ab}$ as a sum of an ASD and SD part
\begin{equation}
    f^{ab} = \xi^{\alpha\beta}\epsilon^{\dot \alpha\dot \beta} + \tilde \xi^{\dot \alpha\dot \beta}\epsilon^{\alpha\beta}\, ,
\end{equation}
using the antisymmetry of $f^{ab}$. With Minkowski reality conditions,  $\tilde \xi_{\dot \alpha\dot\beta} = \bar\xi_{\dot \alpha\dot\beta}$, but for now, we keep the discussion complex. The equation \eqref{CKY} on $f^{ab}$ becomes  the valence-2 twistor equations on $\xi^{\alpha \beta}$ and $\tilde\xi^{\dot \alpha\dot \beta}$
\begin{equation}\label{TwistorEquations}
    \nabla^{(\alpha}_{\,\,\,\dot \alpha}\xi^{\beta\gamma)} = 0 , \qquad \nabla_\alpha^{\,\,\,(\dot \alpha}\tilde \xi^{\dot \beta\dot \gamma)} = 0 .
\end{equation} 
We can ignore the $\tilde \xi^{\dot \alpha\dot \beta}$ term  and take the real part later if desired.\footnote{In fact the imaginary part will often automatically vanish \cite{Penrose:1986uia} chapter 6.} If we decompose the Riemann curvature 2-form using spinors and assume Ricci flatness  we can write 
$$
R_{\alpha\beta}= \psi_{\alpha\beta\gamma\delta }\Sigma^{\gamma\delta}, \qquad \Sigma^{\alpha\beta}:=\theta^{(\alpha}_{\dot\alpha} \wedge \theta^{\beta)\dot\alpha},
$$ 
in terms of the ASD Weyl spinor $\psi_{\alpha\beta\gamma\delta}$ which satisfies the vacuum Bianchi identity
\begin{equation}\label{Bianchi Psi}
    \nabla^{\dot\alpha \alpha} \psi_{\alpha\beta\gamma\delta} =0 .
\end{equation} 
The charge \eqref{PenroseCharge} becomes the real part of
\begin{equation}\label{Weyl Charge}
    H[\xi] = \oint_\mathcal{S}\psi_{\alpha\beta\gamma\delta}\xi^{\alpha\beta}\Sigma^{\gamma\delta}
\end{equation}
which can now be seen to be conserved  as a consequence of \eqref{TwistorEquations} and \eqref{Bianchi Psi}.
Such  conserved charges correspond to  solutions to the twistor equation. On a general curved background, solutions $\xi^{\alpha\beta}$ and in turn $f^{ab}$ do not exist in much the same way that Killing vectors generally do not exist. Indeed  a solution to the Killing-Yano equation, one can construct a complex Killing vector $K^a$ by defining 
\begin{equation}
   K^{\dot \alpha(\alpha}\epsilon^{\beta)\gamma}= \nabla^{\dot \alpha\gamma}\xi^{\alpha\beta}  \, ,\qquad   \Im K_a:=\epsilon_{abcd}\nabla^b f^{cd}\, , \qquad \Re K^a=\nabla_b f^{ab}\, ,
\end{equation}
and a calculation shows that $K_a$ satisfies Killing's equations as  a consequence of \eqref{TwistorEquations}, see \S6.4 \cite{penrose1984spinors}. In the linearized case, there are, in general, 20 solutions to \eqref{CKY}, while only half of these give non-vanishing Killing vectors that generate the real Poincaré group. 

Following \cite{Penrose:1982wp}, any solution to \eqref{TwistorEquations} can be decomposed as a symmetric product of solutions to the valence $1$ twistor equation
\begin{equation} \label{twistor 1 equation}
    \nabla^{(\alpha}_{\,\,\,\dot \alpha}\xi^{\beta)} = 0  \quad \text{with}\,\,\,\ \xi^{\alpha} = \xi_0 \,\iota^{\alpha} + \xi_1 o^{\alpha} .
\end{equation}
There are six equations for only two functions $\xi_0$ and $\xi_1$, resulting in an overdetermined system. We can consider instead the twistor equation involving only derivatives tangential to $\mathcal{S}$ given by
\begin{equation}
    \eth_{\mathcal{C}} \xi_1 + \xi_0 \lambda = 0, \qquad \bar{\eth}_{\mathcal{C}}\xi_0 - \xi_1 \sigma = 0 .\label{2-surface-twistor}
\end{equation}
Using the Atiyah-Singer index theorem, if $\mathcal{S}$ is topologically a sphere, there exist at least four independent solutions and generically just four. Having established existence, the space of all solutions to these equations is called the 2-surface twistor space. However, due to the curvature of spacetime, the 2-surface twistor equations are incompatible with the rest of the valence one twistor equation on a general background, resulting in the charges no longer being conserved. Non-conserved charges are a general feature of curved spacetimes. In fact, at null infinity, the non-conservation of BMS charges is a result of gravitational radiation. 

The alternative to the definition \eqref{2-surface-twistor} is that exploited in \cite{Dougan:1991zz,Dougan_1992}.  This takes holomorphic or anti-holomorphic local twistors on the 2-surface.  In vacuum the holomorphic case leads to the equations
\begin{align}
    \bar\eth \xi^\alpha + io^\alpha\iota^{\dot\alpha} \chi_{\dot\alpha}=0\, , \qquad \bar \eth \chi_{\dot\alpha}=0\, . \label{2-surface-twistor-DM}
\end{align}
This  definition yields the Bondi definitions on large 2-surfaces going out to future null infinity.  The anti-holomorphic version diverges in this limit when there is radiation,  but yields the Bondi definitions as spheres are taken out to $\scri^-$.  Both definitions lead to well-defined mass, energy-momentum and angular momentum in general, with a positivity theorem for the energy.  These issues remain open for the Penrose definition.

Putting these ingredients together, the quasi-local charge becomes
\begin{equation}
    H[\xi] =\oint_{\mathcal{S}} R_{\alpha\beta}\xi^\alpha\xi^{\prime\beta}= \oint_{\mathcal{S}}dS \left(\psi_3 \xi_0 \xi_0'  - \psi_2 \left(\xi_0 \xi_1' + \xi_1 \xi_0'\right) + \psi_1\xi_1\xi_1 '\right)
\end{equation}
where $dS$ is the volume form on the $2$-surface $\mathcal{S}$, and $\xi^\alpha, \xi'^\alpha$ are generally two different solutions to the 2-surface twistor equations. At this stage, we are left with a complex charge and need to identify the mass and angular momentum. To proceed, one introduces an inner product on the 2-surface twistor space and requires it to be constant in order to select a suitable reality structure. Although such an inner product is not constant on general curved backgrounds, Penrose’s definition of quasi-local mass nevertheless succeeds in capturing mass and energy appropriately in several important examples \cite{Tod:1983waa, Bailey:1990qn, Huggett_Tod_1994}, suggesting that it contains essential physical insight, even if further refinements are needed in the general case.

To relate this definition to Hamiltonian and Noether arguments, first note that the twistor equation \eqref{twistor 1 equation} allows us to define $\chi_{\dot\alpha}$ via the covariant exterior derivative 
\begin{equation}    D\xi^\alpha:=\theta^{\beta\dot\alpha}\nabla_{\beta\dot\alpha}\xi^\alpha= \theta^{\beta\dot\alpha} \chi_{\dot\alpha}\delta_\beta{}^\alpha= \chi_{\dot\alpha}\theta^{\alpha\dot\alpha} \,.\label{chi-def}
\end{equation}
With this, it is easy to construct what in flat space would be a Killing vector 
\begin{equation}    K^{\dot\alpha\alpha}=\xi^\alpha\chi^{\dot\alpha}\, .\label{kv-def}
\end{equation}
Defining $\chi_{\dot\alpha}$ on a 2-surface by the components of \eqref{chi-def} tangent to the 2-surface, one can use the curvature identity $D^2 \xi^\alpha= R^\alpha_{\,\,\beta}\xi^\beta$  for covariant exterior derivative $D$ to rewrite the charge with $\xi^{\prime \alpha}=\xi^\alpha$ for simplicity as
\begin{equation}
    H[\xi]=\oint_S   \xi_\alpha D^2\xi^\alpha =\oint_S \chi_{\dot\alpha}\theta^{\alpha\dot\alpha}\wedge D\xi_\alpha 
\end{equation}
where we have used  \eqref{chi-def}   restricted to the 2-surface.  The right hand side is the Witten-Nestor 2-form used in the Witten positive energy proof \cite{Witten:1981mf,penrose1984spinors}. It is the boundary term for the total Hamiltonian generating diffeomorphisms in the direction $K^a$, here used quasi-locally rather than asymptotically, see \cite{Mason_1989,Mason:1990Fra}  for more details.  This is automatically real for real $K^a$.  Thus the issue of too many charges and identifying which charges should correspond to momenta, or angular momenta and reduces to  the issues of constructing real $K^a$ and identifying their nature as putative translations or rotations. In general for 2-surface twistors this issue has not been resolved without assumptions on the space-time and/or 2-surface.  However, more progress can be made using local twistors as in 
\cite{Dougan:1991zz,Dougan_1992} where closely related definitions are provided based on (anti-) holomorphic local twistors and these definitions lead to definitions of real 4-momenta with proofs of positivity of energy  and so on.

\subsection{Higher-spin extension}

We now introduce a higher-spin generalization of Penrose's quasi-local mass formula. 
The form of \eqref{Weyl Charge} motivates a higher spin generalization
\begin{equation} \label{higher spin ql}
    H_s = \oint_{\mathcal{S}} \phi_{\alpha_1 \cdots \alpha_{s+1}\beta\gamma}\xi^{\alpha_1\cdots \alpha_{s+1}}\Sigma^{\beta\gamma} .
\end{equation} 
 Here $\phi_{\alpha_1\cdots\alpha_{s+3}}$ is a valence $s+3$ zero-rest-mass (z.r.m.) field required to satisfy
\begin{equation} \label{zrm eq}
    \nabla^{\alpha_1}_{\,\,\,\,\dot \alpha}\phi_{\alpha_1\cdots\alpha_{s+1}\beta\gamma} = 0, 
\end{equation} which is the higher-spin generalization of \eqref{Bianchi Psi}.  In addition, conservation of the charge $H_s$ would require $\xi^{\alpha_1\cdots \alpha_{s+1}}$ to be a valence $s+1$ twistor satisfying 
\begin{equation}  \label{Highertwistorequation} \nabla^{(\alpha_0}_{\dot\alpha}\xi^{\alpha_1\cdots\alpha_{s+1})} = 0,
\end{equation} hence generalizing the first of \eqref{TwistorEquations}.
With these equations satisfied, the charge $H_s$ is easily seen to be conserved and independent of the 2-surface $\mathcal{S}$ as 
\begin{equation} \label{cons 2.20}
    \nabla^\beta_{\dot\beta}(\phi_{\alpha_1 \cdots \alpha_{s+1}\beta\gamma}\xi^{\alpha_1\cdots \alpha_{s+1}})=0
\end{equation}
follows directly from these equations. Thus these charges are exactly conserved.
On the one hand, this formula indicates that charges of higher spin fields should be thought of as dual to higher valence Killing spinors.  On the other, given a procedure to generate a higher spin field from  the gravitational data, this will be shown to be related to multipole formulae  and the $Lw_{1+\infty}$ charges  
in \S\ref{sec:Space-time definitions of celestial charges and multipole moments} and \S\ref{sec:Celestial charges from twistor space}.

\section{Multipole moments}
\label{sec:Space-time definitions of celestial charges and multipole moments}


Equation \eqref{higher spin ql}  was already used by Curtis \cite{Curtis:1978} in 1978 in his discussion of multipoles for stationary fields in linear electromagnetism and gravity.
Focusing first on linear theory, both  the z.r.m. equation \eqref{zrm eq}  and the higher valence twistor equation \eqref{Highertwistorequation} are consistent on a conformally flat background.  The higher spin z.r.m.\ equations again propagate two degrees of freedom from initial data. Furthermore, the higher valence solutions to the twistor equation can be obtained as symmetrized products of solutions to the valence-one twistor equation \eqref{twistor 1 equation}.  This has a four-dimensional solution space on a conformally flat background, and so  the valence $n$ solutions to the twistor equation \eqref{Highertwistorequation} have a solution space that is  $(n+3)!/n!3!$-dimensional, see chapter 6 of \cite{penrose1984spinors}. 

 In the first subsection we review the Geroch, Hansen, Curtis construction of multipoles in the stationary case, where it is possible to introduce a recursion  that generates higher spin fields from the gravitational data.  In the next subsection we go on to review and extend Curtis's  analysis of the construction at null infinity and this allows us to connect his analysis from the 1970s to the recent  definitions of celestial charges at $\scri$ that are also based on \eqref{higher spin ql} but now on a nonlinear background at $\scri$.

In Section \ref{sec:curved case} we explain how, on a general curved background, these equations are overdetermined and become inconsistent. However,  on restriction to a choice of general  null hypersurface, we can find a consistent propagating subset that extends  these definitions  to curved space-times  on choices such as incoming null hypersurfaces from a cut of $\scri$ or horizons or even a strictly quasi-local definition,

\subsection{Multipole moments for stationary fields in linear theory}
To obtain some intuition as to what these quantities measure, we first investigate them in linear theory about a flat background where our higher spin charges will be conserved.  In particular, we will see that in the time independent case they yield standard multipole moments in a form identified by Geroch \cite{Geroch:1970cc} but reformulated into a spinor and twistor version by Curtis \cite{Curtis:1978} which we now review.

To make contact with multipole moments for linear gravity, we must first 
 generate the higher valence solution to the z.r.m.\ equations from the linearized metric. 
To relate to the Geroch, Hansen, Curtis definitions,  we start by taking our linearized metric to be stationary.  Thus, following \cite{Curtis:1978}, we assume that the linearized Weyl spinor is constant along some constant timelike vector $t_a$ so that
\begin{equation}
    t\cdot \nabla \psi_{\alpha\beta\gamma\delta}=0\,.\label{Stationary}
\end{equation}
We take $t^a$ to be a constant vector of length $t^2=2$ so that in a normalized spin-frame it can be expressed as the identity matrix.  This will allow us to use $t_{\alpha\dot\alpha}$ to eliminate dotted spinor indices in favour of undotted ones. A vector projected into the three-dimensional space not spanned by $t_a$ can be defined by
\begin{equation}
    V^i = V^{\alpha\beta} = t^{(\alpha}_{\,\,\,\dot \alpha}V^{\beta)\dot \alpha}\, ,
\end{equation}
where the index $i = 1,2,3$. This projection can be inverted explicitly via the relation 
\begin{equation}
    V^{\alpha\dot \alpha} = t_{\beta}^{\,\,\dot \alpha}V^{\alpha\beta} + \frac{1}{2}(t\cdot V)t^{\alpha\dot \alpha}\, ,
\end{equation}
and we can define the spatial derivative by
\begin{equation}
D_{\alpha\beta}=t_{(\alpha}^{\dot\alpha}\nabla_{\beta)\dot\alpha}\, .
\end{equation}

Given a time-independent Weyl spinor, Geroch defines a sequence of higher spin fields by
\begin{equation}
    {}_{(n)}    \phi_{\alpha_1\ldots \alpha_{2n}}= D_{\alpha_{2n}\alpha_{2n-1}}\ldots D_{\alpha_6\alpha_{5}} \psi_{\alpha_4\alpha_3\alpha_2\alpha_1}\, .
\end{equation}
It is easy to see that this is totally symmetric and that it satisfies \eqref{zrm eq} for all $n$  as a consequence of the field  on $\psi_{\alpha\ldots \delta}$, and \eqref{Stationary}; $n=2$  gives back the Weyl spinor.  
In fact such a time-independent Weyl spinor is automatically curl-free as a consequence of the linearized Bianchi identity, and so is a second derivative of a complex potential
\begin{equation}   \psi_{\alpha\beta\gamma\delta}=D_{\alpha\beta}D_{\gamma \delta} \phi\, , 
\end{equation}
which is the real Newtonian potential in the static case, but more generally complex with a magnetic imaginary part.  It is a time-independent solution to the wave equation $t\cdot \nabla \phi=0$,  $\nabla^2\phi=0$, which of course reduces to Laplace's equation. With this, we see 
\begin{equation}
 {}_{(n)}   \phi_{\alpha_1\ldots \alpha_{2n}}= D_{\alpha_1\alpha_2}\ldots D_{\alpha_{2n-1}\alpha_{2n}} \phi\, .
\end{equation}

If we flip the dotted index of \eqref{Highertwistorequation}
with $t_{\alpha\dot\alpha}$ and symmetrize over all spinor indices  with $s+1 =2n$ is even we obtain
\begin{equation}
    D^{(\alpha\beta}\xi^{\alpha_1\ldots \alpha_{2n})}=0\, .
\end{equation}
Identifying symmetric pairs of spinor indices with 3-vector indices as above we obtain $\xi^{i_1\ldots i_n}$ symmetric and trace-free over all its indices whose trace-free symmetrized derivative vanishes so that it satisfies 
 the 3-dimensional conformal Killing tensor equation 
\begin{equation}
    D^{(j}\xi^{i_1\cdots i_n)} = g_{3d}^{(j\,i_1}K^{i_2\cdots i_{n})}
\end{equation}
for some $K^{i_2 \ldots i_n}$.  Here the three dimensional metric is given by $g_{3d}^{ij} = g_{3d}^{\alpha\beta\gamma\delta} = \epsilon^{\alpha(\gamma\vert}\epsilon^{\beta\vert\delta)}$ when written in terms of spinor indices. With these definitions, we see that \eqref{higher spin ql} reduces to the Geroch definition of multipole moments as given and applied in \cite{Geroch:1970cc,Geroch:1970cd,Hansen:1974zz}
\begin{equation}
    H_s = \oint_{\mathcal{S}}\xi^{i_1 \cdots i_{n}}D_j D_{i_1}\cdots D_{i_n}\phi\, dS^j \, .
\end{equation}

\subsection{Expansions at null infinity, from multipoles to celestial charges} 

We now wish to see how these definitions translate to $\scri$. In order to do this we must introduce Bondi coordinates by setting
\begin{equation} \label{BondiCoordinatesSpinor}
   \,  x^{\alpha\dot\alpha}=u t^{\alpha\dot\alpha} + r o^\alpha\bar o^{\dot\alpha}\, , \qquad o_\alpha =\frac{(1,z)}{\sqrt{1+|z|^2}}\, , \qquad z=\frac{x^1+ix^2}{1-x^3}\, .
\end{equation}
Here $z$ is a stereographic coordinate on the sphere of null directions and we complete $o_\alpha$ to the normalized spinor dyad $(o_\alpha,\iota_\alpha):= (o_\alpha, t_\alpha^{\dot\alpha}\bar o_{\dot\alpha})$. We can further introduce the associated tetrad with 
\begin{equation}
    D=o^\alpha\bar o^{\dot\alpha}\nabla_{\alpha\dot\alpha} =\p_r \, , \qquad \delta = o^\alpha\bar \iota^{\dot\alpha}\nabla_{\alpha\dot\alpha}=\frac{1+|z|^2}{r}\p_{\bar z}\, .
\end{equation}

In this spin frame, the standard peeling results for the components 
$$
{}_{(n)}\phi_p:={}_{(n)}\phi_{\alpha_1\ldots \alpha_{2n} }\iota^{\alpha_1}\ldots \iota^{\alpha_p}o^{\alpha_{p+1}}\ldots o^{\alpha_{2n}}
$$
of a massless field of helicity $n$ that is conformally smooth at $\scri$ lead to the radial expansion  
\begin{equation}
    {}_{(n)}\phi_p=\sum_{m\geq 0}\frac{{}_{(n)}\phi_p^m(u,z,\bar z)}{r^{m+2n-p+1}}
    \,.\label{falloff}
\end{equation}
The symmetry along $t^a$ implies $u$-independence of ${}_{(n)}\phi_p^0(u,z,\bar z)$ at $\scri$ and the asymptotic Bianchi identities then imply $\bar\eth {}_{(n)}\phi_p^0(u,z,\bar z)=0$.  The spin weights are such that for $p>n$ $_{(n)}\phi_p^0(z,\bar z)=0$ and for $p\leq n$ the $_{(n)}\phi_p^0(z,\bar z)$ are spin-weighted spherical harmonics of total spin $n-p$ corresponding to the coefficients of the $n-p+1$th multipole moment.

We loosely follow Curtis's analysis of the extension to $\scri$ of the Geroch recursion
\begin{equation}\label{MultipoleRecursion}
{}_{(n)}\phi_{\alpha_1\ldots\alpha_{2n}}=D_{\alpha_1\alpha_2}\,\, {}_{(n-1)}\phi_{\alpha_3\ldots\alpha_{2n}}\, ,
\end{equation}
for stationary solutions to the higher-spin massless field.  Note that it follows from the field equations and time symmetry that this is symmetric in its spinor indices).  For integral helicity, he shows\footnote{Curtis introduces factors of $T := t^{\alpha\dot \alpha}o_\alpha \bar o_{\dot \alpha}$ which  is 1 in our spin frame, but allows us to keep track of boost or conformal weights that arise when $o_\alpha$ is not normalized.  It encodes the dependence of the constructions on a timelike vector. } for $p = 1,\dots,n$ that \eqref{MultipoleRecursion} gives a relationship of asymptotic data
\begin{equation}
    {}_{(n)}\phi^0_p = 
   2(2n-p) {}_{(n-1)}\phi^0_{p-1} \, . \label{scri-stat-recursion}
\end{equation}
This can be seen by contracting $o^{\alpha_1}\iota^{\alpha_2}$ into \eqref{MultipoleRecursion} and observing that $D_{01}$ gives the radial derivative $D=\p_r$ so that our equation follows from the leading order in \eqref{falloff}. This does not determine ${}_{(n)}\phi_0^0$ where  we find more generally that
\begin{equation}
    {}_{(n)}\phi^m_0 =- 
    \bar\eth  \, 
    {}_{(n-1)}\phi^{m+1}_{0}
    \, .\label{scri-stat0-recursion}
\end{equation}
This follows by saturating the recursion with $o_\alpha$s and expanding in $1/r$. This can be iterated to make a connection to the subleading terms of linearized Weyl spinor
\begin{equation}
    {}_{(n)}\phi^0_0 =(-1)^{n-2}
    \bar\eth^{n-2} \psi_0^{n-2}\, .
\end{equation}

\paragraph{Celestial charges and multipoles beyond linear theory.}
We now  relate the Geroch-Hansen-Curtis definition of multipoles to that of the higher celestial charges \cite{Freidel:2021ytz,Geiller:2024bgf,Kmec:2024nmu}. A similar connection was provided in \cite{Compere:2022zdz}, so we here  extend this  to Geroch's \cite{Geroch:1970cc} and Curtis' \cite{Curtis:1978} frameworks.

The $Lw_{1+\infty}$ quasi-local charges defined in Equation \eqref{higher spin ql}, can also be evaluated at null infinity to recover the results of \cite{Freidel:2021ytz,Geiller:2024bgf,Kmec:2024nmu}, see Section \ref{sec: Charges at Null Infinity}. In particular, the z.r.m. field can be expressed in terms of the higher-spin charge aspects $Q_s$ ($s=-1, 0, 1, 2, \ldots$) introduced in these references, see also Section \ref{sec:Covariant Spcaetime Expressions}. These formulae at $\scri$ lead to extensions beyond the stationary and linearized cases.
Following \cite{Freidel:2021ytz, Geiller:2024bgf}, one makes the following expansion of $\psi_0$
\begin{equation}
    \psi_0 = \frac{\psi_0^0}{r^5} + \sum_{s=3}^{\infty} \frac{1}{r^{s+3}}\frac{(-1)^s}{(s-2)!}\left(\eth^{s-2}Q_{s}^0 +\Phi_{s-2} \right)\, ,\label{psi-0-exp}
\end{equation}
where in the full theory of gravity $\Phi_{s-2}$ contains non-local terms involving the asymptotic shear and overleading Weyl spinor components. In the linearised theory where the asymptotic shear vanishes $\Phi_{s-2} = 0$ and the equations satisfied by $Q_{s}^0$ are given by
\begin{equation}
    \partial_u Q_{s}^0 = \bar\eth Q_{s-1}^0 \, .
    \label{int Bianch}
\end{equation}
In the stationary case, $\partial_u Q_s^0=0$, and the
asymptotic Bianchi identities imply $\bar{\eth}Q_s^0=0$.
In our conventions, $\bar{\eth}$ raises spin weight and
has a trivial kernel on globally smooth fields of negative
spin weight on the sphere (see §4.15 of \cite{Penrose:1984uia}). Hence $Q_{-1}^0=Q_{-2}^0=0$. Now using the commutator on the sphere $[\eth, \bar\eth ]\eta_s = s \eta_s$ where $\eta_s$ is a field of spin weight $s$ we find the following relationship between the Curtis z.r.m. field components
\begin{equation}
    {}_{(n)}\phi_0^0 = \frac{(-1)^n (2n)!}{2^{n-2}(n+2)!}  Q_{n}^0 \, .
\end{equation}
In what follows, for general curved backgrounds, we will work with the higher spin field given by the asymptotic data $Q_{n}^0$. In general, while the charges will have the same decomposition into components of the z.r.m. field and the twistor, the exact relation between the components of the z.r.m. field and the subleading orders of the Weyl spinor $\psi_{\alpha\beta\gamma\delta}$ will be different depending on the recursion procedure used. We will now outline the recursion relations that we wish to use in the rest of the text. 

\subsection{Curved space definitions on null hypersurfaces}
\label{sec:curved case}

In this section we focus mostly on \emph{semi-local} definitions, where the definition of a celestial charge or multipole measured at a 2-surface $\mathcal{S}$ depends on constructions of $\phi_{\alpha_1\ldots \alpha_n} $ and $\xi^{\alpha_1\ldots\alpha_m}$ on a null hypersurface $\mathcal{N}$ going out to infinity. At the end we give some briefer discussion of possible genuinely quasi-local definitions but that discussion will be more tentative.

As emphasised earlier, in curved space, Equations \eqref{zrm eq} and \eqref{Highertwistorequation} are overdetermined and become inconsistent. The higher valence z.r.m. equation \eqref{zrm eq} is $2(s+2)$ equations on $s+4$ unknowns and the twistor equation \eqref{Highertwistorequation} are $2(s+3)$ equations on $s+2$ unknowns. Solutions to \eqref{zrm eq} must satisfy the Buchdahl condition 
\begin{equation}
   \psi^{\alpha_1\alpha_2\alpha_3}_{\qquad \,\,\,(\beta_3}\phi_{\beta_4\cdots\beta_{s+3}) \alpha_1\alpha_2\alpha_3} = 0
\end{equation}
which can be derived using \eqref{Commutator}. For gravity, the Weyl spinor satisfies the Buchdahl condition due to the algebraic relation
\begin{equation}
\psi^{\alpha_1\alpha_2\alpha_3}_{\qquad \,\,\,\beta}\psi_{\gamma\alpha_1\alpha_2\alpha_3} = \left[\psi_0\psi_4 - 4\psi_1\psi_3 + 3(\psi_2)^2\right]\epsilon_{\gamma\beta} . 
\end{equation}
However, for higher spin z.r.m. fields, no such natural geometric relations exists and the Buchdahl conditions further over-determine the system. We will circumvent this in two ways. In the rest of this subsection, we only use those equations tangent to a null hypersurface generalizing the description of the celestial charges at $\scri$.  In the next section we work on a SD background where $\psi_{\alpha\beta\gamma\delta} = 0$. 


A similar argument holds for the higher valence twistor equation which is even more over-determined. This means  that we cannot impose even all the components of the tangential equation that are tangent to a 3-surface; thus, in general,  we will lose exact conservation leaving us with a flux law for the charges. This will be the case even on a SD background, as the twistor equation is overdetermined there too in general. 

To be specific, we work on a null hypersurface $\mathcal{N}$  with a boundary $\p \mathcal{N}= C\cup \Sigma$ where $C$ is a cut of $\scri $ and  $\Sigma$ is a finite quasi-local 2-surface, both of topology $S^2$.  We assume that $\Sigma$ is close enough to $\scri$ that $\mathcal{N}$  can be taken to be smooth, i.e., with no caustics,  with tangent $n^a=\iota^\alpha\bar \iota^{\dot\alpha}$.  
We circumvent the integrability conditions for higher spin fields and higher valence solutions to the twistor equations so as to construct charge integrals in general on $\Sigma$.  To do this, we define our $\phi_{\alpha_1\ldots\alpha_n}$ and  $\xi^{\alpha_1\ldots \alpha_m}$ by choosing judicious subsets of the equations \eqref{zrm eq} and \eqref{Highertwistorequation} tangent to the null hypersurface.

We first remark that   
\begin{equation}
\bar \iota^{\dot\alpha}\nabla_{\dot\alpha}^{\alpha_1}\phi_{\alpha_1\ldots \alpha_{n}} =0\, .
\end{equation}
These are $n$ equations for the $n+1$ components of
$\varphi_{\alpha_1\ldots\alpha_n}$. Using the conformally
covariant GHP operators defined in Appendix~A.2,
Eqs.~\eqref{GHP derivative} and~\eqref{A27}, they read
 \begin{equation}
     \text{\th}'_{\mathcal{C}} \phi_{m} - \bar\eth_{\mathcal{C}}\phi_{m+1} = (n-m-1)\sigma \phi_{m+2}\, , \label{recurs} 
 \end{equation}
noting that because null hypersurfaces are geodesic, the $\nu$-term that is generally there automatically vanishes. This higher spin field $\phi_{\alpha_1\ldots \alpha_n}$ can be seeded from the gravitational field on the null hypersurface, represented by the `null datum' $\psi_4$ \cite{Penrose1980GoldenON},  the component of the Weyl spinor tangent to the surface  of the hypersurface which defines the characteristic data for the gravitational field on $\mathcal{N}$. If we identify $\psi_4$ with $\phi_n$, the remaining components of $\phi_{\alpha_1\ldots \alpha_n}$ are obtained by solving ODEs up the generators of  $\mathcal{N}$. There remains freedom  in the boundary conditions for the $\phi_m$, $m< n$, and for a semi-local definition are fixed as $r\rightarrow\infty$ at $\scri$ so as to be identified with the charge aspects defined there by \cite{Freidel:2021ytz,  Kmec:2024nmu,Geiller:2024bgf}.  These are in turn fixed  by the corresponding equations on $\scri$  which are seeded by the asymptotic components of the Weyl spinor and determined by their asymtotics at $i^0$. More precisely, the identification $\phi_n=\psi_4$
uses a valence-$n$ spinor
$\phi_{\alpha_1\cdots\alpha_n}$ carrying additional
GHP weight $(n-4,0)$, consistently with Section~5.1.
Under $\iota^\alpha\mapsto t^{-1}\iota^\alpha$ and
$o^\alpha\mapsto t\,o^\alpha$, this spinor transforms
as $\phi_{\alpha_1\cdots\alpha_n}\mapsto
t^{n-4}\phi_{\alpha_1\cdots\alpha_n}$.
Consequently,
$\phi_n=\phi_{\alpha_1\cdots\alpha_n}
\iota^{\alpha_1}\cdots\iota^{\alpha_n}$
has GHP weight $(-4,0)$, exactly as $\psi_4$.
Their equality therefore specifies the characteristic
datum directly. Covariant derivatives of these weighted
fields include the corresponding GHP weight connection,
as in (5.2). The reference spin frame can be fixed at
$\mathscr I$ and propagated along $\mathcal N$,
consistently with the correspondence in (4.14).\footnote{\label{foot7}Alternatively, for an unweighted valence-$n$ field
with even $n\geq4$, one may identify
$\psi_4=\phi_{2+n/2}$, whose spin weights agree. However,  this latter choice will, in general, involve solving $\bar \eth$-equations rather than simple parallel propagation so the PDE theory is more subtle and a  more elaborate discussion is needed. One is then interested in inverting $\bar\eth$, but this in general can have some global obstruction or ambiguity when $\mathcal{S}$ is topologically a sphere. The problem $\bar\eth f = g$ where $f$ has spin weight $s$ and $g$ is given has no obstruction if $s>-1$, but is obstructed for $\phi_m$ for $m>2+\frac{n}{2}$. One can still find a closest-fit solution.  The optimal choices will depend on the applications and simply  change the multipole moment or celestial charge that is to be measured.}

The tangential components of the twistor equation \eqref{twistor 1 equation} are overdetermined when $\mathcal{S}$ is embedded into the null hypersurface $\mathcal{N}$. 
For our approximate Killing spinors $\xi^{\alpha_1\cdots\alpha_{s+1}}$ (with complementary GHP weight $(1-s,0)$,
so that the charge integrand has zero additional GHP weight) we can proceed by solving the equations
\begin{equation}
\iota_{\alpha_0}\bar \iota^{\dot\alpha}\nabla_{\dot\alpha}^{(\alpha_0}\xi^{\alpha_1\ldots \alpha_{s+1})} =0\, . \label{xi-def}
\end{equation}
These are $s+2$ evolution equations on the $s+2$ unknowns determining $\xi^{\alpha_1\ldots \alpha_{s+1}}$ up the generators of the null hypersurface. Thus, they determine these components in terms of boundary data again either at the cut $C$ of $\scri$ or at our finite boundary $\Sigma$ of $\mathcal{N}$.  At $\scri$ the boundary data can naturally be identified with the corresponding quantities defined in \cite{Kmec:2024nmu}, whereas on $\Sigma$ they can be constructed from solutions to either the 2-surface twistor equation \eqref{2-surface-twistor} or the holmorphic \eqref{2-surface-twistor-DM} or anti-holmorphic twistors of \cite{Dougan:1991zz}.

If we now introduce the 2-surface charge integral
\begin{equation}
    H_s=\int_{\mathcal{S}}  \phi_{\alpha_1\ldots \alpha_{s+1}\alpha\beta}\xi^{\alpha_1\ldots \alpha_{s+1}}\Sigma^{\alpha\beta}
\end{equation}
we see that the difference between the charge integral on different slices of the null hypersurface is given by  the integral of the flux
\begin{equation}
\mathcal{F}=\iota^{\alpha} \bar\iota^{\dot\alpha}\nabla_{\,\,\,\dot\alpha}^{\beta}     (\phi_{\alpha_1\ldots \alpha_{s+1}\alpha\beta}\xi^{\alpha_1\ldots \alpha_{s+1}})=\psi_4\dot \xi
\end{equation}
where we define $\dot \xi$ by 
\begin{equation}
\bar \iota^{\dot\alpha}\nabla_{\dot\alpha}^{(\alpha_0}\xi^{\alpha_1\ldots \alpha_{s+1})} =\dot \xi \iota^{\alpha_0}\ldots \iota^{\alpha_{s+1}}    
\end{equation}
where the form of the right-hand side follows from \eqref{xi-def}. Thus, we see that when the radiation through the hypersurface, as measured by $\psi_4$, vanishes the charge is conserved. Otherwise, the flux is measured by $\psi_4 \dot \xi$. Thus we have


\begin{propn}
Given cross-sectional spherical 2-surfaces $S_1$  and $S_2$ on $\mathcal{N}$ then the difference between the charges on the two 2-drufaces is given by  the flux integral
\begin{equation}
    H_s\vert_{\mathcal{S}_2}-H_s\vert_{\mathcal{S}_1} =\int_{\mathcal{N}'} \psi_4 \dot\xi d^3\mathcal{N}
\end{equation}
where $d^3\mathcal{N}=o_\alpha\bar o_{\dot\alpha} \Sigma^{3\alpha\dot\alpha}|_{\mathcal{N}}$ is the natural weighted volume form on $\mathcal{N}$ and $\mathcal{N}'$ is the region bounded by $\mathcal{S}_1$ and $\mathcal{S}_2$. 
\end{propn}

To illustrate the lack of conservation, we remark that for conservation to hold between two cuts $\mathcal{S}$ and $\mathcal{S}'$ of $\mathcal{N}$, in the one index case, one must solve a system of three PDEs
\begin{equation} \label{twistor equation null}
    \text{\th}'_{\mathcal{C}}\xi_1 + \nu \,\xi_0 = 0 \quad, \quad \text{\th}'_{\mathcal{C}} \xi_0 - \bar{\eth}_{\mathcal{C}}\xi_1= 0\quad ,\quad \bar{\eth}_{\mathcal{C}} \xi_0 - \sigma\,\xi_1 = 0 ,
\end{equation} which are obtained by contracting \eqref{twistor 1 equation} with $\bar{\iota}^{\dot \alpha}$. This is still an overdetermined system. To get some physical intuition on this obstruction, consider the system at null infinity using the Newman-Unti solution space \cite{Newman:1962cia}. At leading order, one has
\begin{equation} 
    \partial_u \xi_1^0 = 0\quad , \quad \partial_u\xi_0^0 - \bar{\eth} \xi_1^0 = 0 \quad, \quad \bar\eth \xi_0^0 - \sigma^0\xi_1^0 = 0 .
\end{equation}
The compatibility of this system is obstructed by the Bondi news $N = \partial_u \sigma^0$, which encodes the gravitational radiation. The insolvability of the system is a direct consequence of gravitational radiation, giving nontrivial fluxes at null infinity. In order to capture this property, we therefore demand that the components of the twistor $\xi^\alpha$ satisfy only the evolution equations along the null direction of $\mathcal{N}$ generated by $n^\mu$, hence selecting only the two first equations of \eqref{twistor equation null} and this generalizes to give \eqref{xi-def}. 

 \paragraph{Tentative fully quasi-local definitions.}  A properly quasi-local definition should only depend on the data at the 2-surface $\mathcal{S}$.  Using the higher spin analogue of the Penrose quasi-local mass integral, we have two tasks, the construction of the 2-surface Killing spinors and the higher spin field $\phi_{\alpha_1\ldots \alpha_n}$ recursed from the Weyl spinor. The quasi-local construction of a 2-surface Killing spinor has already been covered earlier based on  symmetric products of 2-surface twistors \cite{Penrose:1986uia} or from the the holomorphic or antiholomorphic local twistors of \cite{Dougan:1991zz}.  In order to construct $\phi_{\alpha_1\ldots \alpha_n}$, we must assume that we are given  the characteristic data on both the ingoing and outgoing null hypersurfaces through $\mathcal{S}$ as the higher charges will need higher derivatives of the field.  At $\mathcal{S}$ itself, this amounts to knowing \th$^r\psi_0$ and \th$^{\prime s}\psi_4$ for $r,s \geq 0$ together with $\psi_1, \psi_2,\psi_3$ on $\mathcal{S}$. For this alternative construction, we use
unweighted spinors $\phi$ and $\xi$, as in
footnote~\ref{foot7}. We have \eqref{recurs} and its unprimed version
\begin{equation}
 \text{\th}'_{\mathcal{C}} \phi_{m} - \bar\eth_{\mathcal{C}}\phi_{m+1} = (n-m-1)\sigma \phi_{m+2}\, , \qquad     \bar\eth_{\mathcal{C}}\phi_{m} -\text{\th}_{\mathcal{C}} \phi_{m+1} =
     m\lambda \phi_{m-1}\, , \label{recurse-prime}
 \end{equation}
where the $\kappa$-term vanishes as $l^a = o^{\alpha}\bar o^{\dot \alpha}$ can be taken to be geodesic. With $\phi$ of even valence $n \ge 4$, we first set $ \phi_{n/2+i}=\psi_{2+i}$ and aim to build the  $\phi_m$ by solving subsets of these equations. Choices must be made particularly as  \eqref{recurse-prime} will be overdetermined just as the higher spin equations are themselves and having identified the 5 middle components with the Weyl spinor, we need to choose $n-4$ equations from \eqref{recurse-prime} that identify the remaining components of $\phi$ to determine these.  The optimal choices will depend on circumstance, and we do not investigate these further here. This will, in particular, involve  solving the $\eth$ operator or its conjugates, then there will in general either be obstructions to solving or freedom in the homogeneous solutions.   Although it should be possible to  find a unique minimal approximate solution, projecting out the obstruction and minimizing the freedom.  Otherwise an exact solution will have some wire singularities and the integrals will need regulating. We note that similar considerations apply asymptotically in the work of \cite{Freidel:2021ytz,Geiller:2024bgf}.  For example, the $\Phi_{s-2}$ of \eqref{psi-0-exp} can be characterized as being that part of the expansion of $\psi_0$ that is orthogonal to the image of $\eth^{s-2}$. In general the details will very much depend on special features of any situation considered.

\section{Celestial charges from twistor space}
\label{sec:Celestial charges from twistor space}

In this section, we first introduce the basics of twistors and the elementary twistor representatives for Killing spinors, celestial symmetries, linear fields and stationary multipole solutions.  This allows us to be very explicit about the duality between local twistors and multipoles and the relationship to celestial symmetries in linear theory.  We go on to
provide a first-principles derivation of the charges associated with the $Lw_{1+\infty}$ symmetries on twistor space. The latter are interpreted as large gauge transformations on twistor space, following \cite{Kmec:2024nmu, Kmec:2025ftx}. To be self-consistent, we include a review on the relevant twistor theory for SD gravity along with ASD perturbations and the action principle describing both.

\subsection{Twistor space,  $Lw_{1+\infty}$, killing spinors and multipoles for linear fields}
\label{SectionTwistorBasics}
First, let us consider flat twistor space $\mathbb{PT}$ corresponding to complexified Minkowski space $\mathbb{C}\mathbb{M}$. $\mathbb{PT}$ is an open subset of the complex projective space $\mathbb{CP}^3$. Using homogeneous coordinates $Z^A = (\mu^{\dot \alpha},\lambda_\alpha)\in\mathbb{CP}^3$ such that $Z^A\sim tZ^A$ for $t\in \mathbb{C}^\times$, we have
\begin{equation}
    \mathbb{PT} = \left\{\left.Z^A\in \mathbb{CP}^3\right| \lambda_\alpha \neq 0\right\} . 
\end{equation}
Another description of this space is given by the total space of the vector bundle over $\mathbb{CP}^1$
\begin{equation}\label{flattwistorspace}
    \mathbb{PT} = \mathcal{O}(1)\oplus \mathcal{O}(1)\rightarrow \mathbb{CP}^1
\end{equation}
where $\mathcal{O}(n)$ denotes the line bundle over $\mathbb{CP}^1$ with Chern class $n$. The base has homogeneous coordinates $\lambda_{\alpha} $ while the fiber coordinates $\mu^{\dot \alpha}$ have the same scaling weight as the base coordinate. 

A point $x$ in Minkowski space is represented in flat twistor space as a Riemann sphere $X \simeq \mathbb{CP}^1$ with homogeneous coordinates $\lambda_\alpha$ via the incidence relation
\begin{equation} \label{FlatIncidenceRelations}
    \mu^{\dot \alpha} = ix^{\alpha\dot \alpha}\lambda_\alpha .
\end{equation}
This gives  a  non-local relationship between twistor space and space-time, with a point in twistor space corresponding to a 2-plane in complexified space-time. 

There is a natural Poincar\'e inavariant holomorphic Poisson structure of weight -2 defined on homogeneous functions on twistor space given by
\begin{equation}
    \{h_1,h_2\}= \varepsilon^{\dot\alpha\dot\beta}\frac{\p h_1}{\p\mu^{\dot\alpha}}\frac{\p h_2}{\p\mu^{\dot\beta}}\, ,\label{Poisson}
\end{equation}
together with a   holomorphic 1-form of homogeneity degree-2 on  the base Riemann sphere
\begin{equation}
     D\lambda = \langle \lambda d\lambda \rangle \, .
\end{equation}
The  holomorphic volume form on $\mathbb{PT}$  given by
\begin{equation}
    D^3 Z = \varepsilon_{ABCD} Z^A dZ^B\wedge dZ^C\wedge dZ^D\, .
\end{equation}
is conformally invariant.

The Poisson structure defines the $Lw_{1+\infty}$ or $LHam(\mathbb{C}^2)$ algebra as the Poisson algebra of functions on twistor space of homogeneity degree 2.  We will allow these to be meromorphic with poles at $\lambda_0=0$ and $\lambda_1=0$; this is the `wedge condition' that requires that the powers of $\mu$ should be non-negative. Generators can be explicitly given by 
\begin{equation}
w_{m,a}^p = \frac{\left(\mu^{\dot 0}\right)^{p+m-1}\left(\mu^{\dot 1}\right)^{p-m-1}}{\lambda_0^{2p-4-a}\lambda_1^a}\, , \qquad p\pm m-1 \geq 0\, . \label{w-def}
\end{equation}
The generators $w_{m,a}^p$ are holomorphic on the patch of twistor space where $\lambda_1 \neq 0 $ and $\lambda_0 \neq 0$ and they satisfy the algebra $LHam(\mathbb{C}^2)$ (also referred to as $Lw_{1+\infty}$, see Footnote \ref{celestial algebra name}), given explicitly as
\begin{equation}
    \{w^p_{m,a} , w^q_{n,b}\} = 2(m(q-1) - n(p-1))w^{p+q-2}_{m+n,a+b} . 
\end{equation}
When $w^p_{m,a}$ is a polynomial, $a\geq 0, 2p-4-1\geq 0$, the Hamiltonian vector field of $w^p_{m,a}$ defines a translation or self-dual rotation.

There is a 1:1 correspondence between solutions $\xi^{\alpha_1\ldots \alpha_m}$ to the Killing spinor equation and homogeneous  polynomials $w(Z)$ on twistor space of homogeneity degree $n$ given by 
\begin{equation}    w(Z)=\lambda_{\alpha_1}\ldots\lambda_{\alpha_n} \xi^{\alpha_1\ldots \alpha_n}\, .
\end{equation}
This follows because twistor functions can be characterized by the equation $\lambda^\alpha\nabla_{\alpha\dot\alpha}w(x,\lambda)=0$ and it's clear that this equation follows from and implies the Killing spinor equation.  The fact that it is global means that it must be a polynomial of degree $n$ in $\lambda_\alpha$.

Because the $w^p_{m,a}$ are rational rather than polynomial, they are not equivalent to Killing spinors.  However, the wedge condition means that they are polynomials in the $\mu^{\dot\alpha}$ and  the   polynomial in $\mu^{\dot\alpha}$ gives a well-defined 
Killing spinor.  This is multiplied by some rational function in $\lambda_\alpha$ generically of negative weight with powers of either or both components of $\lambda_\alpha$ in the denominator. The Killing spinor arising from the numerator $(\mu^{\dot 0})^{p+m-1}(\mu^{\dot 1})^{p-m-1}$ can be easily seen to be 
\begin{equation}
\xi^{\alpha_1\ldots\alpha_{2p-2}}_{p,m}=x^{(\alpha_1|\dot 0|}\ldots x^{\alpha_{p+m-1}|\dot 0|}x^{\alpha_{p+m}|\dot 1|}\ldots x^{\alpha_{2p-2})\dot 1}\, .
\end{equation}

Linear massless fields of helicity $n/2$ are represented by cohomology classes $f_{n-2}\in H^1(\PT,\cO(n-2))$ with for example the positive helicity graviton $h \in H^1(\PT,\cO(2))$ and negative helicity graviton $g\in H^1(\PT,(\cO(-6))$ etc..  These cohomology classes can either be represented by holomorphic functions with singularities, \emph{\v Cech cocycles}, or by $\bar\partial$-closed $(0,1)$-forms modulo exact ones, i.e., \emph{Dolbeault cocycles}.  In either case their corresponding massless fields are given on space-time by the formulae
\begin{equation} \label{4.11}
    \phi_{\alpha_1\ldots \alpha_{n}} = \frac{1}{2\pi i}\int_X D\lambda\wedge\lambda_{\alpha_1}\ldots \lambda_{\alpha_{n}}\, f_{-n-2}\vert_X . 
\end{equation}
for helicity $-n/2$ and 
\begin{equation}\label{PenroseTransformPositiveHelicity}
  \phi_{\dot\alpha_1\ldots \dot\alpha_{n}} = \frac{1}{2\pi i}\int_X D\lambda\wedge\frac{\partial}{\partial\mu^{\dot\alpha_1}}\ldots \frac{\partial}{\partial\mu^{\dot\alpha_{n}}}\, f_{n-2}\vert_X . 
\end{equation}
for positive helicity. These integral formulae are known as the Penrose transform
and realize the correspondence between twistor cohomology
classes and spacetime z.r.m. fields of the corresponding
helicities.

\paragraph{Celestial charges and multipoles:} We will see later that the Noether celestial charge for a spin-2 ASD field $\psi_{\alpha\beta\gamma\delta}$  with twistor representative $g\in H^1(\PT,\cO(-6))$ 
is given by
\begin{equation}
H_{w^p_{m,a}}^\mathcal{S}=\int_{\mathcal{S}\times \mathbb{CP}^1}  w^p_{m,a} \, g \wedge D^3Z\, .\label{twis-charge}
\end{equation}
The poles in $w^p_{m,a}$ mean that this formula is not completely cohomologically natural.  A choice of representative for $g$ that vanishes at the zeros of the two components of $\lambda_\alpha$ can be found so as to regulate the integral, but as in the recursion above, the answer depends on the choice. Our strategy to understand this integral on space-time will be to separate the $w^p_{m,r}$ into 
\begin{equation}
    w^p_{m,a}= \frac{w^p_m}{\lambda_0^{2p-4-a}\lambda_1^a}\, , \qquad w^p_m= (\mu^{\dot 0})^{p+m-1} (\mu^{\dot 1})^{p-m-1} \label{Lw-to-KS}
\end{equation}
so that we can evaluate $w^p_m$ as a Killing spinor where $\xi^{\alpha_3\ldots \alpha_{2p}}$, and $ g/\lambda_0^{2p-4-a}\lambda_1^a$ as defining a  higher spin massless field $\phi_{\alpha_1\ldots \alpha_{2p}} $ and we will see that it can be defined by the recursion \eqref{recurs1}
below.  It is then straightforward to see that our formula \eqref{twis-charge} evaluates to give
\begin{equation}
    H_{w^p_{m,a}}^\mathcal{S}=\int_{\mathcal{S}}  \xi^{\alpha_3\ldots \alpha_{2p}}\phi_{\alpha_1\ldots \alpha_{2p}} \Sigma^{\alpha_1\alpha_2}\, . \label{Cel-charge}
\end{equation}

The recursion we used in \eqref{recurs} of \S\ref{sec:curved case} is most simply understood from the \v Cech cohomology  version of these formulae. For this we take a function $g_{-n-2}$ to be a \v Cech representative defined on the complement of $\lambda_0=0$ and $\lambda_1=0$ with gauge freedom consisting of functions holomorphic on the complement of just one of these two sets.  On inserting the incidence relations \eqref{FlatIncidenceRelations},  we can expand in powers of $\lambda=\lambda_0/\lambda_1$ to obtain
\begin{equation}
g(ix^{\alpha\dot{\alpha}}\lambda_\alpha,\lambda_\beta)
=
\sum_{m=-\infty}^{\infty}
\frac{\phi(x)_m\,\lambda^{m+1}}{\lambda_0^{n+2}}.
\end{equation}
With this it is easily seen that $\phi_0,\ldots ,\phi_n$ are the standard components of $\phi_{\alpha_1\ldots \alpha_n}$ and that these components are gauge invariant.  For $i$ outside this range,   the condition that $f$ is a twistor function $\lambda^\alpha\nabla_{\alpha\dot\alpha}f=0$ gives the recursion
\begin{equation}
    \iota^\alpha\nabla_{\alpha\dot\alpha}\phi_m =o^\alpha\nabla_{\alpha\dot\alpha} \phi_{m+1}\label{recurs1}
\end{equation}
The  $\phi_m$ therefore satisfy the field equation of the higher spin equation $\nabla^{\alpha_1}_{\dot\alpha}\phi_{\alpha_1\ldots \alpha_{n+r}}=0$.  It can now be seen that the range of the components $\phi_{-s}, \ldots \phi_{n+r}$  gives the massless field corresponding to the twistor function $g'=g/\lambda_0^r\lambda_1^s$.  The gauge dependence of the $\phi_i$ for $i$ outside the range $0,\ldots , n$  is reflected in the freedom in constants of integration required to solve the recursion.\footnote{In the Dolbeault framework, a  choice of representative for $g$ that vanishes to all orders at the zeros of the two components of $\lambda_\alpha$ can be found so as to regulate the integral, but the answer still depends on the gauge choice; this can be identified with that of the \v Cech representative using bump functions to extend the \v Cech gauge choices to give Dolbeault gauge transformations.} This can then provide the definition up to this ambiguity for the higher spin massless field appearing in \eqref{Cel-charge}.

\paragraph{Stationary solutions and Geroch multipoles:} The Geroch recursion and corresponding formulae work somewhat differently to the above  with no ambiguities in the recursion, but relying on the fact that we were working with stationary solutions. On twistor space the time translation symmetry  is generated by the vector field $\lambda_\alpha t^{\alpha\dot\alpha}\p/\p\mu^{\dot\alpha}$ and invariant functions depend on $\mu^{\dot\alpha}$ through 
\begin{equation}
    \mu=\lambda_\alpha t^\alpha_{\dot\alpha}\mu^{\dot\alpha}\, .
\end{equation}
The basic $k$th multipole fields are homogeneous powers of $\mu^{\dot \alpha}$ with Cech cocycles 
\begin{equation}
     f_{-2n-2} = \frac{I_{2k}(\lambda)}{\mu^{n+k+1}}\, ,\label{multipole}
\end{equation}
where
\begin{equation}
     I_{2k}(\lambda)=I_{2k}^{\alpha_1\ldots \alpha_{2k}}\lambda_{\alpha_1}\ldots \lambda_{\alpha_{2k}}\, ,
\end{equation} and $I_{2k}^{\alpha_1\cdots\alpha_{2k}}$ is a constant
totally symmetric spinor. (Note, however, that a factor of $\log \mu$ is also required when $k\leq 0$.)  The integrals are simply evaluated on space-time by introducing the spin frame $(o_\alpha,\iota_\beta)$ such that 
 \begin{equation}
     x^{\alpha\beta}=t^{(\alpha}_{\dot\beta}x^{\beta)\dot\alpha}=r \bo^{(\alpha}\bi^{\beta)}\, ,
 \end{equation}
 by setting
 \begin{equation}
 \qquad \bo_\alpha=\frac{(1, z)}{\sqrt{1+|z|^2}}\, , \quad \bi_\alpha=\frac{(\bar z, -1)}{\sqrt{1+|z|^2}}\, , \quad z =\frac{x^1+ix^2}{x^3}\, .     
 \end{equation}
Expressing $\lambda_\alpha$, $\phi$, and $I$ in this spin frame, gives $\mu=r\langle\bi \lambda\rangle\langle \bo\lambda\rangle$ and $I$ a polynomial. The integral \eqref{4.11} reduces to a simple residue calculation at  $\langle\bi \lambda\rangle/\langle \bo\lambda\rangle=0$ for each component of the field, giving
\begin{equation}
    {}_{(n)}\phi_p={}_{(n)}\phi_{\alpha_1\cdots\alpha_{2n}}
\bi^{\alpha_1}\cdots\bi^{\alpha_p}\bo^{\alpha_{p+1}}\cdots \bo^{\alpha_{2n}} = \frac{(-1)^{n+k-p}I_{2k}^p}{r^{n+k+1}} 
\end{equation}
where $n+k\geq p \geq n-k$ otherwise the contraction vanishes. The numerator is a spin-weighted spherical harmonic \cite{penrose1984spinors} given in the form
\begin{equation}
    I_{2k}^p=I_{2k}^{\alpha_1\cdots \alpha_{2k}}\bi_{\alpha_1}\cdots \bi_{\alpha_{p+k-n}}\bo_{\alpha_{p+k-n+1}}\ldots \bo_{\alpha_{2k}}\, .
\end{equation}
To obtain a nontrivial charge, the combined fall-off of ${}_{(n)}\phi_{\alpha_1\ldots \alpha_{2n}}\xi^{\alpha_3\ldots \alpha_{2n}}$ should be $1/r^2$ otherwise the integral will vanish either for large or small $r$.  

A solution to the higher valence twistor equation, which has growth $r^{n+k}$ can be constructed from the twistor representative
\begin{equation} \label{4.25}
    \xi_{2n} = \tilde\xi_{\dot \alpha_{1}\cdots \dot \alpha_{n+k}}^{\beta_1\cdots\beta_{n-k}}\mu^{\dot \alpha_1}\cdots \mu^{\dot \alpha_{n+k}}\lambda_{\beta_1}\cdots\lambda_{\beta_{n-k}}
\end{equation}
where $\tilde \xi^{\cdots}_{\cdots}$ is a constant mixed spinor. Note that this definition requires $0\leq k\leq n$. Using the incidence relations \eqref{FlatIncidenceRelations} and Bondi coordinates given by \eqref{BondiCoordinatesSpinor}, we find the following large $r$ expansion
\begin{equation}
    \xi_{2n} = r^{n+k} \hat \xi^{\alpha_1 \cdots\alpha_{2n}} \lambda_{\alpha_1}\cdots\lambda_{\alpha_{2n}} + O(r^{n+k-1})\, ,
\end{equation}
where
\begin{equation}
    \hat \xi^{\alpha_1 \cdots\alpha_{2n}} = \bo^{(\alpha_1}\cdots \bo^{\alpha_{n+k}}\tilde\xi_{\dot \alpha_{1}\cdots \dot \alpha_{n+k}}^{\alpha_{n+k+1}\cdots\alpha_{2n})} \bar \bo^{\dot \alpha_1}\cdots \bar \bo^{\alpha_{n+k}}\, .
\end{equation}
The integrated expression for the charge at infinity is given by
\begin{equation}
     H_{s=2n-1} =\int_{S}{}_{(n+1)}\phi_{\alpha_1\cdots\alpha_{2n}\alpha\beta}\xi^{\alpha_1\cdots\alpha_{2n}}\Sigma^{\alpha\beta}  = \frac{4\pi (-1)^{n+k}}{n+k+1} I_{2k}^{\alpha_{n-k+1}\cdots \alpha_{n+k}} \tilde \xi^{\alpha_1\cdots\alpha_{n-k}}_{\alpha_1 \cdots\alpha_{n+k} } \label{multipole-duality}
\end{equation}
where the dotted spinor indices have been converted to undotted ones through contraction of the time-like vector. The representative \eqref{4.25} selects the $k$th multipole
contribution displayed in \eqref{multipole-duality}. More generally, a
polynomial representative of homogeneity $2n$ gives a
finite linear combination of multipoles $I_{2k}$ with
$0\leq k\leq n$, even when the field has an infinite
multipole expansion. For regular stationary data, these
contributions are finite: by \eqref{cons 2.20}, the charge integrand
is a closed two-form, so its integral is independent of
the enclosing two-surface in the source-free exterior.
Angular singularities associated with meromorphic
representatives are a separate issue, discussed below.

Note that the contractions of $\xi$ appearing in \eqref{multipole-duality} mean that the duality is not perfect, and there are plenty of different choices of $\xi$ that pick out the same multipole components; there are more  $Lw_{1+\infty}$ charges than multipole components.

\paragraph{Celestial charges and recursion.} The details for celestial charges work quite differently from the case of Geroch recursion and it is not so straightforward to land on concrete formulae. The Geroch recursion takes a derivative in increasing the spin of a field $\phi$ by one and hence increases the fall-off in $r$ by one.  However, the recursion of \eqref{recurs1} is essentially of degree zero, with a differentiation followed by an integration. Thus multipoles of different fall-offs will appear in different places.  Rather more seriously, the integration implicit in the recursion \eqref{recurs1} will introduce wire singularities when there are sources. This can be seen explicitly by taking the recursion of \eqref{multipole} which gives say  $f_{-2n-2}/\lambda_0^r $ and then evaluating on space-time using a contour that surrounds the pole in $\langle \bo \lambda\rangle$ so that  the new denominator factors $\lambda_0^{-r}$ yields an extra  $(\sqrt{1+|z|^2}/z)^{-r}$ factor. The freedom in  choice of different contours reflects ambiguities in the solution to \eqref{recurs1}.    The singularities mean that the remaining $\mathcal{S}$-integrals will in general need regulation to make sense for those $w^p_m$ that do not entail sufficient cancellation.   These are very much analogous to the poles found  for example, in super-rotations.

\subsection{Twistor theory for SD spacetimes} 
Having obtained the twistor definitions  for flat spacetimes, Penrose's non-linear gravitation theorem \cite{Penrose:1976jq, Penrose:1976js} allows us to extend these considerations to curved SD spacetimes by deforming the complex structure. The complex structure of $\mathbb{PT}$ is a splitting of holomorphic and anti-holomorphic coordinates such that the exterior derivative splits as $d = \partial + \bar\partial$, where
\begin{equation}
    \partial = d Z^A \frac{\partial}{\partial Z^A} \qquad \bar \partial = d \bar Z^A \frac{\partial}{\partial \bar Z^A} . 
\end{equation}
A deformation of complex structure can be represented by a change in the Dolbeault operator
\begin{equation}
    \bar \partial \longrightarrow  \bar \nabla =\bar  \partial +V ,\qquad V = V^A_{\,\,\, B} \,d\bar Z^B \otimes\frac{\partial}{\partial Z^A}
\end{equation}
such that $V \in \Omega^{0,1}(\mathbb{PT}, T^{1,0}\mathbb{PT})$. A general deformation only defines an almost complex structure on twistor space, as one must consider global issues. The deformed Dolbeault operator $\bar \nabla$ must satisfy an integrability condition, $\bar \nabla^2 = 0$, to be a complex structure. This condition gives the Maurer-Cartan equation on the deformation
\begin{equation}
    \bar\partial V +\frac{1}{2}[V,V] = 0
\end{equation}
where $[\cdot,\cdot]$ is the Schouten–Nijenhuis bracket defined as the commutator between the vector fields combined with the wedge product on $(0,1)$-forms. The new complex manifold built using this complex structure is denoted $\mathcal{PT}$. In general, such complex structure deformations provide twistor spaces for spacetimes with SD conformal structure, i.e., they only guarantee that $\psi_{\alpha\beta\gamma\delta} = 0$. One must add extra structures on the deformed twistor space to give the further Ricci-flat restriction on spacetime for the non-linear graviton theorem. The first  is the preservation of the holmorphic fibration over $\mathbb{CP}^1$ of the deformed twistor space so that $D\lambda$ remains a holomorphic $(1,0)$-form, in particular $\lambda_\alpha$ should remain  holomorphic. The second  is the preservation  of the holmorphic Poisson structure on the fibers of the   fibration given in local coordinates by
\begin{equation}
    \{f,g\}= \varepsilon^{\dot\alpha\dot\beta}\frac{\p f}{\p\mu^{\dot\alpha}}\frac{\p g}{\p\mu^{\dot\beta}}\, ,\label{Poisson}
\end{equation}
and this extends so that  $f,g\in \Omega^{p,q}_\mathbb{PT}(\cO(n))$.
More generally $\p/\p\mu^{\dot\alpha}=\partial_{\dot \alpha}$ is the dual vector field to $e^{\dot \alpha}$ such that $\partial_{\dot \alpha}\lrcorner \, e^{\dot \beta} = \delta^{\dot \beta}_{\dot \alpha}$ and all other contractions vanishing. 
The preservation of the Poisson structure means that the complex structure deformation is generated from a Hamiltonian $h\in\Omega^{0,1}_\PT(\cO(n))$. Therefore, $V = \{h, \cdot \}$ and the integrability condition reduces to
\begin{equation}
    \bar\partial h + \frac{1}{2}\{h,h\} = 0 . 
\end{equation}

Under the new complex structure, one defines a new basis of $(1,0)$-forms by the condition
\begin{equation}
    \bar\nabla\lrcorner D\lambda = 0, \qquad \bar\nabla \lrcorner\,\theta ^{\dot \alpha} = 0, 
\end{equation}
where $D\lambda$ is undeformed, but $\theta^{\dot \alpha}$ are given by
\begin{equation}
    \theta ^{\dot \alpha} = e^{\dot \alpha} - \epsilon^{\dot \alpha\dot \beta}\mathcal{L}_{\partial_{\dot \beta}}h . 
\end{equation}
In the deformed complex structure, the holomorphic volume form is given by
\begin{equation}
    \Omega = \frac{1}{2}D\lambda \wedge \theta^{\dot \alpha}\wedge \theta_{\dot \alpha} = D^3Z - D\lambda \wedge e^{\dot \alpha}\wedge \mathcal{L}_{\partial_{\dot \alpha}} h + \frac{1}{2}D\lambda \wedge \{h,h\} . 
\end{equation}

To recover the spacetime, one must construct the incidence relations on the deformed twistor space $\mu^{\dot \alpha} = F^{\dot \alpha}(x,\lambda)$ where $F^{\dot \alpha}$ is homogeneous $1$ in $\lambda_\alpha$ but no longer holomorphic. On each twistor line, the functions $F^{\dot \alpha}$ must satisfy
\begin{equation}\label{curvedincidence}
    \bar\partial\vert_X F^{\dot \alpha} = \epsilon^{\dot \alpha\dot \beta}\left.\mathcal{L}_{\partial_{\dot \beta}}h\right|_X
\end{equation}
which defines the holomorphic curves in twistor space. The holomorphic functions $F^{\dot \alpha}$ are then used to define the map 
\begin{equation}
\begin{split}
      p&: \mathbb{PS}\rightarrow \mathcal{PT} \\
      &(x,\lambda)\mapsto (\mu^{\dot \alpha} = F^{\dot \alpha}(x,\lambda), \lambda_\alpha)
\end{split}
\end{equation}
where $\mathbb{PS}$ is the projective spin bundle which locally looks like $\mathcal{M}\times \mathbb{CP}^1$. On the spin bundle, the condition for the holomorphic curves can be translated to the condition
\begin{equation}
    L_{\dot \alpha} F^{\dot \beta} = \lambda^{\alpha}\nabla_{\alpha\dot\alpha }F^{\dot \beta} = 0 .
\end{equation}
The differential operators $L_{\dot \alpha}$ are called the Lax pair, which satisfy $[L_{\dot \alpha}, L_{\dot \beta}] = 0$, modulo terms proportional to the Lax pair. The holomorphic volume form $\Omega$ has a simple form when pulled back from $\mathcal{PT}$ to $\mathbb{PS}$. Using \eqref{curvedincidence}, one can show 
\begin{equation}
    \mathcal{L}_{\bar\partial_0}p^* \Omega = 0 
\end{equation}
where $\bar \partial_0$ is the anti-holomorphic vector field on $\mathbb{CP}^1$. Since $p^*\Omega$ is valued in $\mathcal{O}(2)$ and holomorphic, by Liouville’s theorem 
\begin{equation}
    p^* \Omega = D\lambda\wedge \Sigma^{\alpha\beta}\lambda_\alpha\lambda_\beta
\end{equation}
where $\Sigma^{\alpha\beta}$ are the basis of ASD 2-forms. 

On the curved SD background, one can construct an ASD perturbation on twistor space using the Penrose transform. In particular, there exists an isomorphism
\begin{equation}
        g\in H^{0,1}(\mathcal{PT}, \mathcal{O}(-6)) \simeq \{ \phi_{\alpha\beta\gamma\delta} \,\,\text{on }\, \mathcal{M}\vert \nabla^{\alpha\dot \alpha} \phi_{\alpha\beta\gamma\delta} = 0\}. 
\end{equation}
The correspondence can be made explicit through the use of an integral formula over the Riemann sphere $X\simeq \mathbb{CP}^1$
\begin{equation}
    \phi_{\alpha\beta\gamma\delta} = \frac{1}{2\pi i}\int_X D\lambda\wedge\lambda_\alpha\lambda_\beta\lambda_\gamma\lambda_\delta\, g\vert_X . 
\end{equation}
The ASD perturbation only gives the perturbations of  the ASD Weyl spinor. Purely ASD  perturbations of the metric are  obstructed by the SD background curvature \cite{Mason:2009afn}. 
However, for the SD perturbations, one can construct the perturbation to the metric explicitly via a potential. The relevant cohomology class giving the SD perturbations is
\begin{equation}
    \delta h \in H^{0,1}(\mathcal{PT},\mathcal{O}(2))\simeq \{\delta g_{ab} \,\,\text{on }\, \mathcal{M}\vert \psi_{\alpha\beta\gamma\delta} = 0\} . 
\end{equation}
If one restricts the perturbation to the holomorphic curve $X$, a short calculation shows that $\delta h\vert_X \in H^{0,1}(\mathbb{CP}^1,\mathcal{O}(2)) $ because
\begin{equation}
    (\bar \nabla \delta h)\vert_X = \bar\partial\vert_X \delta h\vert_X - \bar\partial\vert_X F^{\dot \alpha}\left.\frac{\partial \delta h}{\partial\mu^{\dot \alpha}}\right|_X + \left.\{h,\delta h\}\right|_X =\bar\partial\vert_X \delta h\vert_X = 0 \, ,
\end{equation}
where we have used \eqref{curvedincidence}. Since $H^{0,1}(\mathbb{CP}^1,\mathcal{O}(2))$ is trivial, the restriction of the SD perturbation is exact
\begin{equation}
    \delta h\vert_X = \bar \partial\vert_X j \, ,
\end{equation}
where $j = j(x,\lambda)$ is a function on $\mathbb{PS}$ with homogeneity $2$. Using the fact that $\bar \nabla\delta h=0$, it follows that $L_{\dot \alpha} \delta h \vert_X = 0$. Therefore, due the holomorphicity of $L_{\dot \alpha}$, the function $L_{\dot \alpha} j $ is globally holomorphic of weight $3$ and by Liouville's theorem can be decomposed
\begin{equation}
    L_{\dot \alpha} j =\lambda^{\alpha}\lambda^{\beta}\lambda^{\gamma}\varphi_{\alpha\beta\gamma\dot \alpha}\, ,
\end{equation}
where $\varphi_{\alpha\beta\gamma\dot \alpha}$ is a spin field on $\mathcal{M}$ satisfying 
\begin{equation}\label{metricpotential}
    \nabla_{(\alpha}^{\,\,\,\dot\alpha}\varphi_{\beta\gamma\delta)\dot \alpha} = 0 . 
\end{equation}
This is a potential for perturbation to the metric
\begin{equation}\label{metricperturbation}
    \delta g_{\alpha\dot \alpha\beta\dot \beta} = \nabla^{\gamma}_{\,\,\,(\dot \alpha}\varphi_{\dot \beta)\alpha\beta\gamma}\, ,
\end{equation}
which gives a vanishing linearised ASD Weyl spinor. We further note $\varphi_{\alpha\beta\gamma \dot \alpha}$ is a potential modulo gauge description for SD perturbation, as \eqref{metricpotential} has a gauge symmetry on a SD background given by $\varphi_{\alpha\beta\gamma \dot \alpha} \sim \varphi_{\alpha\beta\gamma \dot \alpha} + \nabla_{\dot \alpha (\alpha}\omega_{\beta\gamma)}$. At the level of the twistor data, this gauge symmetry can be seen from the ambiguity of defining $j$ as $j\sim j + \omega$ where $\omega$ is a holomorphic weight $2$ function that would give the same $\delta h\vert_X$. Acting on $\omega$ with the Lax operators reproduces the gauge symmetry. As a result, one can choose a gauge in which all the undotted indices point in the same direction $\varphi_{\alpha\beta\gamma \dot \alpha} = \iota_{\alpha}\iota_{\beta}\iota_{\gamma}\varphi_{\dot \alpha}$ for some reference spinor $\iota_{\alpha}$ as an example.

We now consider the description of solutions to the twistor equation using twistor space data. A solution to the valence $n$ twistor equation \eqref{Highertwistorequation} corresponds to elements of cohomology
\begin{equation}
    \xi_n \in H^0(\mathcal{PT}, \mathcal{O}(n))\simeq\left\{\xi^{\alpha_1\cdots\alpha_n} \,\,\, \text{on}\, \,\mathcal{M}\vert \nabla^{(\alpha_0}_{\dot\alpha}\xi^{\alpha_1\cdots\alpha_{s+1})} = 0\right\} . 
\end{equation}
A similar short calculation shows
\begin{equation}
    (\bar\nabla \xi_n)\vert_X = \bar\partial\vert_X \xi_n\vert_X - \bar\partial\vert_X F^{\dot \alpha}\left.\frac{\partial \xi_n}{\partial\mu^{\dot \alpha}}\right|_X + \left.\{h,\xi_n\}\right|_X =\bar\partial\vert_X \xi_n\vert_X\, ,
\end{equation}
where we have used \eqref{curvedincidence}. Therefore, if $\xi_n$ is holomorphic with respect to the deformed complex structure, $\bar\nabla\xi_s= 0$, then on each twistor line $\xi_s\vert_X$ is holomorphic in $\lambda_\alpha$ with homogenous weight $n$. Therefore,
\begin{equation}
    \xi_n\vert_X  = \xi^{\alpha_1\cdots\alpha_n}\lambda_{\alpha_1}\cdots\lambda_{\alpha_n} . 
\end{equation}
Since $\xi_n\vert_X = \xi_n(F^{\dot \alpha}(x,\lambda),\lambda_{\alpha})$ the action of the Lax operator gives the twistor equation
\begin{equation}
    L_{\dot \alpha}\xi_n\vert_X = \nabla^{(\alpha_0}_{\quad\dot \alpha}\xi^{\alpha_1\cdots\alpha_n)}\lambda_{\alpha_0}\cdots\lambda_{\alpha_n} = 0 . 
\end{equation}
However, on a curved twistor space, such $\xi_n$ generically do not exist unless they are constant (constant undotted spinors always exist on a self-dual Ricci-flat manifold). Only special twistor spaces admit sections of $H^0(\mathcal{PT}, \mathcal{O}(n))$ \cite{Ward1978, Dunajski:2000iq, Dunajski:2003gp} corresponding to non-constant Killing spinors. Since the twistor equation is an overdetermined system even on a SD background, the existence of such functions on $\mathcal{PT}$ is not guaranteed beyond homogenous polynomials on the $\mathbb{CP}^1$ base. We will therefore consider a weaker version of the
twistor equation associated with residual gauge
transformations on twistor space. These are restricted
so that their induced spacetime perturbations preserve
the chosen gauge for the metric potential, as
explained in Section~\ref{sec:Large Gauge Transformations}.

\subsection{Twistor action and charges}

The action describing SD gravity with an ASD perturbation is given by \cite{Mason:2007ct}
\begin{equation} \label{TwistorAction}
    S[g,h] = \int_{\mathbb{PT}}D^3Z\wedge g \wedge \left(\bar \partial h + \frac{1}{2}\{h,h\}\right) . 
\end{equation}
The equations of motion for this action are given by
\begin{equation}
    \bar\partial h + \frac{1}{2}\{h,h\} = 0 \qquad\quad \bar\partial g + \{h,g\} = 0 . 
\end{equation}
The first is the integrability condition for $h$ to be the Hamiltonian generating complex structure deformations of $\mathbb{PT}$. The second states that $g$ is $\bar\p$-closed $(0,1)$-form for the deformed complex structure. Along with the equations of motion, the symmetries of this action are given by
\begin{equation}
    \delta_{\xi} h = \bar \nabla \xi , \qquad \qquad \delta_{\xi,\chi} g = \{g, \xi\} + \bar\nabla \chi . 
\end{equation}
The $\xi$ symmetry describes symplectomorphisms on twistor space, while the $\chi$ symmetry makes $g$ valued in Dolbeault cohomology. 

Using standard phase-space methods, e.g.\ \cite{Iyer:1994ys,Wald:1999wa,Barnich:2001jy,Barnich:2003xg}, the twistor action \eqref{TwistorAction} yields the symplectic potential 
\begin{equation}
    \theta[\delta] = \Omega \wedge \delta h \wedge g\,  . 
\end{equation}
The symplectic current arises by taking a further variation to give
\begin{equation}
\begin{split}
    \omega[\delta_1,\delta_2] &= \delta_1\theta[\delta_2] - \delta_2\theta[\delta_1]\\
    &= \delta_1\Omega \wedge \delta_2h\wedge g + \Omega\wedge \delta_2 h\wedge \delta_1 g - (1\leftrightarrow2) . 
\end{split}
\end{equation}
On shell, the symmetries contracted into the symplectic current give exact terms
\begin{equation}
    \begin{split}
        \delta_\chi \lrcorner\, \omega &= d\left(\Omega \wedge \chi \delta h\right),\\
        \delta_\xi \lrcorner\, \omega &= d\left(\Omega\wedge \xi \delta g - D\lambda \wedge \theta^{\dot \alpha}\wedge \mathcal{L}_{\partial_{\dot \alpha}}(\xi \delta h)\wedge g\right) . 
    \end{split}
\end{equation}
We now integrate over a co-dimension 1 hypersurface $\Sigma$ on twistor space, resulting in charge expressions that are integrals over the boundary of $\Sigma$. In general, the contraction of the symmetries will not give a $\delta$-exact term but will split into an integrable and non-integrable part 
 \begin{equation}
     \delta_\alpha\lrcorner\, \int_\Sigma \omega = \delta H_\alpha - \Xi_\alpha[\delta]\, ,
 \end{equation}
where $\alpha = (\xi , \chi)$ and the charges decompose into two parts $H_\alpha = (H_\xi,\tilde H_\chi)$. The integrable part gives the charges for each symmetry as integrals over $\partial \Sigma$
\begin{equation}
    H_{\xi} = \frac{1}{2\pi i}\int_{\partial \Sigma}\Omega \wedge \xi g, \quad \quad H_\chi = \frac{1}{2\pi i}\int_{\partial\Sigma}\Omega\wedge \chi h . 
\end{equation}
The non-integrable part of the charges is given by 
\begin{equation}
\begin{split}
    \Xi_\xi[\delta] &= \int_{\partial \Sigma} \delta\Omega \wedge \xi g + \Omega \wedge \delta \xi g +  D\lambda \wedge \theta^{\dot \alpha}\wedge \mathcal{L}_{\partial_{\dot \alpha}}(\xi \delta h)\wedge g ,\\
    \Xi_\chi[\delta] &=\int_{\partial\Sigma} \delta\Omega \wedge \chi h + \Omega \wedge \delta \chi h . 
\end{split}
\end{equation}
From these expressions, we observe that the non-integrability of charges comes not only from the possibility of the gauge parameters being field dependent, but also because we work on a curved spacetime background. This is in contrast to what occurs in the gauge theory case, where only the variation of the gauge parameters appears \cite{Kmec:2025ftx}. 

In the presence of non-integrable charges, a new charge bracket must be defined \cite{Barnich:2011mi} that utilises the non-integrable part
\begin{equation}\label{ChargeBracket}
    \{H_\alpha ,H_{\alpha'}\}_\star = \delta_\alpha H_{\alpha'} - \Xi_\alpha [\delta_{\alpha'}] . 
\end{equation}
The charge algebra is given by
\begin{equation}
    \begin{split}
        \{H_\xi , H_{\xi'}\} &= H_{\{\xi , \xi'\}_*} , \\
         \{H_\xi , \tilde H_{\chi}\} &= - \{\tilde H_\chi ,  H_{\xi}\}=\tilde H_{\{\xi, \chi\}_*} +  K_{\xi,\chi} , \\
          \{\tilde H_\chi , \tilde H_{\chi'}\} &= 0 ,
    \end{split}
\end{equation}
where the modified bracket for the gauge parameters is given by \cite{Barnich:2010xq, Barnich:2010eb}
\begin{equation}\label{ModifiedBrackets}
    \begin{split}
        \{\xi,\xi'\}_* &= \{\xi,\xi'\}+\delta_\xi \xi' - \delta_{\xi'}\xi , \\
         \{\xi,\chi\}_* &= \{\xi,\chi\}+\delta_\xi \chi \, . 
    \end{split}
\end{equation}
The charge algebra picks up a field-dependent cocycle given by 
\begin{equation}
    \begin{split}
        K_{\xi,\chi}& =\int_{\partial \Sigma} \Omega\wedge \chi \bar\partial \xi + D\lambda \wedge \theta^{\dot \alpha}\mathcal{L}_{\partial_{\dot \alpha}}\xi\wedge \bar \nabla(\chi h) . 
    \end{split}
\end{equation}
In general, field-dependent cocycles must satisfy the generalized cocycle condition \cite{Barnich:2011mi} given by
\begin{equation}
    K_{\{\alpha_1,\alpha_2\},\alpha_3} - \delta_{\alpha_3}K_{\alpha_1,\alpha_2} + \text{cyclic}(1,2,3) = 0 . 
\end{equation}
Finally, we note that the splitting of integrable and non-integrable is not unique, as we can make the redefinition
\begin{equation}
    H_\alpha \rightarrow H'_\alpha = H_\alpha + N_\alpha , \qquad \qquad \Xi_\alpha \rightarrow \Xi'_\alpha = \Xi_\alpha + \delta N_\alpha  . 
\end{equation}
This redefinition would affect the charge algebra by modifying the field-dependent cocycle
\begin{equation}
    \{H'_\alpha, H'_{\alpha'}\}_\star = H'_{\{\alpha,\alpha'\}_*} + K'_{\alpha,\alpha'} ,\qquad  K'_{\alpha,\alpha'} = K_{\alpha,\alpha'}  \,+ \delta_\alpha N_{\alpha'} - \delta_{\alpha'}N_{\alpha} -N_{\{\alpha,\alpha'\}_*} . 
\end{equation}
In this way, the new part of the cocycle automatically satisfies the generalized cocycle condition, and is trivial, as it can be eliminated by a redefinition of the charge. Due to the subtleties of field dependence, the charges cannot be seen directly as generating variation of the fields when using the charge bracket \eqref{ChargeBracket}. However, one can show that the variations form a representation of the modified brackets \eqref{ModifiedBrackets}
\begin{equation}
    [\delta_\xi, \delta_{\xi'}] = \delta_{\{\xi,\xi'\}_*} , \qquad [\delta_\xi, \delta_\chi] = \delta_{\{\xi, \chi\}_*} , \qquad [\delta_\chi, \delta_{\chi'}] = 0 .  
\end{equation}

\subsection{Large gauge transformations}\label{sec:Large Gauge Transformations}
We will now choose a gauge in which to work, such that the SD perturbations when restricted to a holomorphic curve $\delta h\vert_X = \bar\partial\vert_X j$ satisfy 
\begin{equation}\label{gaugechoice}
  L_{\dot \alpha} j = \langle\iota \lambda\rangle^3 \varphi_{\dot \alpha} . 
\end{equation}
This gauge choice uses the redundancy for the potential for the metric perturbation on spacetime $\varphi_{\alpha\beta\gamma \dot \alpha} = \iota_\alpha \iota_\beta\iota_\gamma \varphi_{\dot \alpha}$. From the action, the variation of the background $ h$ is given explicitly by
\begin{equation}
    \delta h\vert_X = \bar\nabla\xi\vert_X = \bar\partial\vert_X \xi\vert_X . 
\end{equation}
Non-trivial SD perturbations correspond to $\xi$ which are not global on twistor space. The $\xi$ that generate such SD perturbations were considered in \cite{Adamo:2021lrv} in the framework of \v Cech cohomology. On twistor space, these generators are explicitly given by 
\begin{equation}
    \check{\xi}_s = \sum_{\vert m\vert \leq p-1}\sum_{a\in \mathbb{Z}}\tau_{m,a}^p w_{m,a}^p  , \qquad w_{m,a}^p = \frac{\left(\mu^{\dot 0}\right)^{p+m-1}\left(\mu^{\dot 1}\right)^{p-m-1}}{\lambda_0^{2p-4-a}\lambda_1^a}\, ,
\end{equation}
where $s=2p-3$. The generators $w_{m,a}^p$ are holomorphic on the patch of twistor space where $\lambda_1 \neq 0 $ and $\lambda_0 \neq 0$ and they satisfy the algebra $LHam(\mathbb{C}^2)$ (also referred to as $Lw_{1+\infty}$, see Footnote \ref{celestial algebra name}), given explicitly as
\begin{equation}
    \{w^p_{m,a} , w^q_{n,b}\} = 2(m(q-1) - n(p-1))w^{p+q-2}_{m+n,a+b} . 
\end{equation}
As we are working in a Dolbeault framework, we wish to translate the holomorphic sections of $\mathcal{O}(2)$ $\check{\xi}_s$ to their smooth counterparts $\xi_s$ also of weight two. To achieve this, we utilize bump functions. On $\mathbb{CP}^1$, there exists an open cover $\{U_+, U_-\}$ where $U_- = \{\lambda_0 \neq0\}$ and $U_+ = \{\lambda_1 \neq0\}$ describe the two hemispheres. We can construct a partition of unity $\{f_-, f _ {+} \}$ where $f_{\pm}$ are functions having compact support in $U_{\pm}$ and satisfy $f_- + f_+ = 1$.  When we pull back $\check{\xi}_s$ to the spin bundle, in general, it will consist of two terms
\begin{equation}
    \check{\xi}_s\vert_X = \check{\xi}_s^+ - \check{\xi}_s^- \, ,
\end{equation}
where $\check{\xi}_s^-$ carries singular terms in $\lambda_0$ and $\check{\xi}_s^+$ carries singular terms in $\lambda_1$. In principle, one can also have terms that are regular everywhere and hence holomorphic. Where one puts these regular terms will determine which gauge one works in for the $\varphi_{\alpha\beta\gamma\dot \alpha}$. The smooth function $\xi_s\vert_X$ is defined as
\begin{equation}
    \xi_s\vert_X = \check{\xi}_s^+ f_+ + \check{\xi}_s^- f_- .
\end{equation}
Therefore, the perturbation to $h$ is given by
\begin{equation}
    \delta h\vert_X  = \bar\partial\vert_X\left(\check{\xi}_s^+ f_+ + \check{\xi}_s^- f_-\right) . 
\end{equation}
Now applying the Lax operator on $\xi_s\vert_X$ and using the fact that $L_{\dot \alpha}\check{\xi}_s\vert_X = 0$, we find
\begin{equation}
    L_{\dot \alpha} \xi_s\vert_X = L_{\dot \alpha}\check \xi^-_s = \langle\iota\lambda\rangle^3 \varphi_{\dot \alpha} . 
\end{equation}
We therefore work in a patch where $\lambda_\alpha \neq \iota_\alpha$, i.e. $\lambda_0 \neq 0$, and allow for arbitrarily high poles when $\lambda_\alpha = \iota_\alpha$. We also place all the regular parts in $\check{\xi}^-_s$ which explicitly can be expand as
\begin{equation} \label{XiExapansion}
    \check{\xi}_s^-= \frac{\xi^{\alpha_1 \cdots\alpha_{s+1}}\lambda_{\alpha_1}\cdots\lambda_{\alpha_{s+1}}}{\langle \iota \lambda\rangle^{s-1}}\, ,
\end{equation}
where, for the moment, we ignore the loop part of the algebra. We note that we retain even the non-singular terms for values of $s=-1,0,1$, as these can be understood on space-time as diffeomorphisms preserving the gauge choice of the metric (see Section \ref{sec:Celestial symmetries of Plebanski's equations}). In other words, the resulting variations of the metric end up being local while the singular gauge parameters give non-local variations. The SD perturbation is given by
\begin{equation}\label{CovRecursionRelations}
    L_{\dot \alpha}\check{\xi}_s^- = \frac{\lambda_{\alpha_0}\cdots\lambda_{\alpha_{s+1}}\nabla^{\alpha_0 }_{\,\,\,\,\dot \alpha}\xi^{\alpha_1 \cdots \alpha_{s+1}}}{\langle\iota\lambda\rangle^{s-1}} . 
\end{equation}
To preserve the gauge choice \eqref{gaugechoice}, the field $\xi^{\alpha_1 \cdots \alpha_{s+1}}$ must satisfy 
\begin{equation} \label{gaugeparameter}
    \iota_{\alpha_0} \nabla^{(\alpha_0}_{\,\,\,\,\,\dot \alpha }\,\xi^{\alpha_1 \cdots\alpha_{s+1})} = 0 ,
\end{equation}
which results in the expression for the potential for the SD metric perturbation to be
\begin{equation}
    \varphi_{\dot \alpha} = o_{\alpha_0}\cdots o_{\alpha_{s+1}}\nabla^{\alpha_0}_{ \,\,\,\,\dot \alpha}\xi^{\alpha_1 \cdots \alpha_{s+1}} . 
\end{equation}
In the GHP formalism, the equations \eqref{gaugeparameter} can be decomposed into two towers of equations called dual recursion relations
\begin{equation} \label{Dual Recursion}
\begin{split}
    \text{\th}_{\mathcal{C}}'\xi_{s,m} - \bar \eth_{\mathcal{C}}\xi_{s,m+1} + (m+2)\sigma\xi_{s,m+2}+ (3-m)\nu \xi_{s,m-1} =0 , \\
    \eth_{\mathcal{C}}\xi_{s,m} - \text{\th}_{\mathcal{C}}\xi_{s,m+1}+(m+2)\kappa\xi_{s,m+2}+(3-m)\lambda\xi_{s,m-1} = 0 ,
\end{split}
\end{equation}
where the index $m = 0, \ldots, s+1$\footnote{We note that the covariant derivative appearing in \eqref{CovRecursionRelations} include the fact that the spin field $\xi^{\alpha_1\cdots\alpha_{s+1}}$ has a GHP weight $(1-s,0)$ since the twistor function $\xi_s$ has no weight. This results in $\xi_{s,m}$ having weight $(2-2m,0)$ and conformal weigth $W=m$.}. These include the heuristic twistor equations displayed in the two first equations of \eqref{twistor equation null} when $s=1$. The components appearing in the dual recursion relations are components of $\xi^{\alpha_1\cdots\alpha_{s+1}}$ explicitly given by
\begin{equation}
    \xi^{\alpha_1\cdots\alpha_{s+1}} = \sum_{m=0}^{s+1}\xi_{s,m}o^{(\alpha_1}\cdots o^{\alpha_m} \iota^{\alpha_{m+1}}\cdots\iota^{\alpha_{s+1})} . 
\end{equation} The two equations that appear in the SD perturbation $\varphi_{\dot \alpha}$ are given by
\begin{equation}\label{PotentialXi}
    \varphi_{\dot \alpha} = \bar o _{\dot \alpha}\left(\bar\eth_{\mathcal{C}}\xi_{s,0} - \sigma \xi_{s,1}\right)  -\bar \iota_{\dot \alpha}\left(\text{\th}_{\mathcal{C}}\xi_{s,0} -\kappa \xi_{s,1} \right)\, .
\end{equation}
As discussed in Section \ref{sec:Celestial charges from twistor space}, the potential $\varphi_{\alpha\beta\gamma\dot\alpha}$ generates a SD perturbation to the metric if \eqref{metricpotential} is satisfied. To check this, denote $\varphi_{\dot \alpha} = \varphi_0 \bar o_{\dot \alpha} - \varphi_1 \bar\iota_{\dot \alpha}$, after which \eqref{metricpotential} gives
\begin{equation}
\begin{split}
    \nu \varphi_1 &- \lambda \varphi_0 = 0 ,  \qquad  \quad \bar\eth\varphi_1 + \bar\rho\varphi_0 - \text{\th}\varphi_0 + \bar \pi\varphi_1 = 0 , \\
    &\text{\th}'\varphi_1 + \bar\tau\varphi_0 - \eth \varphi_0 + \bar\mu \varphi_1 + 3\pi \varphi_0 - 3\mu\varphi_1 = 0 . 
\end{split}
\end{equation}
The first equation requires us to set $\lambda = 0 = \nu$, which is a consequence of choosing to work in the gauge $\varphi_{\alpha\beta\gamma\dot \alpha} = \iota_\alpha\iota_\beta\iota_\gamma\varphi_{\dot \alpha}$. This means that the intrinsic shear $\lambda$ of the null
hypersurface $\mathcal N$ generated by the vector $n^\mu$
of the null tetrad vanishes, and hence $\mathcal N$ is
foliated by $\alpha$-planes. Furthermore, $\nu=0$ in the
real Minkowski setting implies that $n^\mu$ is geodesic
and hence generates a shear-free null geodesic congruence. In more general signatures this needs to be supplemented by $\bar\nu = 0$. The last two equations are identically zero when the explicit expressions for $\varphi_0,\varphi_1$ appearing in \eqref{PotentialXi} are used, assuming the spacetime is SD, the vanishing of the spin coefficients $\lambda, \nu$, and the $\xi_{s,m}$ satisfy the dual recursion relations \eqref{Dual Recursion}. The fact that the $\xi_{s,m}$ satisfy \eqref{metricpotential} requires the use of various spin coefficient identities and commutators of GHP operators, see Appendix \ref{sec:Useful formulae}, making the calculation non-trivial. This calculation becomes significantly simpler when a special gauge is used; see Section \ref{sec:Celestial symmetries of Plebanski's equations}. The general metric perturbation can be written explicitly as 
\begin{equation}
\begin{split}
    \delta g_{\alpha\dot \alpha\beta\dot\beta} = \iota_\alpha\iota_\beta&\left[\bar o _{\dot\alpha}\bar o_{\dot \beta}\left(\bar\nu\varphi_1 - \text{\th}'\varphi_0 - \mu\varphi_0 \right) + \bar\iota_{(\dot \alpha}\bar o_{\dot \beta)}\left( \eth\varphi_0 + 2\mu \varphi_1 - \bar\mu \varphi_1 - \pi\varphi_0\right) \right.\\
    &\qquad\left.-\iota_{\dot \alpha}\iota_{\dot \beta}\left(\eth\varphi_1 + \pi \varphi_1 + \bar\sigma \varphi_0\right) \right]\, .
\end{split}
\end{equation}

The loop part of the algebra only changes the number of spinor indices. If we pick out a mode from the loop, we obtain
\begin{equation}
    \check{\xi}^-_{s,a} = \frac{\xi^{\alpha_1\cdots\alpha_{s+a+1}}\lambda_{\alpha_1}\cdots \lambda_{\alpha_{s+a+1}} }{\langle\iota \lambda\rangle^{s+a-1}} . 
\end{equation}
For values of $a< -s-1$, the field will vanish and hence have no action on spacetime. While the loop affects the spin indices for a given value of $s$, it does not change the growth of the field as
\begin{equation}
    \xi^{\alpha_1\cdots\alpha_{s+a+1}} \sim O\left(x^{s+1}\right) . 
\end{equation}
Therefore, the loop in practice labels how many more times one must solve the dual recursion relations \eqref{Dual Recursion} while keeping the same seed solution with boundary conditions labeled by $s$. In what follows, we will suppress the loop part of the algebra, as many of the considerations follow along similar lines. See Appendix \ref{sec: Linear SD Perturbation} for further discussion of the asymptotic growth of the SD perturbation in linear theory.

\section{Quasi-local higher-spin charges}
\label{sec:Quasi-local higher-spin charges}

In this section, we translate the above spacetime results using Penrose’s transform. We recover the general covariant higher-spin expression heuristically constructed in Section \ref{sec:Penrose's Quasi-Local Mass Formula}, hence providing a first-principles derivation of the quasi-local celestial charges. We also discuss the charges associated with the other graviton helicity. We then specialize the discussion to null infinity to connect with \cite{Freidel:2021ytz, Geiller:2024bgf, Kmec:2024nmu}, and to a finite-distance null hypersurface to connect with \cite{Ruzziconi:2025fct}.

\subsection{Covariant spacetime expressions} \label{sec:Covariant Spcaetime Expressions}

To begin, the charge containing the ASD perturbation $H_s$ can be translated to spacetime by defining a higher spin field. 
This requires extra boundary conditions to be imposed on $g$ so as to give a well-defined integral formula. The Penrose transform for this field is given by
\begin{equation}
    \phi_{\alpha_1 \cdots \alpha_{s+3}} = \frac{1}{2\pi i}\int_{X}\frac{D\lambda}{\langle \iota \lambda\rangle^{s-1}}\wedge \lambda_{\alpha_1}\cdots\lambda_{\alpha_{s+3}}g\vert_{X} . 
\end{equation}
To make the projective weights work\footnote{We note that by requiring the projective weight to work, the field $\phi_{\alpha_1\cdots\alpha_{s+3}}$ obtains a GHP weight of $(s-1,0)$ while the components $Q_n$ have a GHP weight of $(2n,0)$ and conformal weight $W=-n-3$. Furthermore, due to the shift in weights, the z.r.m. equation changes to
\begin{equation}
    \nabla^{\alpha_1}_{\,\, \dot\alpha} \phi_{\alpha_1 \cdots \alpha_{s+3}} + (s-1)\Gamma_{\alpha_0\dot \alpha \alpha_1 \beta}\,\iota^{\beta}\phi^{\alpha_0 \alpha_1}_{\quad \,\,\,\alpha_2\cdots\alpha_{s+3} } = 0 . 
\end{equation}}, we have had to introduce a singularity of order $s-1$ when $\lambda_\alpha = \iota_\alpha$. As a consequence, $g$ must vanish to sufficiently high order at $\lambda_\alpha = \iota_\alpha$ to make the integral well-defined. Therefore, to capture arbitrarily large values of $s$, one can use Schwartzian falloffs to $g$ at this point. By considering the components of this z.r.m field, one finds 
\begin{equation} \label{HigherSpinPenroseTransform}
    Q _n = \frac{1}{2\pi i}\int_{X}q^{n+2} dq \wedge g'\vert_X ,\qquad  q = \frac{\langle o \lambda \rangle}{\langle\iota \lambda \rangle}\, ,
\end{equation}
where $g' = \langle\iota \lambda \rangle^6 g $. As we will see in Section 7, the components $Q_n$ of the higher valence z.r.m. field are built from the repeated action of the recursion operator. The exact relationship between the $Q_n$ and contractions of the z.r.m. field is given by
\begin{equation}
    Q_n = \phi_{\alpha_1\cdots \alpha_{s+3}}o^{\alpha_1}\cdots o ^{\alpha_{n+2}}\iota^{\alpha_{n+3}}\cdots \iota^{\alpha_{s+3}} . 
\end{equation}
In the GHP formalism, the z.r.m equations can be decomposed into equations living on a null hypersurface $\mathcal{N}$ and equations propagating the field away from the null hypersurface. These are given respectively by 
\begin{equation} \label{recusion relations charge aspects}
    \begin{split}
        \text{\th}'_{\mathcal{C}} Q_{m} - \bar\eth_{\mathcal{C}}Q_{m-1} = (2-m)\nu Q_{m+1} + (m+1)\sigma Q_{m-2},\\
        \eth_{\mathcal{C}}Q_{m} - \text{\th}_{\mathcal{C}}Q_{m-1} = (2-m)\lambda Q_{m+1}+(m+1)\kappa Q_{m-2} . 
    \end{split}
\end{equation}
To translate the charge $H_s$ to spacetime, we use the pullback of the map $p: \mathbb{PS}\rightarrow\mathcal{PT}$ from the spin bundle to twistor space and apply all the definitions
\begin{equation}
\begin{split}
    H_s &= \frac{1}{2\pi i }\int_{\partial\Sigma}\Omega_Z\wedge \xi_s g\\
    &= \frac{1}{2\pi i }\int_{\mathbb{CP}^1\times \mathcal{S}} D\lambda \wedge \Sigma^{\alpha\beta}\wedge\lambda_\alpha\lambda_\beta\frac{\xi^{\alpha_1\cdots\alpha_{s+1}}\lambda_{\alpha_1}\cdots\lambda_{\alpha_{s+1}}}{\langle \iota\lambda \rangle^{s-1}}g\vert_X \\
    &=\int_{\mathcal{S}} \phi_{\alpha_1\cdots\alpha_{s+1}\alpha\beta}\xi^{\alpha_1\cdots\alpha_{s+1}}\Sigma^{\alpha\beta} . 
\end{split} \label{higher spin expression derived}
\end{equation}
When pulled back to the spin bundle $p^*(\Sigma) = \mathbb{CP}^1\times\mathscr{N}$ and hence as $g\vert_X$ is a global form on $\mathbb{CP}^1$, the only boundary contribution arises from $\partial\mathscr{N} = \mathcal{S}$ where the surface $\mathcal{S}$ is a slice of the null hypersurface $\mathscr{N}$. Notice that the last line of \eqref{higher spin expression derived} reproduces the heuristic expression given in \eqref{higher spin ql}. We can further reduce this charge into components using the expansion of the SD two-form
\begin{equation}
    \Sigma^{\alpha\beta} = o ^{\alpha}o^{\beta}  \boldsymbol{n} \wedge \boldsymbol{\bar m}- \iota^{(\alpha} o^{\beta)}\boldsymbol{ m} \wedge \boldsymbol{\bar m} +\iota^{(\alpha}o^{\beta)} \boldsymbol{ l}\wedge \boldsymbol{ n} +\iota^\alpha \iota^\beta \boldsymbol{m}\wedge \boldsymbol{ l} \, .
\end{equation}
where $\{\boldsymbol{l},\boldsymbol{n},\boldsymbol{m},\boldsymbol{\bar m}\}$ is the co-tetrad basis. The null hypersurface can be defined by the condition that $\boldsymbol{n}\vert_{\mathcal{N}} = 0$ such that the vector field $n^{\mu}$ is the generator of the null direction. Therefore, on $\mathcal{S}$, the only non-zero elements of the co-tetrad basis are $\boldsymbol{m}$ and $\boldsymbol{\bar m}$. The charge in terms of spin fields and components is given 
\begin{equation}
    H_s = \int_{\mathcal{S}} \phi_{\alpha_1\cdots\alpha_{s+1}\alpha\beta}\xi^{\alpha_1\cdots\alpha_{s+1}}\Sigma^{\alpha\beta} = \sum_{m=0}^{s+1}\int_{\mathcal{S}}\boldsymbol{m}\wedge \boldsymbol{\bar m} \,\xi_{s,m}Q_{m-1} . 
\end{equation}
We can also consider the flux, which is defined as the difference between two cuts $\mathcal{S}$ and $\mathcal{S}'$ of $\mathcal{N}$
\begin{equation}
\begin{split}
     F_s &= \int_{\mathcal{S}} - \int_{\mathcal{S}'}\left[\phi_{\alpha_1\cdots\alpha_{s+1}\alpha\beta}\xi^{\alpha_1\cdots\alpha_{s+1}}\Sigma^{\alpha\beta}\right]\\
     &=\int_{\mathcal{N}} d\left[\phi_{\alpha_1\cdots\alpha_{s+1}\alpha\beta}\xi^{\alpha_1\cdots\alpha_{s+1}}\Sigma^{\alpha\beta}\right]\\
     &= -\frac{2}{3}\int_{\mathcal{N}}\left(\nabla^{\alpha }_{\,\,\dot \alpha}\phi_{\alpha_1\cdots\alpha_{s+1}\alpha\beta} \xi^{\alpha_{1}\cdots\alpha_{s+1}}+\phi_{\alpha_1\cdots\alpha_{s+1}\alpha\beta}\nabla^{\alpha}_{\,\,\dot\alpha}\xi^{\alpha_1\dots\alpha_{s+1}}\right)\theta_{\gamma}^{\,\,\,\dot \alpha}\wedge\Sigma^{\gamma\beta} ,
\end{split}
\end{equation}
where Stokes' theorem was used. We see that if the hypersurfaces in question were general, the conservation of the charges would require the full z.r.m. equation and twistor equation to hold. On a SD background, the z.r.m. equation holds for arbitrarily high spin, and with the choice of gauge parameter satisfying \eqref{gaugeparameter}, we find a non-zero flux
\begin{equation}
    F_s = \int_{\mathcal{N}}\boldsymbol{l}\wedge \boldsymbol{m}\wedge \boldsymbol{\bar m} \,Q_{-2}\left(\sigma \xi_{s,1} - \bar\eth_{\mathcal{C}}\xi_{s,0}\right) . 
\end{equation}
As we can see, the vanishing of the flux is obstructed by $Q_{-2}$ and if it were to vanish the charges $H_s$ would be conserved on the null hypersurface. We note that by choosing a $\mathcal{N}$ the choice of the vector field $n^\mu$ is fixed, as it generates the null direction. Therefore, the residual Lorentz transformation of the tetrad does not change the definition of $Q_{-2}$ and therefore the condition for conservation of the charge $Q_{-2}=0$ is preserved under these transformations.

Having translated the charges $H_s$, one can ask to translate the charges $\tilde H_s$ associated with the other symmetry. The obstruction to translating the charge is in how the non-linear field $h$ translates to spacetime. Only when working in special gauges is there an explicit formula for $h$ in terms of spacetime fields. We will discuss the Plebanski gauge in Section 7, where explicit calculations can be done. In the present case, we wish to work in a generic gauge on spacetime, so we consider 
\begin{equation}
    \int_{\Sigma}\delta_\chi \lrcorner\,\omega = \int_{\Sigma} d\left(\Omega \wedge \chi \delta h\right) .
\end{equation}
To translate the charge to spacetime, we can work in a gauge (Woodhouse gauge) where $\delta h$ only has spacetime components on the spin bundle. Hence, if $\chi$ is a global function on twistor space, the charge will vanish as there would be no boundary contribution, signaling that these transformations on the phase space are small gauge transformations. To obtain non-vanishing charges, such that they generate a non-trivial action on the phase space, the gauge parameter $\chi$ must be singular on twistor space, such that 
\begin{equation} \label{ChiGenerators}
    \chi_{s,n} = \frac{1}{2\pi i}\oint_{\Gamma} q^{n+2}dq \,\chi_s'\vert_X \, ,
\end{equation}
with $\chi' = \langle\iota \lambda\rangle^6\chi$ giving non-vanishing contour integrals. This line of reasoning amounts to identifying $\chi_s$ as the Čech representative for the cohomology $H^1(\mathcal{PT},\mathcal{O}(-6))$. The index $s$ labels a truncation such that $\chi_{s,n} = 0$ when $n <-s-3$ with $s\geq -1$. Therefore, we can translate the contracted symplectic structure to spacetime
\begin{equation}   
    \int_{\Sigma}\delta_\chi \lrcorner\,\omega = \int_{\mathscr{N}}\boldsymbol{l}\wedge \boldsymbol{m}\wedge\boldsymbol{\bar m} \,\chi_{s,-2} \,\varphi_0 = \delta \tilde H_s - \Xi_{\chi_s}[\delta] . 
\end{equation}
The difficulty now is to extract the $\delta$ inside the $\varphi_0$ by identifying how the SD perturbation affects the components of the metric. In general, this is difficult, but in certain circumstances, this is possible, as will be discussed in the next sections. 

Using the same representative for the SD perturbation on the spin bundle, the symplectic potential can also be translated to spacetime
\begin{equation} \label{symplectic structure general}
    \int \theta[\delta] = \int_{\mathscr{N}}\boldsymbol{l}\wedge \boldsymbol{m }\wedge \boldsymbol{\bar m} \,\varphi_{\alpha\beta\gamma \dot\alpha} \bar \iota^{\dot \alpha} \iota_{\delta}\phi^{\alpha\beta\gamma\delta} = \int_{\mathscr{N}}\boldsymbol{l}\wedge \boldsymbol{m }\wedge \boldsymbol{\bar m} \, \varphi_0Q_{-2} .
\end{equation}

So far, the $\xi_s$ symmetry on the SD metric has been discussed. To write down its action on the ASD higher-spin field, we use \eqref{HigherSpinPenroseTransform} and the explicit expansion of the $\xi_s$ parameter when pulled back to the spin bundle \eqref{XiExapansion}. Using the distinguished $\iota_\alpha$ spinor, we find
\begin{equation}\label{PhiVariation}
    \delta_{\xi_s}\phi_{\alpha_{1}\cdots \alpha_{s'+3}} = \iota^\alpha \iota^\beta \nabla_{\beta}^{\,\,\,\dot \alpha}\phi_{\alpha_1\cdots\alpha_{s'+3}\beta_1 \cdots \beta_{s+1}}\nabla_{\alpha\dot \alpha} \xi^{\beta_1 \cdots\beta_{s+1}} . 
\end{equation}
An interesting feature, which will be made more explicit, is that the $\xi_s$ symmetry naturally introduces higher spin objects.

\subsection{From null infinity...}\label{sec: Charges at Null Infinity}
The discussion so far has been general, keeping the gauge on spacetime arbitrary. We now wish to push the generic null hypersurface $\mathscr{N}$ to null infinity $\mathscr{I}$. To this end, it is convenient to work in the Newman-Unti (NU) gauge \cite{Newman:1962cia}. We find complete agreement with \cite{Kmec:2024nmu}. 

In the coordinate system $x^\mu = (r,u,x^A)$ where $x^A = (z,\bar z)$, the tetrad in NU gauge is given by
\begin{equation} \label{NU tetrad}
    l = \frac{\partial}{\partial r}  , \qquad n = \frac{\partial }{\partial u} + U \frac{\partial}{\partial r} + X^A \frac{\partial}{\partial x^A} , \qquad m = \omega\frac{\partial }{\partial r}+ L^A \frac{\partial}{\partial x^A} , 
\end{equation}
along with the constraint on the spin coefficients given by
\begin{equation} \label{NU gauge}
    \pi = \epsilon = \kappa = 0 ,  \qquad \tau = \bar \alpha+ \beta ,  \qquad \rho = \bar\rho . 
\end{equation}
We note that the complex conjugate is seen to be independent in general in the complexified setting. In what follows, we further require that the falloffs of the spin coefficients be satisfied
\begin{equation}
    \mu = O\left(r^{-2}\right) , \quad  \quad \lambda = O\left(r^{-2}\right), \quad \quad \nu = O\left(r^{-1}\right) . 
\end{equation}
On shell, along with the SD constraint, this results in these spin coefficients to vanish $\mu = \lambda = \nu = 0$. The vanishing of these spin coefficients is crucial for the metric perturbation to stay in the SD sector. In other words, the equation \eqref{metricpotential} is only satisfied if these spin coefficients vanish. The metric has the expansion
\begin{equation}
    ds^2 = -2dudr + 2r^2 dzd\bar z + r\left(\sigma^0 d\bar z^2 + \bar \sigma^0dz^2\right) + O(r^0) . 
\end{equation}
The details can be found in \cite{Newman:1962cia}. The SD perturbation expanded around null infinity will have a fall-off of the form
\begin{equation}
     \varphi_0 =  \frac{\partial_u^{-1}\delta\sigma^0}{r}+O\left(r^{-2}\right)\, ,
\end{equation}
where the leading contribution can be identified with the perturbation of the asymptotic shear. Furthermore, since $\phi_{\alpha_1\cdots \alpha_{s+3}}$ is a z.r.m. field with conformal weight $W=-1$, it will have a peeling theorem, such that the components have the fall-offs given by
\begin{equation}
    Q_{n} = \frac{Q_{n}^0}{r^{n+3}} + O\left(r^{-n-4}\right)\, .
\end{equation}
Using these fall-offs, the symplectic potential at null infinity is given by
\begin{equation}
    \int\theta[\delta] = \int_{\mathscr{I}} dudzd\bar z \, Q_{-2}^0\partial_u^{-1}\delta \sigma^0 \, ,
\end{equation}
which matches the result found in \cite{Kmec:2024nmu}. Furthermore, using the NU gauge and the peeling theorem, the asymptotic z.r.m. equation along $\mathscr{I}$ on the leading order components gives
\begin{equation}
    \partial_u Q^0_m = \bar\eth^0 Q^0_{m-1} + (m+1)\sigma^0 Q^0_{m-2}\, ,
\end{equation}
where the $\bar \eth$ operator appearing is the leading order contribution given by
\begin{equation}
    \bar \eth  = \frac{\bar\eth ^0}{r}+ O\left(r^{-2}\right) \qquad \bar\eth^0 = \delta^0 -p\beta^0 - q \bar \alpha^0 . 
\end{equation}
The second tower of equations equates the subleading orders of the field to the overleading parts. Turning to the field $\xi^{\alpha_1 \cdots\alpha_{s+1}}$, the components are overleading such that 
\begin{equation}
    \xi_{s,m} = r^m \xi^0_{s,m} + O\left(r^{m-1}\right)\, ,
\end{equation}
where the dual recursion relations \eqref{Dual Recursion} at leading order are given by
\begin{equation}
\label{leading recursions}
    \partial_u \xi^0_{s,m} - \bar\eth^0 \xi^0 _{s,m+1} + (m+2)\sigma^0 \xi^0_{s,m+2} =0 . 
\end{equation}
The second tower of equations relates the subleading orders to the overleading orders as before, together with the condition $\eth^0 \xi^0_{s,s+1} = 0$. Under the identification $\xi_{s,m}^0=\tau_{m-1}$,
$D=\bar{\eth}^0$ and $C=\sigma^0$, equation \eqref{leading recursions}
reproduces the dual evolution equations of
\cite{Cresto:2024mne} for $m\geq1$.
Here $\tau_j$ denotes the components of the parameter
sequence used in that reference, truncated at $j=s$.
The $m=0$ equation reproduces the evolution of
$\tau_{-1}$ used for the covariant charge
\cite{Cresto:2024mne}.
Our angular constraint imposes an additional restriction
on this sequence.

Similarly, the spacetime fields generated by $\chi_s$ satisfy the recursion relation
\begin{equation}
\label{dual recurion chi leading}
    \partial_u \chi_{s,m}^0 = \bar\eth\chi^0_{s,m-1} + (m+1)\sigma^0 \chi_{s,m-2}^0\, ,
\end{equation}
along with $\eth^0 \chi_{s,-s-3} = 0 $. The $\chi_s$ sector is analogous
to the second symmetry sector of self-dual Yang--Mills
theory constructed in \cite{Kmec:2025ftx},
whose gravitational counterpart is anticipated in
Section~6 of that reference. The variation generated by the $\xi_s$ and $\chi_s$ generators at null infinity is given by
\begin{equation}
\begin{split}
    \delta_{\xi(\tau_s),\chi(\rho_s)}Q_n^0 &= \chi^0_{s,n}+\sum_{m=0}^{s+1}m\xi^0_{s,m}\partial_u Q^0_{n+m-1} + (n+m+2)Q^0_{n+m-1}\partial_u \xi^0_{s,m} ,\\
    \delta_{\xi(\tau_s)}\sigma^0 & = \partial_u\left(\sigma^0\xi_{s,1}^0 - \bar\eth^0 \xi_{s,0}^0\right) . 
\end{split}
\end{equation}
Here $\tau_s(z,\bar z):=\xi_{s,s+1}^0$ and
$\rho_s(z,\bar z):=\chi_{s,-s-3}^0$ are
$u$-independent celestial seed fields satisfying
$\eth^0\tau_s=\eth^0\rho_s=0$.
The notation $\xi(\tau_s)$ and $\chi(\rho_s)$ denotes
the solutions generated from these seeds using \eqref{leading recursions}
and \eqref{dual recurion chi leading}, respectively, with a fixed choice of
integration constants. Their remaining components
generally depend on $u$ and the background shear
$\sigma^0$. The explicit variations generated by the $\xi(\tau_s)$ can be found in \cite{Kmec:2024nmu}. The first few examples of the $\chi$ generators are given by
\begin{equation}
    \begin{split}
        \delta_{\chi(\rho_{-1})}Q_{-2} &= \rho_{-1},\\
        \delta_{\chi(\rho_{0})}Q_{-2} &= u\bar\eth\rho_0 ,\\
        \delta_{\chi(\rho_{1})}Q_{-2} &= \frac{u^2}{2}\bar\eth^2 \rho_1 -\rho_1 \partial_u^{-1}\sigma^0 ,
    \end{split}
\end{equation}
where the expressions for the higher $Q_n$ are derived through the use of their recursion relations. 

The charge $H_s$ can now be pushed to null infinity, where one obtains a finite result; the exact form is given explicitly by
\begin{equation}
    H_s = \sum_{m=0}^{s+1}\int_{S^2} \text{vol}(S^2) \xi_{s,m}^0 Q_{m-1}^0 , 
\end{equation}
\begin{equation}
    \tilde H_s = \int dudzd\bar z \,\chi^0_{s,-2}(\rho_s) \partial_u^{-1}\sigma^0  . 
\end{equation}

\subsection{...to a bulk null hypersurface}

Following \cite{Ruzziconi:2025fct, Ruzziconi:2025fuy}, it is also convenient to work with a NU tetrad around a generic null hypersurface $\mathcal{N}$ in the bulk, which satisfies \eqref{NU tetrad} and \eqref{NU gauge}. The null hypersurface is situated at $r=0$. At leading order, the metric around $\mathcal{N}$ is given by 
\begin{equation}
    ds^2 = q_{AB}dx^Adx^B + r((\chi_{AB} + q_{AB}\chi)dx^Adx^B - V_1 dv^2 - P_A dx^A ) + O(r^2) .
\end{equation}
The metric can be decomposed in the NU co-tetrad. We have $q_{AB} = m^0_{(A}\bar m^0_{B)}$, and the other terms in the metric expansion are related to the leading spin coefficients via
\begin{equation}
\begin{split}
    \chi_{AB} &= 2(\sigma^0 \bar m^0_A\bar m^0_B + \bar \sigma^0  m^0_A m^0_B) ,\\
    P_A &= 2(\tau^0 \bar m^0_A + \bar\tau^0 m^0_A) ,\\
    V_1 &= \lambda^0- \bar\lambda^0 - 4\gamma^0 ,\\
    \chi &= 4\rho^0 .  
\end{split}
\end{equation} We refer to \cite{Ruzziconi:2025fuy} for a complete dictionary between metric and NP formalism. The variation of the metric can be expanded as 
\begin{equation}
    \delta q_{AB} = m^0_{(A}\bar m^0_{B)}\left(\bar m^{0\,C}\delta m^0_C + m^{0\,C}\delta \bar m^0_C \right) + \bar m^0_{A}\bar m^0_{B} m^{0\,C}\delta m^0_C +  m^0_{A} m^0_{B} \bar m^{0\,C}\delta \bar m^0_C\, ,
\end{equation} together with
\begin{equation}
    m^{0\,C}\delta m^0_C = -(\text{\th}'^0+\mu^0) \varphi^0_0 . 
\end{equation}
This then gives the leading order relationship to the variation coming from twistor theory 
\begin{align}
    \varphi_0 =& -(\text{\th}'^0+\mu^0) ^{-1}m^{0\,C}\delta m^0_C \\
    &-r(\text{\th}'^0+\mu^0)^{-1}\left(2\delta\sigma^0 + m^{0A}\delta \bar m^0_A -(\text{\th}'^1+\mu^1) \varphi_0^0 + \bar L_A^1 m^{0A}(\text{\th}'^0+\mu^0)\varphi_0^0\right)+O\left(r^2\right) \nonumber\, ,
\end{align}
where we have assumed that $\tau^0 = \bar\nu^1 =0 $. We see if we impose the constraints
\begin{equation} \label{additional constraints}
    \delta m^0_A = 0, \qquad \delta \text{det}(q) = \delta q = 0 \, ,
\end{equation} as in \cite{Ruzziconi:2025fct,Ruzziconi:2025fuy}, the leading order contribution vanishes. Under this assumption, the phase space becomes subleading, and the variation is given by
\begin{equation}
     \varphi_0 = -2r(\text{\th}'^0+\mu^0)^{-1} \delta\sigma^0 + O(r^2) .
    \label{subleading phi}
\end{equation} 

We assume the following falloffs for $Q_n$ around $\mathcal{N}$:
\begin{equation} \label{falloffs Q}
    Q_{n} = Q^0_{n} + O\left(r\right) .
\end{equation} In general, the symplectic potential \eqref{symplectic structure general} admits the following expansion:
\begin{equation}
    \int\theta[\delta] =\int \theta^0[\delta] + r\theta^1[\delta] + \mathcal{O}(r^2) . 
\end{equation} The leading-order phase space encoded in $\theta^0[\delta]$ provides variations of a single component of the hypersurface metric, while the subleading order encoded in $\theta^1[\delta]$ contains the variation of the shear. If the leading order is not fixed, there will be corrections involving the variations of the hypersurface metric in the subleading orders. However, if we impose the constraints \eqref{additional constraints}, the phase space becomes subleading, i.e. $\theta^{0}[\delta]=0$, and the symplectic potential simplifies to
\begin{equation}
    \int\theta[\delta] = -2r\int \, dv dz d\bar z \, \sqrt{q}Q^0_{-2}(\text{\th}'^0+\mu^0)^{-1} \delta\sigma^0 + O(r^2)\, .
 \end{equation} Upon defining $Q^0_{-2} =-(\text{\th}'+ \mu^0) \tilde \lambda^0$ and integrating by parts, this reproduces the subleading phase space considered in \cite{Ruzziconi:2025fct,Ruzziconi:2025fuy}. The latter is formally similar to the Ashtekar-Streubel phase space at null infinity \cite{Ashtekar:1981bq}, and allows the heuristic construction of celestial charges as, it was done at null infinity \cite{Freidel:2021ytz, Geiller:2024bgf}.

We now focus on the recursion relations. Taking into account the falloffs \eqref{falloffs Q}, the leading order recursion relations read as
\begin{equation}
     \text{\th}_{\mathcal{C}}^{\prime 0}  Q^0_m = \bar\eth_{\mathcal{C}}^0 Q^0_{m-1} + (m+1)\sigma^0 Q^0_{m-2} . 
\end{equation} which correctly reproduces the recursion relations of \cite{Ruzziconi:2025fct,Ruzziconi:2025fuy}. Similarly, the parameters $\xi_{s,m}$ can be expanded
around $\mathcal N$ as
\begin{equation}
    \xi_{s,m} = \xi_{s,m}^0 + r\xi_{s,m}^1 + O(r^2)\, .
\end{equation} The leading order dual recursion relations read as \eqref{leading recursions}. Again, imposing the conditions \eqref{additional constraints} implies $\xi_{s,m}^0 =0$ by consistency, and therefore, the subleading parameters $\xi_{s,m}^1$ will satisfy the dual recursion relations. These are the celestial symmetry parameters considered in \cite{Ruzziconi:2025fct,Ruzziconi:2025fuy}, where an explicit solution is provided in terms of a ``dressed time'' on the null hypersurface. Similarly, the parameters $\chi_{s,m}$ can be expanded around $\mathcal{N}$ as
\begin{equation}
    \chi_{s,m} = \chi_{s,m}^0 + \mathcal{O}(r)\, ,
\end{equation} and the $\chi_{s,m}^0$ satisfy \eqref{dual recurion chi leading}. Notice that these leading parameters will appear both in the leading and subleading phase space.

At leading order, the celestial charges integrated over a cut $v = \text{constant}$ of $\mathcal{N}$ are given explicitly by 
\begin{equation}
    H_s = \sum_{m=0}^{s+1}\int dzd\bar z \sqrt{q} \xi_{s,m}^0 Q_{m-1}^0 , 
\end{equation}
and the associated integrated fluxes are
\begin{equation}
    F_s = \int_{\mathcal{N}}dvdzd\bar z\,\sqrt{q}Q^0_{-2}\left(\sigma^0 \xi^0_{s,1} - \bar\eth^0_{\mathcal{C}}\xi^0_{s,0}\right) . 
\end{equation}
The dual charges are in the form of an integrated flux, but as discussed previously, only the contracted symplectic structure can be written down
\begin{equation}
    \delta\tilde H_s - \Xi_{\chi_s}[\delta] = \int dvdzd\bar z \,  \sqrt{q}\chi^0_{s,-2}(\rho_s)(\text{\th}'^0+\mu^0)^{-1}\left( m^{0\, A}\delta m^0_A\right)\, .
\end{equation} 
 Imposing the conditions \eqref{additional constraints}, the leading phase space and charges completely trivialize. The subleading celestial charges read as
\begin{equation}
    H_s = r \sum_{m=0}^{s+1}\int dzd\bar z \sqrt{q} \xi_{s,m}^1 Q_{m-1}^0 , 
\end{equation}
and the integrated fluxes read as 
\begin{equation}
    F_s = r \int_{\mathcal{N}}dvdzd\bar z \, \sqrt{q}Q^0_{-2}\left(\sigma^0 \xi^1_{s,1} - \bar\eth^0_{\mathcal{C}}\xi^1_{s,0}\right) . 
\end{equation}
The dual charge under these assumptions can be written down explicitly and is given by
\begin{equation}
    \tilde H_s = r \int dvdzd\bar z \, \sqrt{q}\chi^0_{s,-2}(\rho_s) (\text{\th}'^0+\mu^0)^{-1} \sigma^0\, .
\end{equation} 
 In particular, this subleading structure is formally completely analogous to that appearing at null infinity. However, this analysis does not
establish a corresponding relation between horizon
memory and soft fluxes. This is consistent with
\cite{Rahman:2019bmk}, which explicitly gives horizon
supertranslation charges and fluxes, but finds no
relation to horizon memory analogous to that at null
infinity.

\section{Celestial symmetries of Plebanski's equations}
\label{sec:Celestial symmetries of Plebanski's equations}

In this section, we will use a special gauge where calculations can be made explicit, and the symmetry algebra can be computed. In the SD sector, two potential formulations exist due to Plebanski \cite{Plebanski:1975wn}. In the present case, we will consider Plebanski's second heavenly formulation.

\subsection{Plebanski's second heavenly equations}
We begin by considering the twistor action \eqref{TwistorAction} when one chooses a special gauge for the field $h$, namely,
\begin{equation}\label{PlebanskiGauge-h}
    h = \langle \iota\lambda\rangle^3 d_{\dot \alpha}\Theta\, \bar e^{\dot \alpha} \, ,
\end{equation}
where $\bar e^{\dot \alpha}$ is the $(0,1)$-form of homogeneity $-1$ on the spin bundle that points along the spacetime directions only and $\Theta$ is a function of spacetime coordinates only. We have also defined the spacetime coordinates $\left(x^{\dot \alpha}=(x,y), \tilde x^{\dot \alpha}=(-\tilde x,\tilde y)\right)$ along with their respective derivatives $d_{\dot \alpha}  = \frac{\partial}{\partial x^{\dot \alpha}}$ and $\tilde d_{\dot \alpha}  = \frac{\partial}{\partial \tilde x^{\dot \alpha}}$. In this gauge, the action reduces to
\begin{equation}
    S[\Theta , Q_{-2}] = \int d^4 x Q_{-2}\left(\tilde d^{\dot \alpha}d_{\dot \alpha}\Theta + \frac{1}{2}d_{\dot \alpha}d_{\dot \beta}\Theta d^{\dot \alpha}d^{\dot \beta}\Theta\right) .
\end{equation}
The equations of motion for this action are given by
\begin{equation}
     \tilde d^{\dot \alpha}d_{\dot \alpha}\Theta + \frac{1}{2}d_{\dot \alpha}d_{\dot \beta}\Theta d^{\dot \alpha}d^{\dot \beta}\Theta = 0 , \qquad  \tilde d^{\dot \alpha}d_{\dot \alpha}Q_{-2}+d_{\dot \alpha}d_{\dot \beta}\Theta d^{\dot \alpha}d^{\dot \beta}Q_{-2} = 0 . 
\end{equation}
The first of which is called Plebanski's second heavenly equation, while the second equation is the linearized version.  In \cite{Plebanski:1975wn}, Plebanski showed that the SD vacuum condition, $R_{\alpha\beta} = 0$, locally implies the existence of a potential $\Theta$ along with a coordinate system $\left(x^{\dot \alpha}, \tilde x^{\dot \alpha}\right)$ that generates a metric
\begin{equation}
\begin{split}
    g &= 2d\tilde  x^{\dot \alpha} d x_{\dot \alpha} - 2d_{\dot \alpha} d_{\dot \beta}\Theta d\tilde x^{\dot \alpha}d\tilde x^{\dot \beta}\\
    &= 2dy d\tilde x + 2dxd\tilde y - 2\Theta_{xx}d\tilde x^2 - 2\Theta_{yy}d\tilde y^2 + 4\Theta_{xy}d\tilde xd\tilde y \, ,
\end{split}
\end{equation}
where the potential $\Theta$ satisfies the second heavenly equation which in explicit coordinates is given by
\begin{equation}
    \Theta_{x\tilde y} + \Theta_{y\tilde x} + \Theta_{xx}\Theta_{yy} - \Theta_{xy}^2 = 0\, .
\end{equation}
When the second heavenly equation holds, one can show that the spin connections are given by
\begin{equation}
    \Gamma_{\alpha\beta} = 0 , \qquad \qquad\tilde\Gamma_{\dot \alpha\dot \beta} = d_{\dot \alpha}d_{\dot \beta}d_{\dot \gamma}\Theta d\tilde x^{\dot \gamma} ,
\end{equation}
which results in the Riemman 2-forms 
\begin{equation}
    R_{\alpha\beta} = 0 , \qquad \qquad \tilde R_{\dot \alpha\dot \beta} = d_{\dot \alpha}d_{\dot \beta}d_{\dot \gamma}d_{\dot \delta}\Theta \tilde \Sigma^{\dot \gamma\dot \delta} . 
\end{equation}
As expected, the ASD Riemann 2-form vanishes; moreover, in this gauge, the ASD spin connection is zero, making the undotted spin bundle trivial. This is the major benefit of this gauge.  

The second heavenly system has a Lax representation by considering the tetrad $e_{\alpha\dot \alpha}$ and a spectral parameter $\lambda_{\alpha} $
\begin{equation}
    L_{\dot \alpha} = \lambda^{\alpha} e_{\alpha\dot \alpha} = \langle\lambda \iota\rangle \left(q d_{\dot \alpha} - \tilde d_{\dot \alpha} + d_{\dot \alpha}d_{\dot \beta}\Theta d^{\dot \beta}\right) . 
\end{equation}
The second heavenly equations are equivalent to the commutator $[L_{\dot \alpha},L_{\dot \beta}] = 0$ for all values of $q$. 

Let us focus on the linearized second heavenly equations given by
\begin{equation}\label{LinearisedHeavenly}
    \tilde d^{\dot \alpha}d_{\dot\alpha} Q_{-2} + d_{\dot \alpha}d_{\dot \beta}\Theta d^{\dot \alpha}d^{\dot \beta}Q_{-2}= 0 . 
\end{equation}
 The linearised equation plays an important role in the integrable hierarchy and the recursion operator. First, consider manipulating the linearised equation
\begin{equation}
\begin{split}
        0 &=  \tilde d^{\dot \alpha}d_{\dot\alpha} Q_{-2} + d_{\dot \alpha}d_{\dot \beta}\Theta d^{\dot \alpha}d^{\dot \beta}Q_{-2}\\
        &=d_{\dot \alpha}\left(\tilde d^{\dot \alpha}Q_{-2} - d^{\dot\alpha}d_{\dot \beta}\Theta d^{\dot \beta}Q_{-2} \right)\\
        &= d_{\dot\alpha}d^{\dot \alpha}Q_{-1}\, ,
\end{split}
\end{equation}
where the last line vanishes identically due to the anti-symmetry of the contraction of indices. Therefore, the linearised heavenly equation locally guarantees the existence of the relations
\begin{equation}
    d_{\dot \alpha} Q_{-1} = \tilde d_{\dot \alpha}Q_{-2} - d_{\dot \alpha}d_{\dot \beta}\Theta d^{\dot \beta}Q_{-2} . 
\end{equation}
One can further check by acting with $\tilde d^{\dot \alpha}$ and using the second heavenly equation that $Q_{-1}$ satisfies the linearised equation \eqref{LinearisedHeavenly}. One can iterate this procedure and construct an infinite tower satisfying the recursion relations
\begin{equation}\label{recursion}
     d_{\dot \alpha} Q_{n} = \tilde d_{\dot \alpha}Q_{n-1} - d_{\dot \alpha}d_{\dot \beta}\Theta d^{\dot \beta}Q_{n-1} . 
\end{equation}
With these considerations in hand, one can define the recursion operator $R: \mathcal{W} \rightarrow \mathcal{W}$ where $\mathcal{W}$ is the space of solutions to \eqref{LinearisedHeavenly} and the map is defined formally by the recursion relations \eqref{recursion}. Furthermore, in this gauge, these are exactly the high valence z.r.m. equations for a z.r.m. field with components $Q_n$. The action of the recursion operator on twistor space has a simple form when considering \eqref{HigherSpinPenroseTransform}. Since $R(Q_{n}) = Q_{n+1}$, the action of the recursion operator on $g$ is simply $R(g) = q \cdot g$, see \cite{Dunajski:2000iq,Dunajski:2003gp} for more details. Furthermore, since these recursion relations are satisfied by both $\xi_{s,n}$ and $\chi_{s,n}$, the recursion operator will also appear in the construction of the symmetry algebra. 

One can also construct the recursion operator on the SD sector of the theory by simply taking a derivative of the Plebanski equation. Define $\tilde Q^{\dot \alpha}_{-2} = d^{\dot \alpha}\Theta$ which satisfies the linearised equation
\begin{equation}
    \tilde d^{\dot \alpha}d_{\dot\alpha} \tilde Q^{\dot \beta}_{-2} + d_{\dot \alpha}d_{\dot \gamma}\Theta d^{\dot \alpha}d^{\dot \gamma}\tilde Q^{\dot \beta}_{-2}= 0 .
\end{equation}
Following the same logic as above, one can construct the so-called integrable hierarchy for SD gravity
\begin{equation}
    d_{\dot \alpha}\tilde Q_{n}^{\dot \beta} = \tilde d_{\dot \alpha}\tilde Q_{n-1}^{\dot \beta} - d_{\dot \alpha}d_{\dot \gamma}\Theta d^{\dot \gamma}\tilde Q_{n-1}^{\dot \beta} , \qquad \tilde Q_{-2}^{\dot \alpha} = d^{\dot \alpha}\Theta , \qquad \tilde Q_{-1}^{\dot \alpha} = \tilde d^{\dot \alpha}\Theta .
\end{equation}
The higher $\tilde Q_{n}^{\dot \alpha}$ can be seen as Hamiltonian flows in higher times associated with the integrable hierarchy \cite{Dunajski:2000iq,Dunajski:2003gp,Takasaki:1988cd, Boyer:1985aj}. There exists a precise relationship between these spacetime fields and twistor space data given by 
\begin{equation}
    \tilde Q_{n}^{\dot \alpha} = \frac{1}{2\pi i}\int_X q^{n+2}dq \wedge \epsilon^{\dot \alpha\dot \beta}\left.\frac{\partial h}{\partial \mu^{\dot \beta}}\right|_X . 
\end{equation}
Once again, the recursion operator can simply be seen as multiplication by the coordinate $q$ on twistor space.

\subsection{Symmetry algebra}
\label{sec:Symmetry Algebra}

We now focus on the explicit form of the symmetry generators in this gauge as they appear in the bulk. In this gauge, the Poisson bracket on twistor space is given by
\begin{equation}
    \{f_1 , f_2\} = \langle \iota\lambda\rangle^{-2}d^{\dot \alpha} f_1 d_{\dot \alpha} f_2 . 
\end{equation}
The gauge parameters $\xi$ and $\chi$ can be constructed as residual gauge transformations. Working on the spin bundle, the pullback $\xi_s\vert_X  = \xi(\tau_s)$ is expanded as discussed in Section 3.3
\begin{equation}
    \xi(\tau_s) = \sum_{n=0}^{s+1} \xi_{s,n}(\tau_s) q^n .
\end{equation}
The coefficients of $\xi(\tau_s)$ must satisfy the recursion relations to preserve the gauge choice \eqref{PlebanskiGauge-h}
\begin{equation}
    d_{\dot \alpha}\xi_{s,n-1} = \tilde d_{\dot \alpha}\xi_{s,n} - d_{\dot \alpha}d_{\dot \beta}\Theta d^{\dot \beta}\xi_{s,n} . 
\end{equation}
These recursion relations are truncated such that $d_{\dot \alpha}\xi_{s,s+1} = 0$ and we define the function $\xi_{s,s+1} = \tau_s(\tilde x^{\dot \alpha})$ depending only on half the spacetime variables. As mentioned previously, the coefficients for $\xi(\tau_s)$ are defined via the recursion operator $R$ and as such each $\xi_{s,n}$ satisfies the linearised Plebanski equation. 

We now turn to the construction of the $\chi_s$ generators. As seen in \eqref{ChiGenerators}, the $\chi_s$ pulls back to give fields on spacetime via the contour integral
\begin{equation}
    \chi_{s,n}=\frac{1}{2\pi i}\oint_{\Gamma}q^{n+2}dq \,\chi_{s}'\vert_X\, ,
\end{equation}
where $\chi_s' = \langle\iota\lambda\rangle^6 \chi_s$. Since $\chi_s$ is the Čech representative, it satisfies $L_{\dot \alpha}\chi_s\vert_X = 0$ which translates to the recursion relations
\begin{equation}
    d_{\dot \alpha}\chi_{s,n+1} = \tilde d_{\dot \alpha}\chi_{s,n} - d_{\dot \alpha}d_{\dot \beta}\Theta d^{\dot \beta}\chi_{s,n} \, ,
\end{equation}
with the truncation $\chi_{s,n} = 0$ for $n< -s-3$ which results in $\chi_{s,-s-3} = \rho_s(\tilde x^{\dot \alpha})$ only depending on half the spacetime coordinates. For future convenience, we denote $\chi_s\vert_X = \chi(\rho_s)$ as the pull back of the $\chi_s$ generator. These parameters form an algebra through the use of the algebroid brackets \eqref{ModifiedBrackets} on the field space due to the inherent field dependence coming from the recursion relations
\begin{equation}
\begin{split}
    \{\xi, \xi'\}_* &= \{\xi, \xi'\} + \delta_\xi \xi' - \delta_{\xi'}\xi , \\
    \{\xi, \chi\}_* &= \{\xi, \chi\} + \delta_\xi \chi .
\end{split} 
\end{equation}
We note that the only field dependence arises from the background metric generated by $h$, and as such, only $\delta_\xi$ terms appear. With these brackets, one can show that the algebra satisfied by the generators is a representation of the algebra between the field-independent parameters $\tau_s$ and $\rho_s$
\begin{equation}
    \begin{split}
        \{\xi(\tau_s) , \xi(\tau_{s'})\}_*  &= \xi\left(\tau_{s+s'-1} = \tilde d^{\dot \alpha}\tau_s \tilde d_{\dot \alpha}\tau_{s'}\right) , \\
        \{\xi(\tau_s) , \chi(\rho_{s'})\}_*  &= \chi\left(\rho_{s+s'-1}=\tilde d^{\dot \alpha}\tau_s \tilde d_{\dot \alpha}\rho_{s'}\right) . 
    \end{split}
\end{equation}
We prove these brackets in Appendix \ref{sec:Proof of Algebra}. If we expand these functions in a basis
\begin{equation}
    \tau[m,n] = \langle \iota\lambda\rangle^2\tilde x^m \tilde y^n , \qquad \rho[m,n] = \langle \iota\lambda\rangle^{-6} \tilde x^m \tilde y^n , 
\end{equation}
such that $m+n = s+1$. We find the algebra satisfied by these generators to be
\begin{equation}
    \begin{split}
        \tilde d^{\dot \alpha}\tau[m,n] \tilde d_{\dot \alpha}\tau[r,l] = (ml - nr)\tau[m+r-1,n+l-1] , \\
        \tilde d^{\dot \alpha}\tau[m,n] \tilde d_{\dot \alpha}\rho[r,l] = (ml - nr)\rho[m+r-1,n+l-1] , 
    \end{split}
\end{equation}
which is the algebra $\text{Ham}(\mathbb{C}^2)$. Since the generators $\xi$ and $\phi$ are linear in $\tau$ and $\chi$ respectively, the structure constant factors through, and we obtain the field-dependent algebra given by
\begin{equation}
    \begin{split}
        \{\xi[m,n], \xi[r,l]\}_* &= (ml-nr)\xi[m+r-1,n+l-1] , \\
         \{\xi[m,n], \chi[r,l]\}_* &= (ml-nr)\chi[m+r-1,n+l-1] ,
    \end{split}
\end{equation}
where $\xi[m,n]  = \xi(\tau[m,n])$ and $\chi[m,n] = \chi(\rho[m,n])$. Therefore, since the charge algebra is a representation of the modified bracket, we find the charges manifest the $\text{Ham}(\mathbb{C}^2)$ algebra.

By using the integral formula \eqref{HigherSpinPenroseTransform}, the action of these symmetries on the spacetime fields can be explicitly written down 
\begin{equation}
\begin{split}
    \delta_{\xi(\tau_s), \chi(\rho_{s'})}Q_n &= \chi_{s',n}+\sum_{m=0}^{s+1}d^{\dot \alpha}Q_{n+m}d_{\dot \alpha}\xi_{s,m} , \\
     \delta_{\xi(\tau_s)}d_{\dot \alpha}\Theta &= \tilde d_{\dot \alpha}\xi_{s,0} - d_{\dot \alpha}d_{\dot \beta}\Theta d^{\dot \beta}\xi_{s,0} . 
\end{split}
\end{equation}
One can show explicitly that these are the symmetries of the recursion relations \eqref{recursion}. We now present several examples of such transformations and highlight their features. The first examples of the variation of the Plebanski potential are given by
\begin{equation}
    \begin{split}
        \delta_{\xi(\tau_{-1})} \Theta &= x^{\dot \alpha}\tilde d_{\dot \alpha}\tau_{-1} , \\
        \delta_{\xi(\tau_{0})} \Theta &= \frac{1}{2!}x^{\dot \alpha}x^{\dot \beta}\tilde d_{\dot \alpha}\tilde d_{\dot \beta} \tau_0 - d_{\dot \alpha}\Theta \tilde d^{\dot \alpha}\tau_0 , \\
        \delta_{\xi(\tau_{1})} \Theta &= \frac{1}{3!}x^{\dot \alpha}x^{\dot \beta}x^{\dot \gamma}\tilde d_{\dot \alpha}\tilde d_{\dot \beta}\tilde d_{\dot \gamma}\tau_1 - x^{\dot \alpha}\tilde d_{\dot \alpha} \tilde d^{\dot \beta} \tau_1 d_{\dot \beta}\Theta + \tilde d^{\dot \alpha}\Theta \tilde d_{\dot \alpha} \tau_1 , \\
        \delta_{\xi(\tau_{2})} \Theta  &= \frac{1}{4!}x^{\dot\alpha}x^{\dot \beta}x^{\dot \gamma}x^{\dot \epsilon} \tilde d_{\dot \alpha}\tilde d_{\dot \beta}\tilde d_{\dot \gamma}\tilde d_{\dot \epsilon}\tau_2 +\frac{1}{2}x^{\dot \alpha}x^{\dot \beta}\tilde d_{\dot \alpha}\tilde d_{\dot \beta}\tilde d_{\gamma}\tau_2 d^{\dot \gamma}\Theta  \\
        & \qquad \qquad \qquad+ x^{\dot \alpha}\tilde d_{\dot \alpha}\tilde d_{\dot \beta}\tau_2 \tilde d^{\dot \beta}\Theta + \frac{1}{2}d_{\dot \alpha}\Theta d_{\dot \beta}\Theta \tilde d^{\dot \alpha}\tilde d^{\dot \beta }\tau_2 + \tilde Q_{0}^{\dot \alpha}\tilde d_{\dot \alpha}\tau_2 . 
    \end{split}
\end{equation}
We note that the first three variations appear as the residual diffeomorphisms that preserve the Plebanski gauge \cite{Dunajski:2000iq,Campiglia:2021srh}. Furthermore, these residual diffeomorphisms were used in \cite{Miller:2025wpq} to identify a $Lw_{1+\infty}$ algebra on spacetime; however, the present higher spin variations differ as can be seen in the last variation where there is a term that is non-local and defined via the integrable hierarchy equations. For higher spins, the variations continue to be non-local as one would expect. Furthermore, note that the $s=-1$ variation changes the potential, but the metric will be left unchanged. The $\xi(\tau_s)$ transformations follow a similar patterns for the $Q_n$ where the first few examples are given by
\begin{equation}
    \begin{split}
        \delta_{\xi(\tau_{-1})}Q_{n} &= 0 , \\
        \delta_{\xi(\tau_{0})}Q_{n} &=\tilde d_{\dot \alpha}\tau_0 d^{\dot \alpha}Q_n , \\
        \delta_{\xi(\tau_{1})}Q_{n} &= x^{\dot\beta}\tilde d_{\dot \alpha}\tilde d_{\dot \beta}\tau_1 d^{\dot \alpha}Q_n + \tilde d_{\dot \alpha}\tau_1\tilde d^{\dot \alpha}Q_n , \\
        \delta_{\xi(\tau_{2})}Q_{n} &= \left(\frac{1}{2} x^{\dot\beta}x^{\dot \gamma} \tilde d_{\dot \alpha}\tilde d_{\dot \beta}\tilde d_{\dot \gamma}\tau_2 + \tilde d_{\dot \alpha}\tilde d_{\dot \beta}\tau_2 d^{\dot \beta}\Theta + \tilde d_{\dot \beta}\tau_2 d_{\dot \alpha}\tilde d^{\dot \beta}\Theta \right)d^{\dot \alpha}Q_n \\
        & \qquad \qquad +x^{\dot \beta}\tilde d_{\dot \alpha}\tilde d_{\dot \beta}\tau_2 \tilde d^{\dot \alpha}Q_n + \tilde d_{\dot \alpha}\tau_2 \tilde d^{\dot \alpha}Q_{n+1} . 
    \end{split}
\end{equation}
We first note that the $s=-1$ variation does not change the field, and as such does not vary the physical fields, only the potential. Secondly, for $s=0,1$ the variation only involves the field itself, while for higher values of $s$, not only does the potential appear, but also the higher spin $Q_n$. Therefore, to make full sense of the higher spin variations, one must include the entire tower of $Q_n$ for all values of $n$. The variations associated with the $\chi(\rho_s)$ symmetry can also be written down in the same fashion, where the first examples are given by
\begin{equation}
    \begin{split}
        \delta_{\chi(\rho_{-1})}Q_{-2} &= \rho_{-1} , \\
        \delta_{\chi(\rho_{0})}Q_{-2} &= x^{\dot \alpha}\tilde d_{\dot \alpha}\rho_{0} , \\
        \delta_{\chi(\rho_1)}Q_{-2} & = \frac{1}{2!}x^{\dot \alpha}x^{\dot \beta}\tilde d_{\dot \alpha}\tilde d_{\dot \beta} \rho_1 - d_{\dot \alpha}\Theta \tilde d^{\dot \alpha}\rho_1 ,
    \end{split}
\end{equation}
and to obtain the variation of the higher spin $Q_{n}$, one has to apply the recursion operator to these variations. 
As mentioned previously, the benefit of Plebanski gauge is the ability to write the non-linear field $h$ in terms of explicit spacetime fields. As such, the two sets of charges are given by
\begin{equation}
    H_s = \sum_{n=0}^{s+1}\int_{\mathcal{S}}dy d\tilde x \, Q_{n-1}\xi_{s,n} \,\, ,  \qquad \tilde H_s = \int_{\mathcal{N}} dx dy  d \tilde x \, \chi_{s,-2}  \partial _{x}\Theta . 
\end{equation}
where the null hypersurface $\mathcal{N}$ is chosen such that $\tilde y$ is a constant. 

\section{Discussion}

In this work, we have proposed  semi-local and quasi-local expressions for the charges associated with the celestial $Lw_{1+\infty}$ symmetries. The expression is a two-surface integral in the bulk, which can be seen as embedded inside a null hypersurface, and is conserved along this null hypersurface in the absence of radiation ($Q_{-2} = 0$). In the linearized theory, we have shown that the 2-surface formulae also captures the Geroch-Hansen definition of multipole moments following Curtis \cite{Curtis:1978}, but with a different definition of recursion. The definition can be extended to the non-linear theory by either restricting the discussion to the bulk null hypersurface, or even a 2-surface,  or by  restricting to the self-dual sector. Its first-principles derivation follows from the interpretation of $Lw_{1+\infty}$ on twistor space as large gauge transformations, the application of covariant phase space methods to the twistor space action, and the subsequent translation to spacetime. This expression has been specified both at null infinity and at finite distance, producing results consistent with earlier analyses. Finally, we have discussed this construction in the Plebański gauge, which is well adapted to self-dual gravity and in which the relation with integrability is particularly transparent.

This work provides a spacetime interpretation of the $Lw_{1+\infty}$ symmetries. Indeed, while the lower-spin symmetries could be interpreted as large gauge diffeomorphisms or asymptotic Killing vectors on spacetime, the status of the higher-spin symmetries was not clear. Here, we see that they should be interpreted in terms of higher-valence spinors on spacetime, satisfying the twistor equation \eqref{Highertwistorequation}.  These in turn generate motions along higher flows or times that arise from the complete integrability of the self-dual Einstein equations, cf \cite{Boyer:1985aj}.  The work in more general spacetimes provides some tentative extension of these motions beyond the self-dual sector. 

In our construction, the null hypersurface along which we study the evolution and flux plays a key role. The recursion relations for the charge aspects \eqref{recusion relations charge aspects}, as well as the dual recursion relations \eqref{Dual Recursion} for the parameters—which are the building blocks of this construction—rely precisely on this null time evolution. In particular, this null hypersurface allows us to define a clear notion of radiation, captured in $Q_{-2}$, which obstructs the null-time conservation of the charges. Identifying radiation at finite distance is usually not an easy task (see, e.g., \cite{Anninos:2023epi,Anninos:2024wpy,Anninos:2024xhc,Adami:2024gdx,Sheikh-Jabbari:2025kjd} for recent discussions on timelike hypersurfaces), and the fact that the hypersurface is null considerably simplifies this analysis. 

Null hypersurfaces are known to carry a Carrollian geometry; see e.g. \cite{Ruzziconi:2026bix} for a review. In this work, celestial symmetries have been identified on arbitrary bulk null hypersurfaces. It would be interesting to further clarify the spacetime geometry associated with these symmetries, and to understand how the universal Carrollian structure should be extended to properly accommodate them. We also refer to \cite{Cordova:2018ygx,Hu:2023geb,Himwich:2025ekg,Sheta:2025oep,Strominger:2026yrh} for similar identifications of the celestial symmetries on null hypersurfaces in CFT.

While the Noether analysis has been performed for self-dual gravity, the same methods could be applied to self-dual Yang–Mills theory, connecting with the analyses of \cite{Freidel:2023gue, Kmec:2025ftx, Cresto:2025bfo}. Finally, our analysis could also be extended to spacetimes with a non-vanishing cosmological constant, where deformed celestial algebras have been found \cite{Taylor:2023ajd, Bittleston:2024rqe, Sheta:2025oep}.

\section*{Acknowledgments}

We thank Atul Sharma and Céline Zwikel for discussions. 

While completing this work, we found that related results were independently obtained in \cite{Freidel:2026iia} for Yang-Mills theory. It would be interesting to see how their 1-form symmetry framework extends to gravity structures addressed in this paper.

AK is supported by the STFC. LJM would like to thank the Simons Collaboration on Celestial Holography CH-00001550-11 and STFC for financial support from grant number ST/T000864/1. RR is supported by the European Union’s Horizon Europe research and innovation programme under the Marie Skłodowska-Curie grant agreement No. 101104845 (UniFlatHolo), hosted at Harvard University and École Polytechnique.

\appendix

\section{Spinor, NP and GHP formalisms} \label{sec:background}

\subsection{Spinor notations}

We will work with a complexified spacetime, where the underlying manifold $\mathcal{M}$ is a complex $4$-dimensional space with a holomorphic metric $g$.
We will assume $\mathcal{M}$ has a spin structure and therefore utilize the isomorphism of the (complexified) holonomy group 
\begin{equation}
    SO(4,\mathbb{C}) = SL(2,\mathbb{C})\times SL(2,\mathbb{C})/\mathbb{Z}_{2}.
\end{equation}
The isomorphism allows us to decompose the complexified tangent bundle into a product of spin bundles
\begin{equation} \label{SpinIso}
    T\mathcal{M}_{\mathbb{C}} = \mathbb{S}\otimes \mathbb{S}'
\end{equation}
where $\mathbb{S}$ and $\mathbb{S}'$ are the anti-self-dual (ASD) and self-dual (SD) spin bundles, respectively. Each spin bundle has a natural symplectic structure given by the Levi-Civita symbol such that the rules for raising and lowering indices are given by $\omega^{\alpha} = \epsilon^{\alpha \beta} \omega_{\beta}$ and $\omega_{\alpha} =  \omega^{\beta}\epsilon_{ \beta\alpha}$, encoding the duals to each spin bundle. The natural inner product is given by $\langle \omega \lambda\rangle =\omega^{\alpha} \lambda_{\alpha} $ and $[\tilde\omega\tilde\lambda] = \tilde\omega^{\dot \alpha}\tilde\lambda_{\dot \alpha} $. In practice, the isomorphism boils down to applying the Van de Waerden symbols (see, e.g., \cite{Penrose:1984uia}) on a vector index to turn it into a matrix 
\begin{equation} \label{VectSpin}
    V^{\alpha \dot \alpha} = \sigma^{\alpha \dot \alpha }_{a}V^{a} = \frac{1}{\sqrt{2}}
    \begin{pmatrix}
        V^{0}+V^{3} & V^{1}+iV^{2} \\
        V^{1}-iV^{2} &  V^{0}-V^{3}
    \end{pmatrix},
\end{equation}
where $\alpha,\dot\alpha = 0,1$ and the norm $2\text{det}(V^{\alpha \dot \alpha}) = \eta_{ab}V^{a}V^{b} = \epsilon_{\alpha\beta}\epsilon_{\dot\alpha\dot\beta}V^{\alpha \dot \alpha}V^{\beta \dot \beta}$. In a co-tetrad basis, the metric is given by
\begin{equation}
    g= \eta_{ab}\theta^a \theta^b = \epsilon_{\alpha\beta}\epsilon_{\dot\alpha\dot\beta}\theta^{\alpha \dot \alpha}\theta^{\beta \dot \beta} ,
\end{equation}
where the co-tetrad satisfies $e_{\alpha\dot\alpha}\lrcorner\,\theta^{\beta\dot\beta} = \delta^\beta_\alpha\delta^{\dot\beta}_{\dot\alpha}$, with $e_{\alpha\dot\alpha}$ being the tetrad. In the torsion free case, the tetrad $\theta^{\alpha\dot \alpha}$ satisfies Cartan's first structure equation
\begin{equation}
    d\theta^{\alpha\dot \alpha} = \Gamma^{\alpha}_{\,\,\,\beta}\wedge\theta^{\beta\dot\alpha} + \tilde\Gamma^{\dot \alpha}_{\,\,\,\dot\beta}\wedge\theta^{\alpha\dot\beta}
\end{equation}
where $\Gamma_{\alpha\beta}$ and $\tilde\Gamma_{\dot \alpha\dot \beta}$ are the spin connection 1-forms on the ASD and SD spin bundles, respectively. The connection 1-forms are symmetric in their indices, $\Gamma_{\alpha\beta} = \Gamma_{\beta\alpha}$. Using the spin connections, the ASD and SD Riemman 2-forms are given by
\begin{equation}
    R_{\alpha\beta} = d\Gamma_{\alpha\beta}+\Gamma_{\alpha}^{\,\,\,\gamma}\wedge\Gamma_{\gamma\beta} , \qquad \tilde R_{\dot\alpha\dot \beta} = d\tilde\Gamma_{\dot\alpha\dot\beta}+\tilde\Gamma_{\dot\alpha}^{\,\,\,\dot\gamma}\wedge\tilde\Gamma_{\dot\gamma\dot\beta} .
\end{equation}
The main benefit of using spinor notation is that there emerges a natural basis of SD and ASD 2-forms on the 4-manifold $\mathcal{M}$. Consider the decomposition of the 2-form 
\begin{equation}
    \theta^{\alpha \dot \alpha} \wedge \theta^{\beta\dot\beta} = \epsilon^{\dot \alpha \dot \beta} \Sigma^{\alpha \beta} + \epsilon^{\alpha \beta}\tilde \Sigma^{\dot \alpha \dot \beta},
\end{equation}
where we have defined $\Sigma^{\alpha \beta} = \theta^{\alpha \dot \alpha} \wedge \theta^{\beta}_{\,\,\,\dot\alpha}$ and $\tilde\Sigma^{\dot\alpha \dot\beta} = \theta^{\alpha \dot \alpha} \wedge \theta^{\,\,\,\dot\beta}_{\alpha}$. Thus, a generic 2-form $F$ is given by
\begin{equation}
    F = F_{\alpha\beta}\Sigma^{\alpha \beta}+ \tilde F_{\dot \alpha \dot \beta} \tilde\Sigma^{\dot\alpha \dot\beta}
\end{equation}
where $F_{\alpha\beta}$ is the ASD part and $\tilde F_{\dot \alpha \dot \beta}$ is the SD part under the Hodge star. Using the various symmetries of the Riemann tensor, it can also be expanded in the basis of ASD and SD 2-forms 
\begin{equation}
    \begin{split}
        R_{\alpha \beta} &=  \psi_{\alpha \beta \gamma\sigma}\Sigma^{\gamma \sigma} + \Phi_{\alpha \beta \dot \alpha \dot \beta} \tilde\Sigma^{\dot \alpha \dot \beta} + \frac{R}{12}\Sigma_{\alpha \beta},\\
        \tilde R_{\dot \alpha \dot \beta} &=  \tilde \psi_{\dot\alpha \dot\beta \dot \gamma\dot\sigma}\tilde\Sigma^{\dot\gamma \dot\sigma} + \Phi_{\alpha \beta \dot \alpha \dot \beta} \Sigma^{ \alpha \beta} + \frac{R}{12}\tilde\Sigma_{\dot\alpha \dot \beta},
    \end{split}
\end{equation}
where $\psi_{\alpha \beta \gamma\sigma}$ and $\tilde \psi_{\dot\alpha \dot\beta \dot \gamma\dot\sigma}$ are a fully symmetric spinors and $\Phi_{\alpha \beta \dot \alpha \dot \beta}  = \Phi_{\beta \alpha\dot \beta \dot \alpha}$. The various quantities appearing above are related to standard objects in Riemannian geometry, i.e.,
\begin{equation}
         C_{abcd} = \psi_{\alpha \beta \gamma \sigma} \epsilon_{\dot\alpha \dot \beta}\epsilon_{\dot \gamma \dot \sigma} + \tilde \psi_{\dot \alpha \dot \beta \dot \gamma \dot \sigma} \epsilon_{\alpha  \beta}\epsilon_{ \gamma  \sigma}, \quad \quad \quad R_{ab} - \frac{1}{4}Rg_{ab} = - 2\Phi_{\alpha \beta \dot \alpha \dot \beta} \, ,
\end{equation}
where $C_{abcd}$ is the Weyl tensor. One can further show that the second Bianchi identity then gives the equations of motion 
\begin{equation}
    \nabla_{\,\, \,\dot \alpha}^{\sigma}\psi_{\alpha \beta \gamma \sigma} = \nabla_{(\alpha}^{\,\,\,\, \,\dot \beta}\Phi_{\beta \gamma)\dot \alpha \dot \beta} \quad , \quad \nabla^{a}\Phi_{ab} + \frac{1}{8}\nabla_{b}R = 0.
\end{equation}
A useful identity involving the commutator of the covariant derivatives on and ASD spinor is given by
\begin{equation}\label{Commutator}
    [\nabla_{\alpha\dot \alpha}, \nabla_{\beta\dot \beta}]\lambda^\gamma = \left[\epsilon_{\dot \alpha\dot \beta}\left(\psi_{\alpha\beta\sigma}^{\quad\,\, \gamma} - \frac{R}{12}\epsilon_{\sigma(\alpha}\delta_{\beta)}^{\gamma}\right)+ \epsilon_{\alpha \beta}\Phi^\gamma_{\,\,\, \sigma\dot \alpha\dot \beta}\right]\lambda^{\sigma} .
\end{equation}
\subsection{NP and GHP formalisms}
The traditional tetrad formalism is set up such that the tetrad is orthonormal and the metric is that of Minkowski. To use the spinor decomposition \eqref{SpinIso} to our advantage, it is convenient to use the Newman-Penrose (NP) formalism \cite{Newman:1961qr}. We first notice that under the isomorphism \eqref{SpinIso}, null vectors turn into matrices by \eqref{VectSpin} which have vanishing determinants. Therefore, a generic null vector can be written as 
\begin{equation}
    V^{\alpha \dot \alpha} = \lambda^{\alpha} \tilde\lambda^{\dot \alpha},
\end{equation}
where the residual little is captured by the automorphism $\lambda \rightarrow r \lambda$ and $\tilde\lambda \rightarrow r^{-1} \tilde\lambda$ for some $r\in \mathbb{C}^{*}$. We therefore wish to use a null tetrad which has the internal metric
\begin{equation}
    \eta_{ab} =  \begin{pmatrix}
        0 & 1 & 0 & 0 \\
        1 & 0 & 0 & 0 \\
        0 & 0 & 0 & -1 \\
        0 & 0 & -1 & 0 
    \end{pmatrix}.
\end{equation}
This tells us that the tetrad
\begin{equation}
    e_{\mu}^{a} = (l_{\mu} , n_{\mu} , m_{\mu} , \bar m_{\mu}),
\end{equation}
has the normalization 
\begin{equation} \label{norm}
    \begin{split}
        l_{\mu}l^{\mu} = n_{\mu}n^{\mu}  = m_{\mu}m^{\mu}  = &\bar m_{\mu}\bar m^{\mu} = l_{\mu}m^{\mu} = l_{\mu}\bar m^{\mu} = n_{\mu}m^{\mu} = n_{\mu}\bar m^{\mu} = 0 ,\\
        &l_{\mu}n^{\mu} = -m_{\mu}\bar m^{\mu} = 1 ,
    \end{split}
\end{equation}
and the metric is given by
\begin{equation}
    g_{\mu\nu} = l_{\mu}n_{\nu} + l_{\nu}n_{\mu} - m_{\mu}\bar m_{\nu}- m_{\nu} \bar m_{\mu}.
\end{equation}
The translation of the null tetrad in the spinor notation is provided by choosing a spin basis (the simplest example being $\iota^{\alpha} = \bar \iota^{\dot\alpha} = (1,0)$ and $o^{\alpha} = \bar o^{\dot \alpha} = (0,-1) $) such that the inner product $\langle \iota o \rangle = [\bar \iota \bar o] = 1$. Therefore, in abstract index notation, we can make the identification
\begin{equation} \label{tetrad NP}
    l_{\mu} = o_{\alpha}\bar o_{\dot \alpha} \quad n_{\mu} = \iota_{\alpha}\bar \iota_{\dot \alpha} \quad 
    m_{\mu} = o_{\alpha}\bar \iota_{\dot \alpha} \quad 
    \bar m_{\mu} = \iota_{\alpha}\bar o_{\dot \alpha},
\end{equation}
which solves the normalization conditions \eqref{norm}. The next task is to build the complex spin coefficients which are linear combinations of various components of the spin connection $1$-form $\Gamma_{\alpha \beta}$. The explicit formulae for the spin coefficients are given in Appendix \ref{sec:Useful formulae}, Equation \eqref{spin coeff}. An important spin coefficient that appears in asymptotically flat spacetimes and null geodesic congruences is called the shear $\sigma$, as it captures the radiative data and null infinity. The other basic objects in the NP formalism are the components of the Weyl tensor displayed in Equation \eqref{Weyl components}. We can further define the components of the covariant derivative in the null tetrad
\begin{equation}
    D = o^{\alpha}\bar o^{\dot \alpha}\nabla_{\alpha \dot \alpha}, \quad  \Delta = \iota^{\alpha}\bar \iota^{\dot \alpha}\nabla_{\alpha \dot \alpha}, \quad \delta = o^{\alpha}\bar \iota^{\dot \alpha}\nabla_{\alpha \dot \alpha}, \quad \bar \delta = \iota^{\alpha}\bar o^{\dot \alpha}\nabla_{\alpha \dot \alpha} ,
\end{equation}
along with the components of $\Phi_{\alpha \beta\dot \alpha \dot \beta}$ which give information on the matter content. The NP formalism we have described in the notations of spinors is known as the GHP (Geroch–Held–Penrose) formalism \cite{Geroch:1973am} and is particularly useful in the study of null 2-surfaces and Penrose's quasi-local momentum and angular momentum. However, in the definition of the dyad $(\iota, o)$, the only requirement was that the inner product satisfy $\langle \iota o \rangle = 1$ to solve the normalization \eqref{norm}. Therefore, there is a scaling ambiguity
\begin{equation}
    (\iota^{\alpha},o^{\alpha}) \rightarrow (r(x)^{-1}\iota^{\alpha}, r(x) o^{\alpha}) ,
\end{equation}
where $r(x)$ is an arbitrary non-vanishing complex function of spacetime. This allows for the definition of scalar fields of weights $\{p,q\}$ to have the transformation under the scaling
\begin{equation}
    \eta_{p,q} \rightarrow r^{p}\bar r^{q} \eta_{p,q},
\end{equation}
and we define the spin weight as $s = \frac{p-q}{2}$ and the boost weight $w = \frac{p+q}{2}$. By looking at all the NP scalars, the only ones that do not have a definite scale are $\{\alpha, \beta, \gamma,\epsilon \}$. This fact can be used however to construct differential operators which send a scalar $\eta$ of weight $\{p,q\}$ to scalars of a new weight 
\begin{equation} \label{GHP derivative}
\begin{split}
    \text{\th} &= D - p\epsilon  - q\bar \epsilon, \quad \quad \,\, \eth = \bar\delta - p\alpha-q\bar \beta, \\
    \text{\th}' &=\Delta - p\gamma -q\bar \gamma, \quad \quad \bar \eth = \delta - p\beta -q\bar \alpha.
\end{split}
\end{equation}
 To analyse Einstein's equations at null infinity, one can go further and define operators that transform homogenously under Weyl rescaling of the metric
 \begin{equation}
     \hat g = \Omega^2 g
 \end{equation}
 where $\Omega = 0$ defines infinity. In spinor notation, the Levi-Civita symbols will transform
 \begin{equation}
     \hat \epsilon_{\alpha\beta} = \Omega \epsilon_{\alpha\beta}\qquad \hat \epsilon_{\dot \alpha\dot\beta} = \Omega \epsilon_{\dot\alpha\dot\beta} .
 \end{equation}
The basis of spinors we use to define our null tetrad will also be rescaled via
\begin{equation} \label{weyl dyad}
\begin{split} 
    \hat o_{\alpha} = o_{\alpha} \quad , \quad \hat \iota_{\alpha} = \Omega \iota_{\alpha}, \\
    \hat o^{\alpha} = \Omega^{-1} o^{\alpha} \quad , \quad  \hat \iota^{\alpha} = \iota^{\alpha},
\end{split}
\end{equation}
such that the inner product $\langle \hat\iota \hat o\rangle = \langle \iota o\rangle =1 $ is invariant  \cite{penrose1984spinors}. Under \eqref{weyl dyad}, a Weyl-weighted scalar transforms as
\begin{equation}
    \hat \eta^{W}_{p,q} = \Omega^{W}\eta^{W}_{p,q}.
\end{equation}
Using the spin coefficients which do not have a definite Weyl weight, we define the covariant operators
\begin{equation} \label{A27}
\begin{aligned}[c]
    \text{\th}_{\mathcal{C}} &= \text{\th}+ W\rho,\\
    \text{\th}'_{\mathcal{C}} &= \text{\th}'- (W+p+q)\mu,
\end{aligned}
\quad \quad
\begin{aligned}[c]
    \eth_{\mathcal{C}} &= \eth -(W+p)\pi,\\  
    \bar \eth_{\mathcal{C}} &=\bar \eth +(W+q) \tau.
\end{aligned}
\end{equation}

\subsection{Useful formulae in NP formalism}
\label{sec:Useful formulae}

In this appendix, we present some further details of the NP and GHP formalism. A full account can be found in \cite{Penrose:1984uia,Penrose:1986uia}. In NP formalism, the spin coefficients expressed in the null tetrad \eqref{tetrad NP} are given by
\begin{equation} \label{spin coeff}
\begin{aligned}[c]
\alpha &= \iota^{\alpha} \iota^{\beta}\bar o^{\dot \beta} \nabla_{\beta \dot \beta} o_{\alpha},\\
\beta &= \iota^{\alpha} o^{\beta}\bar \iota^{\dot \beta} \nabla_{\beta \dot \beta} o_{\alpha},\\
\gamma &= \iota^{\alpha} \iota^{\beta}\bar \iota^{\dot \beta} \nabla_{\beta \dot \beta} o_{\alpha},\\
\epsilon &= \iota^{\alpha} o^{\beta}\bar o^{\dot \beta} \nabla_{\beta \dot \beta} o_{\alpha},
\end{aligned}
\quad
\begin{aligned}[c]
\lambda &= \iota^{\alpha} \iota^{\beta}\bar o^{\dot \beta} \nabla_{\beta \dot \beta} \iota_{\alpha},\\
\mu &= \iota^{\alpha} o^{\beta}\bar \iota^{\dot \beta} \nabla_{\beta \dot \beta} \iota_{\alpha},\\
\nu &= \iota^{\alpha} \iota^{\beta}\bar \iota^{\dot \beta} \nabla_{\beta \dot \beta} \iota_{\alpha},\\
\pi &= \iota^{\alpha} o^{\beta}\bar o^{\dot \beta} \nabla_{\beta \dot \beta} \iota_{\alpha},
\end{aligned}
\quad
\begin{aligned}[c]
\rho &= o^{\alpha} \iota^{\beta}\bar o^{\dot \beta} \nabla_{\beta \dot \beta} o_{\alpha},\\
\sigma &= o^{\alpha} o^{\beta}\bar \iota^{\dot \beta} \nabla_{\beta \dot \beta} o_{\alpha},\\
\tau &= o^{\alpha} \iota^{\beta}\bar \iota^{\dot \beta} \nabla_{\beta \dot \beta} o_{\alpha},\\
\kappa &= o^{\alpha} o^{\beta}\bar o^{\dot \beta} \nabla_{\beta \dot \beta} o_{\alpha}.
\end{aligned}
\end{equation} The components of the Weyl tensor read as
\begin{equation} \label{Weyl components}
    \begin{split}
        \psi_{0} &= C_{\mu\nu\rho\sigma} l^{\mu} m^{\nu} l^{\rho} m^{\sigma} = \psi_{\alpha\beta \gamma \delta} o^{\alpha} o^{\beta}o^{\gamma}o^{\delta}, \\
        \psi_{1} &= C_{\mu\nu\rho\sigma} l^{\mu} n^{\nu} l^{\rho} m^{\sigma}  = \psi_{\alpha\beta \gamma \delta} \iota^{\alpha} o^{\beta}o^{\gamma}o^{\delta}, \\
        \psi_{2} &= \frac{1}{2}(C_{\mu\nu\rho\sigma} l^{\mu} n^{\nu} l^{\rho} n^{\sigma} - C_{\mu\nu\rho\sigma} l^{\mu} n^{\nu} m^{\rho} \bar m^{\sigma}) = \psi_{\alpha\beta \gamma \delta} \iota^{\alpha} \iota^{\beta}o^{\gamma}o^{\delta},\\
        \psi_{3} &= C_{\mu\nu\rho\sigma} l^{\mu} n^{\nu} n^{\rho} \bar m^{\sigma} = \psi_{\alpha\beta \gamma \delta} \iota^{\alpha} \iota^{\beta}\iota^{\gamma}o^{\delta},\\
        \psi_{4} &= C_{\mu\nu\rho\sigma} n^{\mu} \bar m^{\nu} n^{\rho} \bar m^{\sigma} = \psi_{\alpha\beta \gamma \delta} \iota^{\alpha} \iota^{\beta}\iota^{\gamma}\iota^{\delta}, \\
    \end{split}
\end{equation}
which we can see are simply obtained from the Weyl spinor decomposition 
\begin{equation}
    \psi_{\alpha\beta \gamma \delta} = \psi_{0}\, \iota_{\alpha} \iota_{\beta} \iota_{\gamma} \iota_{\delta} -4 \psi_{1} \,  o_{(\alpha} \iota_{\beta} \iota_{\gamma} \iota_{\delta)} + 6\psi_{2} \,  o_{(\alpha} o_{\beta} \iota_{\gamma} \iota_{\delta)} - 4\psi_{3} \,  o_{(\alpha} o_{\beta} o_{\gamma} \iota_{\delta)} + \psi_{4} \,  o_{\alpha} o_{\beta} o_{\gamma} o_{\delta},
\end{equation}
and similarly for the complex conjugate $\tilde \psi_{\dot \alpha \dot\beta \dot \gamma\dot \delta}$.

Following the notations and definitions in Section \ref{sec:Penrose's quasi-local mass}, there are a total of 36 equations that relate the spin coefficients to the components of the Weyl tensor. They are given by
\begin{equation}
    \begin{split}
        \text{\th} \rho - \eth \kappa &= \rho^2 + \sigma \bar\sigma - \bar\kappa \tau + \pi \kappa + \phi_{00}\\
        \text{\th} \sigma - \bar\eth \kappa &= (\rho + \bar\rho )\sigma - (\tau - \bar\pi)\kappa + \psi_0\\
        \text{\th} \tau - \text{\th}'\kappa &= (\tau + \bar\pi)\rho + (\bar\tau + \pi)\sigma + \psi_1\\
        \bar\eth \rho - \eth\sigma  &= (\rho - \bar\rho)\tau + (\mu-\bar\mu)\kappa - \psi_1 + \phi_{01}\\
        \bar\eth \tau - \text{\th}' \sigma &= \mu\sigma + \bar\lambda\rho + \tau^2 - \nu \kappa + \phi_{02}\\
        \text{\th}' \rho - \eth\tau &= \kappa \nu - \rho \bar\mu - \sigma\lambda - \tau\bar\tau - \psi_2 - 2\Lambda\\
        \bar \eth \nu - \text{\th}' \mu &= \mu^2 + \lambda\bar\lambda - \bar\nu \pi + \tau \nu + \phi_{22}\\
        \eth \nu - \text{\th}' \lambda & = (\mu+ \bar\mu)\lambda + (\bar\tau - \pi)\nu + \psi_4\\
        \text{\th}\nu - \text{\th}'\pi &= (\pi + \bar\tau)\mu + (\bar\pi + \tau)\lambda + \psi_3 + \phi_{21}\\
        \bar\eth \lambda - \eth \mu &= (\mu - \bar\mu)\pi - (\bar\rho - \rho)\nu - \psi_3 + \phi_{21}\\
        \text{\th}\lambda - \eth\pi &= \rho\lambda + \bar\sigma \mu + \pi^2 - \nu\bar\kappa + \phi_{20}\\
        \bar\eth \pi - \text{\th}\mu &= \nu\kappa - \mu\bar\rho - \lambda\sigma - \pi \bar\pi - \psi_2 - 2\Lambda
    \end{split}
\end{equation}
along with their complex conjugates. The fields $\phi_{ij}$ are related to contractions of the spin field $\Phi_{\alpha\beta\dot \alpha\dot\beta}$. The remaining spin coefficient identities are encoded in the commutators of the GHP operators. Given a $(p,q)$-scalar $\eta$ the commutator relations are given by
\begin{equation}
    \begin{split}
        & [\text{\th},\text{\th}']= (\bar\tau + \pi)\bar\eth + (\tau+\bar\pi)\eth - p(\pi\tau - \kappa\nu + \psi_2+\phi_{11}-\Lambda) - q(\bar\pi\bar\tau - \bar\kappa\bar\nu + \bar\psi_2+\phi_{11}-\Lambda) \\
        &[\text{\th}, \bar\eth] = \bar\rho\bar\eth + \sigma\eth + \bar\pi \text{\th}- \kappa \text{\th}' - p\left(\pi\sigma - \mu\kappa+\psi_1\right) - q(\bar\rho\bar\pi - \bar\lambda\bar\kappa + \phi_{10})\\
        &[\bar\eth, \eth] = (\mu - \bar\mu)\text{\th} + (\rho - \bar\rho)\text{\th}' + p(\sigma\lambda - \rho\mu + \psi_2-\phi_{11}-\Lambda) - q(\bar\sigma\bar\lambda - \bar\rho\bar\mu + \bar\psi_2-\phi_{11}-\Lambda) 
    \end{split}
\end{equation}
The last three commutators can be found by complex conjugation and making the replacement $\iota^\alpha,\bar\iota^{\dot \alpha}\leftrightarrow o^\alpha, \bar o^{\dot \alpha}$.

\section{Proof of the algebra}
\label{sec:Proof of Algebra}

In this appendix, we prove the following, as seen in Section \ref{sec:Symmetry Algebra}
\begin{equation}
    \begin{split}
        \{\xi(\tau_s) , \xi(\tau_{s'})\}_*  &= \xi\left(\tau_{s+s'-1} = \tilde d^{\dot \alpha}\tau_s \tilde d_{\dot \alpha}\tau_{s'}\right)\\
        \{\xi(\tau_s) , \chi(\rho_{s'})\}_*  &= \chi\left(\rho_{s+s'-1}=\tilde d^{\dot \alpha}\tau_s \tilde d_{\dot \alpha}\rho_{s'}\right)
    \end{split}
\end{equation}
when working in Plebanski gauge. To begin, we focus on the algebra between the $\xi(\tau_s)$ generators. The $\xi(\tau_s)$ are constructed through the condition
\begin{equation}
    L_{\dot \alpha} \xi(\tau_s) = \langle\iota\lambda\rangle^3\delta_{\xi_s}d_{\dot \alpha}\Theta
\end{equation}
Therefore, for $\{\xi(\tau_s),\xi(\tau_{s'})\}_{*}$ to give another generator of the same type, it must also satisfy this condition. A short calculation shows that
\begin{equation}
    \begin{split}
        L_{\dot \alpha}\{\xi(\tau_s),\xi(\tau_{s'})\}_{*} &= \{L_{\dot \alpha}\xi(\tau_s),\xi(\tau_{s'})\} +L_{\dot \alpha}\delta_{\xi(\tau_s)}\xi(\tau_{s'}) - \left(s\leftrightarrow s'\right)\\
        &= \langle \iota\lambda\rangle^3\{d_{\dot \alpha}\Theta, \xi(\tau_{s'})\}+L_{\dot \alpha}\delta_{\xi(\tau_s)}\xi(\tau_{s'}) - \left(s\leftrightarrow s'\right)\\
        &= \delta_{\xi(\tau_s)}L_{\dot \alpha}\xi(\tau_{s'}) - \delta_{\xi(\tau_{s'})}L_{\dot \alpha}\xi(\tau_{s})\\
        &=\langle\iota\lambda \rangle^3[\delta_{\xi(\tau_s)},\delta_{\xi(\tau_{s'})}]d_{\dot \alpha}\Theta  = \langle\iota\lambda \rangle^3 \delta_{\{\xi(\tau_s),\xi(\tau_{s'})\}_{*}} d_{\dot \alpha}\Theta
    \end{split}
\end{equation}
which shows that the new $\xi(\tau_{s''})$ generator also satisfies the same recursion relations. By considering the expansion in $q$ of the generators, the highest order term determines $\tau_{s''}$, which we find to be field independent and given by
\begin{equation}
    \{\xi(\tau_s),\xi(\tau_{s'})\}_{*} = q^{s+s'}\tilde d^{\dot \alpha}\tau_s \tilde d_{\dot \alpha}\tau_{s'} + \cdots
\end{equation}
where we have used the recursion relations and $d_{\dot \alpha}\tau = 0$. Therefore, we find that 
\begin{equation}
    \tau_{s''} = \tau_{s+s'-1} = \tilde d^{\dot \alpha}\tau_s \tilde d_{\dot \alpha}\tau_{s'}
\end{equation} proving the first claim.

Next, we consider the bracket between a $\xi$ generator and a $\chi$ generator. The condition on the pullback of $\chi\vert_X = \chi(\rho)$ is that the Lax operator annihilates it $L_{\dot \alpha}\chi(\rho) = 0$. Another short calculation shows
\begin{equation}
    \begin{split}
        L_{\dot \alpha}\{\xi(\tau_s), \chi(\rho_{s'})\}_* & = \{L_{\dot \alpha}\xi(\tau_s), \chi(\rho_{s'})\} + \{\xi(\tau_s),L_{\dot \alpha}\chi(\rho_{s'})\} + L_{\dot \alpha}\delta_{\xi(\tau_s)}\chi(\rho_{s'})\\
        &= \{\xi(\tau_s),L_{\dot \alpha}\chi(\rho_{s'})\} + \delta_{\xi(\tau_s)}L_{\dot \alpha}\chi(\rho_{s'}) \\
        & = 0
    \end{split}
\end{equation}
where we have used $L_{\dot \alpha}\chi(\rho_{s'}) = 0$. Therefore, the modified bracket gives another $\chi(\rho_{s''})$ generator. The lowest power of $q$ in the integral that gives a non-zero result for $\chi(\rho_{s'})$ is given by
\begin{equation}
    \rho_{s'} = \frac{1}{2\pi i}\oint_{\Gamma}\frac{dq}{q^{s'+1}} \, \chi(\rho_{s'})
\end{equation}
which is importantly field independent. Therefore, since the highest power in $\xi(\tau_s)$ is of order $s+1$ in $q$ and $d_{\dot \alpha}\tau_s = 0 = d_{\dot \alpha}\rho_{s'}$, the new generator of the same type is given by
\begin{equation}
    \rho_{s''} = \frac{1}{2\pi i}\oint_{\Gamma} \frac{dq}{q^{s+s'}}\chi(\rho_{s''}) = \tilde d^{\dot \alpha}\tau_s \tilde d_{\dot \alpha}\rho_{s'}
\end{equation}
where we have used the recursion relations for both $\xi$ and $\chi$. Therefore, we find 
\begin{equation}
    \rho_{s''} = \rho_{s+s-1} = \tilde d^{\dot \alpha}\tau_s \tilde d_{\dot \alpha}\rho_{s'}
\end{equation} proving the second bracket.

\section{The linearised nonlinear graviton and the SD fields}\label{sec: Linear SD Perturbation}
In this appendix, we supplement the discussion given in Section \ref{sec:Large Gauge Transformations} where the SD perturbation was treated on a curved background where only a potential modulo gauge framework is available. On a flat background, there exists a Penrose transform integral formula \ref{PenroseTransformPositiveHelicity} which we can also consider. The key point for the SD field is that the $w^p_{m,r}$ above can be interpreted not just as local diffeomorphism symmetries of twistor space, but as \v Cech cohomology twistor representatives in $H^1(\mathbb{PT},\cO(2))$ for the SD part of the gravitational data. On a flat space background, this can be expressed by the twistor integral formula
\begin{equation}
    \Psi_{\dot\alpha\dot\beta\dot\gamma\dot\delta}(x)=\oint \frac{\p^4 w^{p}_{m,r}}{\p\mu^{\dot\alpha}\partial\mu^{\dot\beta} \partial\mu^{\dot \gamma}\p\mu^{\dot\delta}} D\lambda\, .
\end{equation}
This can be evaluated to give a field that is a polynomial of degree $2p-6$ in $x^{\dot\alpha\alpha}$ as follows.  

In order to expedite the integral, we introduce some notation.  As a polynomial in $\mu^{\dot\alpha}$, the numerator of $g^{p}_{m,r}$ can be expressed as
\begin{equation}
    L^m_{\dot\alpha_1\ldots \dot\alpha_{2p-2}}\mu^{\dot\alpha_1}\ldots\mu^{\dot\alpha_{2p-2}}\, , \qquad L^m_{\dot\alpha_1\ldots \dot\alpha_{2p-2}}=\iota_{(\dot\alpha_1}\ldots \iota_{\dot\alpha_{p+m-1}} o_{\dot\alpha_{p+m}}\ldots o_{\dot\alpha_{2p-2})}\, .
\end{equation}
After differentiating $w^{p}_{m,r}$ four times with respect to $\mu^{\dot\alpha}$ we obtain a polynomial of degree $2p-6$ in in $\mu^{\dot\alpha}=x^{\alpha\dot\alpha}\lambda_\alpha$.  The remaining $\lambda$-integral, stripped of the factors of $x^{\dot\alpha\alpha}$ is then
\begin{equation}
    L^r_{\alpha_1\ldots\alpha_{2p-6}}=\frac{1}{2\pi i}\oint \frac{\lambda_{\alpha_1}\ldots\lambda_{\alpha_{2p-6}}}{\lambda_0^{2p-4-r}\lambda_1^{r}}D\lambda= -\begin{pmatrix}
        2p-6\\
        r-1
    \end{pmatrix}\iota_{(\alpha_1}\ldots\iota_{\alpha_{r-1}}o_{\alpha_{r}}\ldots o_{\alpha_{2p-6})}\, .
\end{equation}
where $r\leq 2p-6$ gives a non-vanishing integral. This yields the formula for the field to be
\begin{equation}
    \Psi_{\dot\alpha\dot\beta\dot\gamma \dot\delta}= L^m_{\dot\alpha\dot \beta\dot\gamma\dot\delta \dot\alpha_1\ldots\dot \alpha_{2p-6}} L^r_{\alpha_1\ldots \alpha_{2p-6}}x^{\dot\alpha_1\alpha_1}\ldots x^{\dot\alpha_{2p-6}\alpha_{2p-6}}\, .
\end{equation}
This field is not asymptotically flat, but such polynomials in $x$ naturally arise when soft limits of momentum eigenstates are taken expanding $\e^{ik\cdot x}$ as $k\rightarrow 0$ and so correspond to the soft expansion.


\addcontentsline{toc}{section}{References}
\bibliographystyle{style}
\bibliography{Biblio}

\providecommand{\href}[2]{#2}\begingroup\raggedright\begin{thebibliography}{10}

\bibitem{Einstein:1916cd}
A.~Einstein, \emph{{Hamilton's Principle and the General Theory of Relativity}}, Sitzungsber. Preuss. Akad. Wiss. Berlin (Math. Phys. ) {\bf 1916} (1916)
1111--1116

\bibitem{Landau:1975pou}
L.~D. Landau and E.~M. Lifschits, {\em {The Classical Theory of Fields}}, vol.~Volume 2 of {\em Course of Theoretical Physics}.
\newblock Pergamon Press, Oxford,
1975

\bibitem{Dirac:1958jc}
P.~A.~M. Dirac, \emph{{Fixation of coordinates in the Hamiltonian theory of gravitation}}, Phys. Rev. {\bf 114} (1959)
924--930

\bibitem{Dirac:1959qwq}
P.~A.~M. Dirac, \emph{{Energy of the Gravitational Field}}, Phys. Rev. Lett. {\bf 2} (1959), no.~8,
368

\bibitem{Arnowitt:1959ah}
R.~L. Arnowitt, S.~Deser  and C.~W. Misner, \emph{{Dynamical Structure and Definition of Energy in General Relativity}}, Phys. Rev. {\bf 116} (1959)
1322--1330

\bibitem{Regge:1974zd}
T.~Regge and C.~Teitelboim, \emph{{Role of Surface Integrals in the Hamiltonian Formulation of General Relativity}}, Annals Phys. {\bf 88} (1974)
286

\bibitem{Penrose:1982wp}
R.~Penrose, \emph{{Quasilocal mass and angular momentum in general relativity}}, Proc. Roy. Soc. Lond. A {\bf 381} (1982)
53--63

\bibitem{penrose1984new}
R.~Penrose, \emph{New improved quasi-local mass and the Schwarzschild solutions}, Twistor Newsletter {\bf 18} (1984)
7--11

\bibitem{penrose1985suggested}
R.~Penrose, \emph{A suggested further modification to the quasi-local formula}, Twistor Newsletter {\bf 1985} (1985), no.~20,
7

\bibitem{penrose1984spinors}
R.~Penrose and W.~Rindler, {\em Spinors and space-time: Volume 2, Spinor and twistor methods in space-time geometry}, vol.~2.
\newblock Cambridge University Press,
1984

\bibitem{Mason_1989}
L.~J. Mason, \emph{A Hamiltonian interpretation of Penrose's quasi-local mass}, Classical and Quantum Gravity {\bf 6} (feb, 1989)
L7

\bibitem{Mason:1990Fra}
L.~Mason and J.~Frauendiener, {\em The Sparling 3-form, Ashtekar Variables and Quasi-local Mass}, p.~189–217.
\newblock London Mathematical Society Lecture Note Series.
\newblock Cambridge University Press,
1990.
\newblock

\bibitem{Witten:1981mf}
E.~Witten, \emph{{A Simple Proof of the Positive Energy Theorem}}, Commun. Math. Phys. {\bf 80} (1981)
381

\bibitem{Bartnik:1989zz}
R.~Bartnik, \emph{{New definition of quasilocal mass}}, Phys. Rev. Lett. {\bf 62} (1989)
2346--2348

\bibitem{Dougan:1991zz}
A.~J. Dougan and L.~J. Mason, \emph{{Quasilocal mass constructions with positive energy}}, Phys. Rev. Lett. {\bf 67} (1991)
2119--2122

\bibitem{Brown:1992br}
J.~D. Brown and J.~W. York, Jr., \emph{{Quasilocal energy and conserved charges derived from the gravitational action}}, Phys. Rev. D {\bf 47} (1993) 1407--1419,
\href{http://www.arXiv.org/abs/gr-qc/9209012}{{\tt gr-qc/9209012}}

\bibitem{Kijowski1997}
J.~Kijowski, \emph{A Simple Derivation of Canonical Structure and Quasi-local Hamiltonians in General Relativity}, General Relativity and Gravitation {\bf 29} (1997), no.~3,
307--343

\bibitem{Liu:2003bx}
C.-C.~M. Liu and S.-T. Yau, \emph{{Positivity of Quasilocal Mass}}, Phys. Rev. Lett. {\bf 90} (2003) 231102,
\href{http://www.arXiv.org/abs/gr-qc/0303019}{{\tt gr-qc/0303019}}

\bibitem{Wang:2008jy}
M.-T. Wang and S.-T. Yau, \emph{{Quasilocal mass in general relativity}}, Phys. Rev. Lett. {\bf 102} (2009) 021101,
\href{http://www.arXiv.org/abs/0804.1174}{{\tt 0804.1174}}

\bibitem{Szabados:2009eka}
L.~B. Szabados, \emph{{Quasi-Local Energy-Momentum and Angular Momentum in General Relativity}}, Living Rev. Rel. {\bf 12} (2009)
4

\bibitem{Iyer:1994ys}
V.~Iyer and R.~M. Wald, \emph{{Some properties of Noether charge and a proposal for dynamical black hole entropy}}, Phys. Rev. D {\bf 50} (1994) 846--864,
\href{http://www.arXiv.org/abs/gr-qc/9403028}{{\tt gr-qc/9403028}}

\bibitem{Wald:1999wa}
R.~M. Wald and A.~Zoupas, \emph{{A General definition of `conserved quantities' in general relativity and other theories of gravity}}, Phys. Rev. {\bf D61} (2000) 084027,
\href{http://www.arXiv.org/abs/gr-qc/9911095}{{\tt gr-qc/9911095}}

\bibitem{Barnich:2001jy}
G.~Barnich and F.~Brandt, \emph{{Covariant theory of asymptotic symmetries, conservation laws and central charges}}, Nucl. Phys. {\bf B633} (2002) 3--82,
\href{http://www.arXiv.org/abs/hep-th/0111246}{{\tt hep-th/0111246}}

\bibitem{Barnich:2003xg}
G.~Barnich, \emph{{Boundary charges in gauge theories: Using Stokes theorem in the bulk}}, Class. Quant. Grav. {\bf 20} (2003) 3685--3698,
\href{http://www.arXiv.org/abs/hep-th/0301039}{{\tt hep-th/0301039}}

\bibitem{Guevara:2021abz}
A.~Guevara, E.~Himwich, M.~Pate  and A.~Strominger, \emph{{Holographic symmetry algebras for gauge theory and gravity}}, JHEP {\bf 11} (2021) 152,
\href{http://www.arXiv.org/abs/2103.03961}{{\tt 2103.03961}}

\bibitem{Strominger:2021mtt}
A.~Strominger, \emph{{$w_{1+\infty}$ Algebra and the Celestial Sphere: Infinite Towers of Soft Graviton, Photon, and Gluon Symmetries}}, Phys. Rev. Lett. {\bf 127} (2021), no.~22,
221601

\bibitem{Adamo:2021lrv}
T.~Adamo, L.~Mason  and A.~Sharma, \emph{{Celestial $w_{1+\infty}$ Symmetries from Twistor Space}}, SIGMA {\bf 18} (2022) 016,
\href{http://www.arXiv.org/abs/2110.06066}{{\tt 2110.06066}}

\bibitem{Penrose:1976jq}
R.~Penrose, \emph{{The Nonlinear Graviton}}, Gen. Rel. Grav. {\bf 7} (1976)
171--176

\bibitem{Penrose:1976js}
R.~Penrose, \emph{{Nonlinear gravitons and curved twistor theory}}, Gen. Rel. Grav. {\bf 7} (1976)
31--52

\bibitem{Freidel:2021ytz}
L.~Freidel, D.~Pranzetti  and A.-M. Raclariu, \emph{{Higher spin dynamics in gravity and $w_{1 + \infty}$ celestial symmetries}},
\href{http://www.arXiv.org/abs/2112.15573}{{\tt 2112.15573}}

\bibitem{Geiller:2024bgf}
M.~Geiller, \emph{{Celestial $w_{1+\infty}$ charges and the subleading structure of asymptotically-flat spacetimes}},
\href{http://www.arXiv.org/abs/2403.05195}{{\tt 2403.05195}}

\bibitem{Cresto:2024fhd}
N.~Cresto and L.~Freidel, \emph{{Asymptotic Higher Spin Symmetries I: Covariant Wedge Algebra in Gravity}},
\href{http://www.arXiv.org/abs/2409.12178}{{\tt 2409.12178}}

\bibitem{Cresto:2024mne}
N.~Cresto and L.~Freidel, \emph{{Asymptotic Higher Spin Symmetries II: Noether Realization in Gravity}},
\href{http://www.arXiv.org/abs/2410.15219}{{\tt 2410.15219}}

\bibitem{Cresto:2025fbc}
N.~Cresto, {\em {Asymptotic Higher Spin Symmetries: Noether Realization {\&} Algebraic Structure in Einstein-Yang-Mills Theory}}.
\newblock PhD thesis, U. Waterloo (main), 2025.
\newblock
\href{http://www.arXiv.org/abs/2509.17137}{{\tt 2509.17137}}.
\newblock

\bibitem{Ashtekar:1981bq}
A.~Ashtekar and M.~Streubel, \emph{{Symplectic Geometry of Radiative Modes and Conserved Quantities at Null Infinity}}, Proc. Roy. Soc. Lond. {\bf A376} (1981)
585--607

\bibitem{Kmec:2024nmu}
A.~Kmec, L.~Mason, R.~Ruzziconi  and A.~Yelleshpur~Srikant, \emph{{Celestial Lw$_{1+\infty}$ charges from a twistor action}}, JHEP {\bf 10} (2024) 250,
\href{http://www.arXiv.org/abs/2407.04028}{{\tt 2407.04028}}

\bibitem{Mason:2007ct}
L.~J. Mason and M.~Wolf, \emph{{Twistor Actions for Self-Dual Supergravities}}, Commun. Math. Phys. {\bf 288} (2009) 97--123,
\href{http://www.arXiv.org/abs/0706.1941}{{\tt 0706.1941}}

\bibitem{Ruzziconi:2025fct}
R.~Ruzziconi and C.~Zwikel, \emph{{Celestial symmetries of black hole horizons}}, Phys. Rev. D {\bf 113} (2026), no.~4, L041504,
\href{http://www.arXiv.org/abs/2504.08027}{{\tt 2504.08027}}

\bibitem{Ruzziconi:2025fuy}
R.~Ruzziconi and C.~Zwikel, \emph{{Celestial Lw1+{\ensuremath{\infty}} symmetries and subleading phase space of null hypersurfaces}}, Phys. Rev. D {\bf 113} (2026), no.~4, 044067,
\href{http://www.arXiv.org/abs/2511.07525}{{\tt 2511.07525}}

\bibitem{Compere:2022zdz}
G.~Comp\`ere, R.~Oliveri  and A.~Seraj, \emph{{Metric reconstruction from celestial multipoles}}, JHEP {\bf 11} (2022) 001,
\href{http://www.arXiv.org/abs/2206.12597}{{\tt 2206.12597}}

\bibitem{Geroch:1970cc}
R.~P. Geroch, \emph{{Multipole moments. I. Flat space}}, J. Math. Phys. {\bf 11} (1970)
1955--1961

\bibitem{Geroch:1970cd}
R.~P. Geroch, \emph{{Multipole moments. II. Curved space}}, J. Math. Phys. {\bf 11} (1970)
2580--2588

\bibitem{Hansen:1974zz}
R.~O. Hansen, \emph{{Multipole moments of stationary space-times}}, J. Math. Phys. {\bf 15} (1974)
46--52

\bibitem{Blanchet:1985sp}
L.~Blanchet and T.~Damour, \emph{{Radiative gravitational fields in general relativity I. general structure of the field outside the source}}, Phil. Trans. Roy. Soc. Lond. A {\bf 320} (1986)
379--430

\bibitem{Blanchet:1986dk}
L.~Blanchet, \emph{{Radiative gravitational fields in general relativity. 2. Asymptotic behaviour at future null infinity}}, Proc. Roy. Soc. Lond. A {\bf 409} (1987)
383--399

\bibitem{Blanchet:1987wq}
L.~Blanchet and T.~Damour, \emph{{Tail Transported Temporal Correlations in the Dynamics of a Gravitating System}}, Phys. Rev. D {\bf 37} (1988)
1410

\bibitem{Blanchet:1992br}
L.~Blanchet and T.~Damour, \emph{{Hereditary effects in gravitational radiation}}, Phys. Rev. D {\bf 46} (1992)
4304--4319

\bibitem{Blanchet:2004re}
L.~Blanchet, T.~Damour  and B.~R. Iyer, \emph{{Surface-integral expressions for the multipole moments of post-Newtonian sources and the boosted Schwarzschild solution}}, Class. Quant. Grav. {\bf 22} (2005) 155--182,
\href{http://www.arXiv.org/abs/gr-qc/0410021}{{\tt gr-qc/0410021}}

\bibitem{Blanchet:2020ngx}
L.~Blanchet, G.~Comp{\`e}re, G.~Faye, R.~Oliveri  and A.~Seraj, \emph{{Multipole expansion of gravitational waves: from harmonic to Bondi coordinates}}, JHEP {\bf 02} (2021) 029,
\href{http://www.arXiv.org/abs/2011.10000}{{\tt 2011.10000}}

\bibitem{Compere:2017wrj}
G.~Comp\`ere, R.~Oliveri  and A.~Seraj, \emph{{Gravitational multipole moments from Noether charges}}, JHEP {\bf 05} (2018) 054,
\href{http://www.arXiv.org/abs/1711.08806}{{\tt 1711.08806}}

\bibitem{Thorne:1980ru}
K.~S. Thorne, \emph{{Multipole Expansions of Gravitational Radiation}}, Rev. Mod. Phys. {\bf 52} (1980)
299--339

\bibitem{Gursel:1983nkl}
Y.~G{\"u}rsel, \emph{{Multipole moments for stationary systems: The equivalence of the Geroch-Hansen formulation and the Thorne formulation}}, Gen. Rel. Grav. {\bf 15} (1983), no.~8,
737--754

\bibitem{Curtis:1978}
C.~G. E., \emph{Twistors and multipole moments}, Proc. R. Soc. Lond. {\bf A359} (1978)
133–149

\bibitem{Penrose:1986uia}
R.~Penrose and W.~Rindler, {\em {Spinors and Space-Time}}, vol.~2 of {\em Cambridge Monographs on Mathematical Physics}.
\newblock Cambridge Univ. Press, Cambridge, UK,
1986

\bibitem{Dougan_1992}
A.~J. Dougan, \emph{Quasi-local mass for spheres}, Classical and Quantum Gravity {\bf 9} (nov, 1992)
2461

\bibitem{Tod:1983waa}
K.~P. Tod, \emph{{Some Examples of Penrose's Quasi-Local Mass Construction}}, Proc. Roy. Soc. Lond. A {\bf 388} (1983), no.~1795,
457--477

\bibitem{Bailey:1990qn}
T.~N. Bailey and R.~J. Baston, eds., {\em {Twistors in mathematics and physics}}, vol.~156.
\newblock
1990

\bibitem{Huggett_Tod_1994}
S.~A. Huggett and K.~P. Tod, {\em An Introduction to Twistor Theory}.
\newblock London Mathematical Society Student Texts. Cambridge University Press, 2~ed.,
1994

\bibitem{Penrose:1984uia}
R.~Penrose and W.~Rindler, {\em {Spinors and Space-Time}}, vol.~1 of {\em Cambridge Monographs on Mathematical Physics}.
\newblock Cambridge Univ. Press, Cambridge, UK,
1984

\bibitem{Penrose1980GoldenON}
R.~Penrose, \emph{Null Hypersurface Initial Data for Classical Fields of Arbitrary Spin and for General Relativity}, General Relativity and Gravitation {\bf 12} (1980)
225--264

\bibitem{Newman:1962cia}
E.~T. Newman and T.~W.~J. Unti, \emph{{Behavior of Asymptotically Flat Empty Spaces}}, J. Math. Phys. {\bf 3} (1962), no.~5,
891

\bibitem{Kmec:2025ftx}
A.~Kmec, L.~Mason, R.~Ruzziconi  and A.~Sharma, \emph{{S-algebra in Gauge Theory: Twistor, Spacetime and Holographic Perspectives}},
\href{http://www.arXiv.org/abs/2506.01888}{{\tt 2506.01888}}

\bibitem{Mason:2009afn}
L.~J. Mason and D.~Skinner, \emph{{Gravity, Twistors and the MHV Formalism}}, Commun. Math. Phys. {\bf 294} (2010) 827--862,
\href{http://www.arXiv.org/abs/0808.3907}{{\tt 0808.3907}}

\bibitem{Ward1978}
R.~S. Ward, \emph{A class of self-dual solutions of Einstein’s equations}, Proceedings of the Royal Society of London. A. Mathematical and Physical Sciences {\bf 363} (1978), no.~1713, 289--295,
\href{http://www.arXiv.org/abs/https://royalsocietypublishing.org/doi/pdf/10.1098/rspa.1978.0170}{{\tt https://royalsocietypublishing.org/doi/pdf/10.1098/rspa.1978.0170}}

\bibitem{Dunajski:2000iq}
M.~Dunajski and L.~Mason, \emph{{HyperKahler hierarchies and their twistor theory}}, Commun. Math. Phys. {\bf 213} (2000) 641--672,
\href{http://www.arXiv.org/abs/math/0001008}{{\tt math/0001008}}

\bibitem{Dunajski:2003gp}
M.~Dunajski and L.~J. Mason, \emph{{Twistor theory of hyperKahler metrics with hidden symmetries}}, J. Math. Phys. {\bf 44} (2003) 3430--3454,
\href{http://www.arXiv.org/abs/math/0301171}{{\tt math/0301171}}

\bibitem{Barnich:2011mi}
G.~Barnich and C.~Troessaert, \emph{{BMS charge algebra}}, JHEP {\bf 12} (2011) 105,
\href{http://www.arXiv.org/abs/1106.0213}{{\tt 1106.0213}}

\bibitem{Barnich:2010xq}
G.~Barnich, \emph{{A Note on gauge systems from the point of view of Lie algebroids}}, AIP Conf. Proc. {\bf 1307} (2010), no.~1, 7--18,
\href{http://www.arXiv.org/abs/1010.0899}{{\tt 1010.0899}}

\bibitem{Barnich:2010eb}
G.~Barnich and C.~Troessaert, \emph{{Aspects of the BMS/CFT correspondence}}, JHEP {\bf 05} (2010) 062,
\href{http://www.arXiv.org/abs/1001.1541}{{\tt 1001.1541}}

\bibitem{Rahman:2019bmk}
A.~A. Rahman and R.~M. Wald, \emph{{Black Hole Memory}}, Phys. Rev. D {\bf 101} (2020), no.~12, 124010,
\href{http://www.arXiv.org/abs/1912.12806}{{\tt 1912.12806}}

\bibitem{Plebanski:1975wn}
J.~F. Plebanski, \emph{{Some solutions of complex Einstein equations}}, J. Math. Phys. {\bf 16} (1975)
2395--2402

\bibitem{Takasaki:1988cd}
K.~Takasaki, \emph{{An Infinite Number of Hidden Variables in Hyperkahler Metrics}}, J. Math. Phys. {\bf 30} (1989)
1515

\bibitem{Boyer:1985aj}
C.~P. Boyer and J.~F. Plebanski, \emph{{An Infinite Hierarchy of Conservation Laws and Nonlinear Superposition Principles for Self-dual Einstein Spaces}}, J. Math. Phys. {\bf 26} (1985)
229--234

\bibitem{Campiglia:2021srh}
M.~Campiglia and S.~Nagy, \emph{{A double copy for asymptotic symmetries in the self-dual sector}}, JHEP {\bf 03} (2021) 262,
\href{http://www.arXiv.org/abs/2102.01680}{{\tt 2102.01680}}

\bibitem{Miller:2025wpq}
N.~Miller, \emph{{Spacetime $Lw_{1+\infty}$ Symmetry and Self-Dual Gravity in Plebanski Gauge}},
\href{http://www.arXiv.org/abs/2504.07176}{{\tt 2504.07176}}

\bibitem{Anninos:2023epi}
D.~Anninos, D.~A. Galante  and C.~Maneerat, \emph{{Gravitational observatories}}, JHEP {\bf 12} (2023) 024,
\href{http://www.arXiv.org/abs/2310.08648}{{\tt 2310.08648}}

\bibitem{Anninos:2024wpy}
D.~Anninos, D.~A. Galante  and C.~Maneerat, \emph{{Cosmological observatories}}, Class. Quant. Grav. {\bf 41} (2024), no.~16, 165009,
\href{http://www.arXiv.org/abs/2402.04305}{{\tt 2402.04305}}

\bibitem{Anninos:2024xhc}
D.~Anninos, R.~Arias, D.~A. Galante  and C.~Maneerat, \emph{{Gravitational observatories in AdS$_{4}$}}, JHEP {\bf 07} (2025) 234,
\href{http://www.arXiv.org/abs/2412.16305}{{\tt 2412.16305}}

\bibitem{Adami:2024gdx}
H.~Adami, M.~Golshani, M.~M. Sheikh-Jabbari, V.~Taghiloo  and M.~H. Vahidinia, \emph{{Covariant phase space formalism for fluctuating boundaries}}, JHEP {\bf 09} (2024) 157,
\href{http://www.arXiv.org/abs/2407.03259}{{\tt 2407.03259}}

\bibitem{Sheikh-Jabbari:2025kjd}
M.~M. Sheikh-Jabbari and V.~Taghiloo, \emph{{AdS$_3$ Freelance Holography, A Detailed Analysis}},
\href{http://www.arXiv.org/abs/2510.10692}{{\tt 2510.10692}}

\bibitem{Ruzziconi:2026bix}
R.~Ruzziconi, \emph{{Carrollian Physics and Holography}},
\href{http://www.arXiv.org/abs/2602.02644}{{\tt 2602.02644}}

\bibitem{Cordova:2018ygx}
C.~C\'ordova and S.-H. Shao, \emph{{Light-ray Operators and the BMS Algebra}}, Phys. Rev. D {\bf 98} (2018), no.~12, 125015,
\href{http://www.arXiv.org/abs/1810.05706}{{\tt 1810.05706}}

\bibitem{Hu:2023geb}
Y.~Hu and S.~Pasterski, \emph{{Detector operators for celestial symmetries}}, JHEP {\bf 12} (2023) 035,
\href{http://www.arXiv.org/abs/2307.16801}{{\tt 2307.16801}}

\bibitem{Himwich:2025ekg}
E.~Himwich and M.~Pate, \emph{{Light-ray Operators and the ${\rm w}_{1+\infty}$ Algebra}},
\href{http://www.arXiv.org/abs/2512.18973}{{\tt 2512.18973}}

\bibitem{Sheta:2025oep}
A.~Sheta, A.~Strominger, A.~Tropper  and H.~Wei, \emph{{Soft Algebras in AdS$_4$ from Light Ray Operators in CFT$_3$}},
\href{http://www.arXiv.org/abs/2601.00096}{{\tt 2601.00096}}

\bibitem{Strominger:2026yrh}
A.~Strominger and H.~Wei, \emph{{EVERY CFT$_3$ HAS AN $ \mathcal{L}_\Lambda w_{1+\infty}$ SYMMETRY}},
\href{http://www.arXiv.org/abs/2603.26459}{{\tt 2603.26459}}

\bibitem{Freidel:2023gue}
L.~Freidel, D.~Pranzetti  and A.-M. Raclariu, \emph{{On infinite symmetry algebras in Yang-Mills theory}},
\href{http://www.arXiv.org/abs/2306.02373}{{\tt 2306.02373}}

\bibitem{Cresto:2025bfo}
N.~Cresto, \emph{{Asymptotic Higher Spin Symmetries III: Noether Realization in Yang-Mills Theory}},
\href{http://www.arXiv.org/abs/2501.08856}{{\tt 2501.08856}}

\bibitem{Taylor:2023ajd}
T.~R. Taylor and B.~Zhu, \emph{{w1+\ensuremath{\infty} Algebra with a Cosmological Constant and the Celestial Sphere}}, Phys. Rev. Lett. {\bf 132} (2024), no.~22, 221602,
\href{http://www.arXiv.org/abs/2312.00876}{{\tt 2312.00876}}

\bibitem{Bittleston:2024rqe}
R.~Bittleston, G.~Bogna, S.~Heuveline, A.~Kmec, L.~Mason  and D.~Skinner, \emph{{On AdS$_{4}$ deformations of celestial symmetries}}, JHEP {\bf 07} (2024) 010,
\href{http://www.arXiv.org/abs/2403.18011}{{\tt 2403.18011}}

\bibitem{Freidel:2026iia}
L.~Freidel and A.~Sharma, \emph{{Celestial 1-form symmetries}},
\href{http://www.arXiv.org/abs/2604.11602}{{\tt 2604.11602}}

\bibitem{Newman:1961qr}
E.~Newman and R.~Penrose, \emph{{An Approach to gravitational radiation by a method of spin coefficients}}, J. Math. Phys. {\bf 3} (1962)
566--578

\bibitem{Geroch:1973am}
R.~P. Geroch, A.~Held  and R.~Penrose, \emph{{A space-time calculus based on pairs of null directions}}, J. Math. Phys. {\bf 14} (1973)
874--881

\end{thebibliography}\endgroup

\end{document}